\documentclass[aps,prd,notitlepage,nofootinbib,preprintnumbers,superscriptaddress,10pt]{revtex4-2}
\usepackage{here}
\usepackage{graphicx}
\usepackage{amsmath,amsthm,amssymb}
\usepackage{bm}
\usepackage{color}

\usepackage{amsfonts}
\usepackage{dcolumn}
\usepackage{hyperref}
\allowdisplaybreaks[1]
\usepackage{stackengine}

\newcommand{\be}{\begin{eqnarray}}
\newcommand{\ee}{\end{eqnarray}}

\newcommand{\bem}{\begin{bmatrix}}
\newcommand{\eem}{\end{bmatrix}}

\newcommand{\bH}{\bar{H}}
\newcommand{\bK}{\bar{K}}

\allowdisplaybreaks
\begin{document}

\preprint{YITP-22-157, IPMU22-0068}

\title{
Gravitational collapse and odd-parity black hole perturbations in Minimal Theory of Bigravity
}

\author{Masato Minamitsuji}
\affiliation{Centro de Astrof\'{\i}sica e Gravita\c c\~ao  - CENTRA, Departamento de F\'{\i}sica, Instituto Superior T\'ecnico - IST, Universidade de Lisboa - UL, Avenida Rovisco Pais 1, 1049-001 Lisboa, Portugal}

\author{Antonio De Felice}
\affiliation{Center for Gravitational Physics and Quantum Information, Yukawa Institute for Theoretical Physics, Kyoto University, 606-8502, Kyoto, Japan}

\author{Shinji Mukohyama}
\affiliation{Center for Gravitational Physics and Quantum Information, Yukawa Institute for Theoretical Physics, Kyoto University, 606-8502, Kyoto, Japan}
\affiliation{Kavli Institute for the Physics and Mathematics of the Universe (WPI), The University of Tokyo Institutes for Advanced Study, The University of Tokyo, Kashiwa, Chiba 277-8583, Japan}

\author{Michele Oliosi}
\affiliation{Center for Gravitational Physics and Quantum Information, Yukawa Institute for Theoretical Physics, Kyoto University, 606-8502, Kyoto, Japan}

\begin{abstract}
We investigate dynamical properties of static and spherically symmetric systems in the self-accelerating branch of the Minimal Theory of Bigravity (MTBG). 
In the former part, we study the gravitational collapse of pressureless dust and find special solutions, where, in both the physical and fiducial sectors, the exterior and interior spacetime geometries are given by the Schwarzschild spacetimes and the Friedmann-Lema\^itre-Robertson-Walker universes dominated by pressureless dust, respectively, with specific time slicings. In the case where the Lagrange multipliers are trivial and have no jump across the matter interfaces in both the physical and fiducial sectors, the junction conditions across them remain the same as those in general relativity (GR). For simplicity, we foliate the interior geometry by homogeneous and isotropic spacetimes. For a spatially flat interior universe, we foliate the exterior geometry by a time-independent flat space, while for a spatiallycurved interior universe, we foliate the exterior geometry by a time-independent space with deficit solid angle. Despite the rather restrictive choice of foliations, we find interesting classes of exact solutions that represent gravitational collapse in MTBG. In the spatially flat case, under a certain tuning of the initial condition, we find exact solutions of matter collapse in which the two sectors evolve independently. In the spatially closed case, once the matter energy densities and the Schwarzschild radii are tuned between the two sectors, we find exact solutions that correspond to the Oppenheimer-Snyder model in GR. 
In the latter part, we study odd-parity perturbations of the Schwarzschild$-$de Sitter solutions written in the spatially flat coordinates. For the higher-multipole modes $\ell\geq2$, we find that, in general, the system reduces to that of four physical modes, where two of them are dynamical and the remaining two are shadowy, i.e., satisfying only elliptic equations. In the case where the ratio of the lapse functions between the physical and fiducial sectors is equal to a constant determined by the parameters of the theory, the two dynamical modes are decoupled from each other but sourced by one of the shadowy modes. Otherwise, the two dynamical modes are coupled to each other and sourced by the two shadowy modes. At least for the cases of collapse described in this paper, we find that the ratio of the lapse functions is determined by the properties of the collapse itself. On giving appropriate boundary conditions to the shadowy modes so as to not strongly backreact/influence the dynamics of the master variables, in the high frequency and short wavelength limits, we show that the two dynamical modes do not suffer from ghost or gradient instabilities. For the dipolar mode $\ell=1$, the two copies of the slow-rotation limit of the Kerr$-$de Sitter metrics cannot be a solution in the self-accelerating branch, unless the mass and spin of black holes and effective cosmological constants are tuned to be the same. Therefore, deviation from GR is expected for rotating black holes in the self-accelerating branch of MTBG. 
\end{abstract}
\maketitle

\section{Introduction}
\label{sec1}

Modified gravity theories have been proposed from theoretical and
observational points of view \cite{Clifton:2011jh,Berti:2015itd}.
While general relativity (GR) has passed all the experimental tests so far in
the weak-field regime~\cite{Will:2014kxa}, a new window for testing
gravitational theories has opened with the dawn of gravitational-wave
astronomy~\cite{Berti:2018vdi,Berti:2018cxi}.  Massive and bigravity
theories are promising candidates to elucidate the origin of the
present day's cosmic acceleration.  The first model of massive gravity
that is free from the ghost instability was formulated in the context
of the linearized gravity by Fierz and Pauli \cite{Fierz:1939ix}.  The
attempts to extend the Fierz-Pauli theory to the nonlinear case have
not been successful for a long time, because of the reintroduction of
a ghost degree of freedom (DOF) through nonlinear higher-derivative
interactions in the scalar sector, which is known as the
Boulware–Deser (BD) ghost \cite{Boulware:1972yco}.  The first
construction of nonlinear massive gravity theory free from the BD
ghost was eventually provided by de Rham, Gabadadze, and Tolley
(dRGT)~\cite{deRham:2010kj}.  dRGT massive gravity was then extended
to a bigravity theory by Hassan and Rosen (HR) \cite{Hassan:2011zd} by
promoting the fiducial metric to be a dynamical field.~\footnote{HR bigravity,
however, still suffers from the BD ghost, when matter is coupled to
both the physical and fiducial metrics \cite{Yamashita:2014fga,deRham:2014fha}.
See also \cite{Gumrukcuoglu:2015nua}.}

The Minimal Theory of Massive Gravity (MTMG)
\cite{DeFelice:2015hla,DeFelice:2015moy} is a different extension of
dRGT massive gravity.  In MTMG, there are two tensorial DOFs as in GR.
While the four-dimensional diffeomorphism invariance is explicitly
broken, as the fiducial metric is given as a function of the
coordinates, the absence of otherwise problematic scalar and vector
DOFs makes it easy for the theory to be consistent with
experimental and observational data.  For instance, while MTMG shares
the same homogeneous and isotropic background cosmology with dRGT
massive gravity, including the existence of two separate branches, it
does not suffer from the instabilities which appear in dRGT massive
gravity \cite{Gumrukcuoglu:2011zh,DeFelice:2012mx,Fasiello:2012rw},
thanks indeed to the absence of the extra scalar and vector DOFs.  The cosmological solutions in MTMG are, as in dRGT, classified into the normal and self-accelerating branches
\cite{DeFelice:2016ufg,Hagala:2020eax,DeFelice:2021trp}.
Any solution in GR with or without matter written in the spatially flat coordinates has been shown to be a solution in the self-accelerating branch of MTMG \cite{DeFelice:2018vza}.

Using a construction similar to MTMG, HR bigravity was then extended
to the Minimal Theory of Bigravity (MTBG) \cite{DeFelice:2020ecp},
where the four-dimensional diffeomorphism invariance is broken down to
the three-dimensional one and time-reparametrization invariance.
While, by construction, MTBG shares the same background cosmological
dynamics with HR bigravity, the number of propagating DOFs are down to
four, two tensorial DOFs in the physical sector and the other two in
the fiducial sector.
The absence of the extra scalar and vector DOFs in MTBG implies the absence of ghost or gradient
instabilities associated with them \cite{Garcia-Saenz:2021uyv,Silva:2021jya,Demirboga:2021nrc}.
In the case where the matter sectors are coupled to the two metrics
separately and independently, cosmology in MTBG has been investigated
in Ref.~\cite{DeFelice:2020ecp}, which showed that in the
self-accelerating branch both background cosmology and dynamics of
the scalar and vector cosmological linear perturbations behave in the same
way as in two copies of GR. On the other hand, two of the tensor modes
acquire a nonzero mass. In the normal branch, another nontrivial
difference arises in addition to the presence of two massive tensor
modes: deviation in the dynamics for both background and the
scalar sector from GR could be already observed 
and possibly lead to nontrivial interesting phenomenology.
The absence of extra scalar and vector modes in MTBG also helps 
in developing a new production scenario of spin-2 dark matter based on the transition from an anisotropic fixed point solution to an isotropic one within axisymmetric Bianchi type-I universe~\cite{Manita:2022tkl}.

MTMG and MTBG are theories with constraints.
By construction, they possess
constraints out of which the unwanted modes can be removed
nonlinearly from the theory. This nice feature is motivated by
the fact that whatever new theory of gravity we introduce should
reproduce the whole range of gravitational phenomenology we already
know. However, the introduction of such constraints is a nontrivial
task. On one hand, the choice of constraints is not unique. Other MTMGs and/or 
MTBGs can be,  in principle, introduced, and {\it a priori} it is not an easy
task to discriminate one choice of constraints from another. 
On the other hand, each set of
constraints we pick up for MTMG, including the one we will study in this
paper, may restrict the space of solutions of the theory in a different way. While this is
to be expected, by the same nature of a constraint, once more, it is
not clear, {\it a priori}, how to determine which configurations are not
allowed to be solutions of the specific choice of theory.

Despite the fact that this fact by itself would not be already introducing new and/or
unexpected phenomenology, there is also another feature
shared by these theories: the presence of shadowy modes \cite{DeFelice:2018ewo}. In fact, as we will
also see in this paper, the constraints lead to the presence of modes
that only exist on a spacelike three-dimensional hypersurface. 
These modes are present not only in MTMG and MTBG, but also in other theories: for instance
Refs. \cite{DeFelice:2020eju,DeFelice:2021hps}, which also allow for the existence of shadowy modes. They all share the property that they
satisfy elliptic equations for which some appropriate
boundary conditions need to be imposed. As a consequence, the very
existence of such modes introduces a preferred frame for the theory,
i.e., the frame on which their equations of motion are manifestly elliptic.
This leads to the consequence that, even for a given GR solution, different
choices of frames may or may not be compatible with the existence
of the shadowy modes; i.e.,\ some slicings of GR solutions may be
allowed as solutions by the theory, but others may not. This fact leads,
for instance, to the absence of the Birkhoff theorem for these
theories. Therefore, it is of primary importance to try to find all the
solutions for a given symmetry or physical configuration.

Some of these foliation-and-branch specific solutions of MTBG have already been found. 
For instance, in a previous work \cite{Minamitsuji:2022vfv}, we have
investigated the static and spherically symmetric solutions in the
self-accelerating and normal branches of MTBG.  We have shown that a
pair of Schwarzschild$-$de Sitter spacetimes with different cosmological
constants and black hole masses written in the spatially flat
[Gullstrand–Painlev\'e (GP)] coordinates is a solution in the
self-accelerating branch of MTBG.  
In the case where the two matter
sectors are coupled to the two metrics separately and independently,
the self-accelerating branch also admits static and spherically
symmetric copies of two GR solutions
written in the spatially flat coordinates, which include neutron stars
with arbitrary equations of state of matter. 
In addition to these solutions we
also find nontrivial solutions that correspond to Schwarzschild$-$de Sitter solutions, 
written in nonstandard coordinates.
It should be added that these solutions have been found by imposing trivial configurations/profiles for the Lagrangian multipliers,
introduced in the theory so as to impose the same above-mentioned
constraints that define the theory itself. The same approach will be
taken in this paper. It is not clear what kinds of new solutions are to
be expected once we move away from this simple ansatz for the
Lagrange multipliers.

On the other hand, in the normal branch, while the spatially flat
coordinates of the Schwarzschild$-$de Sitter metrics cannot be
solutions, those written in the coordinates with $D^i D_i K=0$, where
$D_i$ is the covariant derivative associated with the spatial metric
and $K$ is the trace of the extrinsic curvature tensor on the constant-time hypersurfaces, could be solutions, provided that the two metrics
are parallel, showing also for this case a severe fine-tuning of the
possible configurations.
An example for such solutions were given, i.e.,\ the ones with
$K={\rm constant}$, which turn out to be also solutions for the case of a single physical metric, in the context of 
the VCDM models that replace the cosmological constant $\Lambda$ in the $\Lambda$CDM ($\Lambda$ cold dark
matter) model with a function  of an auxiliary field $\phi$, $V(\phi)$
\cite{DeFelice:2020eju,DeFelice:2020cpt,DeFelice:2020onz,DeFelice:2022uxv}.

As the next step, in this paper we will study the dynamical processes
only in the self-accelerating branch of MTBG and investigate whether
any deviation from the GR predictions can be observed with the
inclusion of time dependence at the level of the background solutions
and perturbations, respectively.

Our analysis will be composed of two
parts.  In the former, we will study spherical gravitational collapse
of pressureless dust in the self-accelerating branch of MTBG.  In the
case where the two species of matter are independently coupled to the
two metrics separately and independently, we will derive special 
solutions describing spherical collapse of pressureless dust in both the
physical and fiducial sectors, where the interior regions of the
matter distribution are described by the spatially flat or closed
Friedmann-Lema\^itre-Robertson-Walker (FLRW) universes and the exterior geometry
is the Schwarzschild spacetime with specific time slicings.
For simplicity, we foliate the interior geometry by homogeneous and isotropic spacetimes. 
For a spatially flat interior universe, we foliate the exterior geometry by a time-independent flat space,
while for a spatiallycurved interior universe, we foliate the exterior geometry by a time-independent space with deficit solid angle. Despite the rather restrictive choice of foliations, we find interesting classes of exact solutions that represent gravitational collapse in MTBG. 
The spatially flat solutions are obtained under certain tuning 
of the initial conditions of the collapse. 
The spatially closed case corresponds to an extension of the
Oppenheimer-Snyder model in GR
\cite{Oppenheimer:1939ue,Kanai:2010ae,Blau} to MTBG.  We will show
that, in this case, gravitational collapse happens in
the physical and fiducial sectors in the same manner as in two copies
of GR independently, under certain tuning of the matter energy densities and 
Schwarzschild radii between the two sectors. 
Needless to say, these tunings reflect the fact that we restrict our considerations to rather specific choice of time slicings.

In the latter part,
we will investigate the odd-parity perturbations of the Schwarzschild$-$de Sitter solutions 
written in the spatially flat coordinates
in the self-accelerating branch of MTBG.
For the higher-multipolar modes $\ell\geq 2$,
where $\ell$ represents the angular multipole moment,
we derive the master equations governing the dynamics of
the odd-parity perturbations in both physical and fiducial sectors in MTBG.
We will find that there are four physical modes,
where two of them are dynamical and 
the remaining two are shadowy,
which obey elliptic equations and are fixed by the spatial boundary conditions
on each step of the time evolution \cite{DeFelice:2018ewo, DeFelice:2021hps}.
In the case where the ratio of the lapse functions in the 
physical and fiducial sectors are equal to a constant determined by the free parameters of theory, 
the two dynamical modes are decoupled and sourced by one of the shadowy modes. 
Otherwise, the two dynamical modes are coupled to each other and sourced by the two shadowy modes.
In the high frequency and short wavelength limits, we also verify the modes are not suffering from ghost or gradient instabilities, provided we can neglect backreactions from the shadowy modes, by giving appropriate boundary conditions to them.
For the dipolar mode $\ell=1$,
we will show that the two copies of the slow-rotation limit of the Kerr$-$de Sitter metrics,
in general, cannot be a solution in MTBG,
unless the mass and spin of the black holes and the effective cosmological constants 
in the two sectors are tuned to be the same.
Therefore, deviation from GR is expected for rotating black holes in the self-accelerating branch of MTBG.
Readers should also refer to the last paragraph in Sec.IV of \cite{DeFelice:2019jxs} for a similar statement in the context of MTMG.

The structure of this paper is as follows:
In Sec.~\ref{sec2}, we review the MTBG theory.
In Sec.~\ref{sec3},
we derive the exact solutions
that describe gravitational collapse of pressureless dust
where the metrics in the interior regions filled with matter are written 
in terms of the spatially flat and spatially closed FLRW universes, respectively.
In Sec.~\ref{sec4},
we investigate the odd-parity gravitational perturbations
of the Schwarzschild$-$de Sitter solutions written in the spatially flat coordinates
in the self-accelerating branch of MTBG.
The last section 
is devoted to giving a brief summary and conclusion.

\section{The Minimal Theory of Bigravity}
\label{sec2}

In this section, we review MTBG following Refs.\ \cite{DeFelice:2020ecp,Minamitsuji:2022vfv}.

\subsection{Metrics}

As in HR bigravity,
MTBG is composed of two metric sectors, 
the physical and fiducial metrics denoted by $g_{\mu\nu}$ and $f_{\mu\nu}$, respectively.
Choosing the unitary gauge,
the two metrics
$g_{\mu\nu}$ and $f_{\mu\nu}$
are expressed in the Arnowitt-Deser-Misner form, respectively, as
\begin{eqnarray}
&&
\label{adm}
g_{\mu\nu}dx^\mu dx^\nu
=-N^2dt^2
+\gamma_{ij}
\left(dx^i+N^i dt\right)
\left(dx^j+N^j dt\right),
\nonumber\\
&&
f_{\mu\nu}dx^\mu dx^\nu
=-M^2dt^2
+\phi_{ij}
\left(dx^i+M^i dt\right)
\left(dx^j+M^j dt\right),
\end{eqnarray}
where 
$t$ and $x^i$ ($i=1,2,3$) are the temporal and spatial coordinates, 
and 
$(N,N^i,\gamma_{ij})$ 
and
$(M,M^i,\phi_{ij})$ 
are the sets of 
the lapse function, shift vector, and spatial metric 
on the constant $t$ hypersurfaces
for the metrics $g_{\mu\nu}$ and $f_{\mu\nu}$,
respectively.
The extrinsic curvature tensors on constant-time hypersurfaces
are defined by 
\begin{eqnarray}
\label{extrinsic}
&&
K_{ij}
:=
\frac{1}{2N}
\left(
\partial_t \gamma_{ij}
-D_i N_j
-D_j N_i
\right),
\\
\label{extrinsic_f}
&&
\Phi_{ij}
:=
\frac{1}{2M}
\left(
\partial_t \phi_{ij}
-{\tilde D}_i M_j
-{\tilde D}_j M_i
\right),
\end{eqnarray}
where $D_i$ and ${\tilde D}_i$ are covariant derivatives 
associated with the spatial metrics $\gamma_{ij}$ and $\phi_{ij}$, respectively.

\subsection{Action and equations of motion}

In the unitary gauge, 
the action of MTBG \cite{DeFelice:2020ecp} is then given by 
\begin{eqnarray}
\label{action}
S
&=&
\frac{1}{2\kappa^2}
\int d^4x 
\left(
{\cal L}_g
\left[N,N^i,\gamma_{ij}; M,M^i,\phi_{ij};
\lambda, {\bar \lambda}, \lambda^i
\right]
+
{\cal L}_m
\left[
N,N^i,\gamma_{ij};
M,M^i,\phi_{ij};
\Psi
\right]
\right),
\end{eqnarray}
where
$\kappa^2$ represents the gravitational constant in the physical sector,
${\cal L}_g$ and ${\cal L}_m$
represent the gravitational and matter parts of the Lagrangian,
respectively,
$\lambda$, ${\bar \lambda}$, and $\lambda^i$
are two scalar and one spatial-vector Lagrange multipliers,
and 
$\Psi$ represents the matter sector.

The gravitational Lagrangian ${\cal L}_g$ of MTBG is further decomposed into
the precursor and constraint parts as 
\begin{eqnarray}
\label{lag}
{\cal L}_g
&:=&
{\cal L}_{\rm pre}
\left[
N,N^i,\gamma_{ij};
M,M^i,\phi_{ij}
\right]
+
{\cal L}_{\rm con}
\left[N,N^i,\gamma_{ij};
M,M^i,\phi_{ij};
\lambda, {\bar \lambda}, \lambda^i
\right],
\end{eqnarray}
with
\begin{eqnarray}
\label{lag2}
{\cal L}_{\rm pre}
&:=&
\sqrt{-g}R[g]
+
\tilde{\alpha}^2
\sqrt{-f} R[f]
-
m^2
\left(
N\sqrt{\gamma}{\cal H}_0
+
M\sqrt{\phi}\tilde{\cal H}_0
\right),
\nonumber
\\
{\cal L}_{\rm con}
&:=&
\sqrt{\gamma}
\alpha_{1\gamma}
\left(
\lambda
+
\Delta_\gamma
{\bar \lambda}
\right)
+
\sqrt{\phi}
\alpha_{1\phi}
\left(
\lambda
-
\Delta_\phi
{\bar \lambda}
\right)
+
\sqrt{\gamma}
\alpha_{2\gamma}
\left(
\lambda
+
\Delta_\gamma
 {\bar\lambda}
\right)^2
+
\sqrt{\phi}
\alpha_{2\phi}
\left(
\lambda
-\Delta_\phi {\bar\lambda}
\right)^2
\nonumber\\
&-&
m^2
\left[
\sqrt{\gamma}
  U^i{}_k D_i \lambda^k
-\beta 
\sqrt{\phi}
{\tilde U}_k{}^i {\tilde D}_i \lambda^k
\right],
\end{eqnarray}
and 
\begin{eqnarray}
\alpha_{1\gamma}
&:=&
-m^2
 U^p{}_q K^q{}_p,
\quad 
\alpha_{1\phi}
:=
m^2 {\tilde U}^p{}_q
    \Phi^q{}_p,
\nonumber\\
\alpha_{2\gamma}
&:=&
\frac{m^4 }{4N}
\left(
U^p{}_q-\frac{1}{2}U^k{}_k \delta^p{}_q
\right)
U^q{}_p,
\quad 
\alpha_{2\phi}
:=
\frac{m^4 }{4M\tilde{\alpha}^2}
\left(
{\tilde U}_q{}^p
-\frac{1}{2}{\tilde U}_k{}^k \delta_q{}^p
\right)
{\tilde U}_p{}^q,
\end{eqnarray}
where the constant $\tilde{\alpha}$ represents the ratio of the two
gravitational constants, $m$ is a parameter with dimensions of mass
that can be related with the graviton mass, $\beta$ is a constant,
and 
$\gamma:={\rm det} (\gamma_{ij})$ and $\phi:={\rm det} (\phi_{ij})$
are the determinants of the three-dimensional spatial metrics
$\gamma_{ij}$ and $\phi_{ij}$ respectively. We also note that $K^q{}_p=\gamma^{qr}K_{rp}$ and $\Phi^q{}_p=\phi^{qr}\Phi_{rp}$. 
Furthermore, ${\cal H}_0$
and $\tilde{\cal H}_0$ are defined by
$ {\cal H}_0 := \sum_{n=0}^3 c_{4-n} e_n( {\cal K})$ and
$ \tilde{\cal H}_0 := \sum_{n=0}^3 c_{n} e_n( \tilde{\cal K})$ with
\begin{eqnarray}
e_0({\cal K})
=1,
\quad
e_1({\cal K})
=
\left[
{\cal K}
\right],
\quad 
e_{2} ({\cal K})
=
\frac{1}{2}
\left(
\left[
{\cal K}
\right]^2
-
\left[
{\cal K}^2
\right]
\right),
\quad
e_{3} ({\cal K})
=
{\rm det} 
({\cal K}),
\end{eqnarray}
and similar for $e_n (\tilde{\cal K})$
with ${\cal K}^i{}_k$ and $\tilde{\cal K}_k{}^i$
characterized by
${\cal K}^i{}_k
{\cal K}^k{}_j
=
{\gamma}^{ik}
\phi_{kj}$
and 
${\cal {\tilde K}}_j{}^k
{\cal {\tilde K}}_k{}^i
=
\gamma_{jk}
{\phi}^{ki}$;
$\Delta_\gamma :=\gamma^{ij} D_i D_j$
and 
$\Delta_\phi :=\phi^{ij} {\tilde D}_i {\tilde D}_j$
are
the Laplacian operators in the physical and fiducial sectors,
respectively,
and 
the spatial tensors
$U^i{}_j$ and ${\tilde U}{}_j{}^i$ are defined by
\begin{eqnarray}
U^i{}_j
&:=&
\frac{1}{2}
\sum_{n=1}^3 c_{4-n}
\left(
U_{(n)}{}^i{}_j
+\gamma^{ik}\gamma_{j\ell}
U_{(n)}{}^\ell{}_k
\right),
\nonumber\\
{\tilde U}_j{}^i
&:=&
\frac{1}{2}
\sum_{n=1}^3 c_{n}
\left(
{\tilde U}_{(n)j}{}^i
+\phi^{ik}\phi_{j\ell}
{\tilde U}_{(n)k}{}^\ell
\right),
\end{eqnarray}
with
$U_{(n)}{}^i{}_k
:=
\frac{\partial e_n ({\cal K})}
        {\partial {\cal K}^k{}_i}$
and  
${\tilde U}_{(n)k}{}^i
:=
\frac{\partial e_n ({\cal {\tilde K}})}
        {\partial \tilde{\cal K}^k{}_i}$,
and $c_j$ ($j=0,1,2,3,4$) being dimensionless coupling constants.
Although one of  $c_j$ ($j=0,1,2,3,4$) may be set to unity
by a redefinition of the mass parameter $m$,
in order to discuss the various limiting cases of each coefficient, 
we keep them as independent parameters.

Variation of the action \eqref{action}
with respect to $N$, $N^i$, $\gamma_{ij}$, $M$, $M^i$, and $\phi_{ij}$
provides the equations  of motion for the two metrics,
and 
variation with respect to $\Psi$
provides the equations of motion for the matter field.
On the other hand,
variation with respect to 
the Lagrange multipliers
$\lambda$, ${\bar \lambda}$, and $\lambda^i$
gives the constraint equations
\begin{eqnarray}
\label{cons2}
&&
  \sqrt{\gamma}\alpha_{1\gamma}
+\sqrt{\phi}\alpha_{1\phi}
+
2\sqrt{\gamma} \alpha_{2\gamma}
\left(
\lambda+
\Delta_\gamma
{\bar \lambda}
\right)
+
2\sqrt{\phi} \alpha_{2\phi}
\left(
\lambda
-
\Delta_\phi
{\bar \lambda}
\right)=0,
\\
&&
\label{cons3}
\sqrt{\gamma}
\Delta_{\gamma} \alpha_{1\gamma}
-
\sqrt{\phi}
\Delta_{\phi}\alpha_{1\phi}
+
2\sqrt{\gamma} 
\Delta_\gamma
\left[
\alpha_{2\gamma}
\left(
\lambda
+
\Delta_\gamma{\bar \lambda}
\right)
\right]
-
2\sqrt{\phi} 
\Delta_\phi
\left[
\alpha_{2\phi}
\left(
\lambda
-
\Delta_\phi
{\bar \lambda}
\right)
\right]
=0,
\\
\label{const}
&&
\sqrt{\gamma} D_p U^p{}_q
-\beta
\sqrt{\phi} {\tilde D}_p {\tilde U}_q{}^p
=0.
\end{eqnarray}

\subsection{Matter}

We also assume that the matter sector $\Psi$ is decomposed into those in
the physical and fiducial sectors
$\Psi_g$ and $\Psi_f$ and the matter Lagrangian ${\cal L}_m$ is given by 
 \be
\label{matter_action}
{\cal L}_m
\left[
N,N^i,\gamma_{ij};
M,M^i,\phi_{ij};
\Psi
\right]
= 
{\cal L}_{m,g}
\left[
N,N^i,\gamma_{ij};
\Psi_g
\right]
+
{\cal L}_{m,f}
\left[
M,M^i,\phi_{ij};
\Psi_f
\right].
\ee
The two matter sectors 
${\cal L}_{m,g}$ and ${\cal L}_{m,f}$
are individually coupled to the metrics $g_{\mu\nu}$ and $f_{\mu\nu}$, respectively,
and they are not directly coupled to each other.
In other words,
the two matter sectors can interact only through gravitation.

\section{Gravitational collapse of pressureless dust}
\label{sec3}

In this section, 
we consider two models of collapsing matter in the self-accelerating branch of MTBG.
We assume the existence of matter surfaces that move inward in both the sectors separately,
and the spacetime metrics inside the matter surfaces are described by
either spatially flat or closed FLRW universes composed of pressureless dust.
The spatially closed case corresponds to an extension of the Oppenheimer-Snyder model in GR 
\cite{Oppenheimer:1939ue,Kanai:2010ae,Blau}
to MTBG.

\subsection{Collapse in the spherically symmetric spacetimes in the spatially flat coordinates}
\label{sec3a}

First, we consider the case in which the spherically symmetric spacetimes
can be expressed in the spatially flat coordinates
\begin{eqnarray}
\label{generalg}
g_{\mu\nu}
dx^\mu dx^\nu
&=&
-
dt^2
+
\left(dr+N^r (t,r)dt\right)^2
+
r^2 
\theta_{ab} d\theta^a d\theta^b,
\\
\label{generalf}
f_{\mu\nu}
dx^\mu dx^\nu
&=&
C_0^2\left[
-b^2 dt^2
+
\left(dr+N^r_{(f)} (t,r) dt\right)^2
+
r^2 
\theta_{ab}d\theta^a d\theta^b
\right],
\end{eqnarray}
where 
$b>0$ and $C_0>0$ are constants,
$t$, $r$, and $\theta^a=(\theta,\varphi)$
are the temporal, radial, and angular coordinates, respectively,
and 
$\theta_{ab}$ represents the metric of the unit two-sphere.
In vacuum GR
with the vanishing cosmological constant, since $N^r_{(f)}=\sqrt{r_{(f)}/r}$,
by a redefinition of ${\tilde r}=\frac{r}{b}$, $C_0\to C_0/b$, and $r_{(f)}\to r_{(f)} b^3$
the metric \eqref{generalf} can be brought to that with $b=1$.
In MTBG, the constant $bC_0$ measures the ratio of the lapse functions between the fiducial and physical sectors.

In the spherically symmetric systems, 
the general ansatz for the Lagrange multipliers is given by 
\be
\lambda=\lambda(t,r),
\qquad 
\bar{\lambda}
=
\bar{\lambda}(t,r),
\qquad 
\lambda^r=\lambda^r (t,r),
\qquad 
\lambda^a=0.
\ee
We assume that in each of the physical and fiducial sectors
the spacetime is divided into the interior and exterior regions,
where the interior regions are filled with pressureless dust. 
The interfaces between the two regions follow the trajectories
on the $(t,r)$ plane
\be
\label{trajectories}
(t,r)=
\left(T_{(g)}(\tau_{(g)}),R_{(g)}(\tau_{(g)})\right),
\qquad 
\left(T_{(f)} (\tau_{(f)}),R_{(f)}(\tau_{(f)})\right),
\ee
where
$\tau_{(g)}$ and $\tau_{(f)}$ represent the proper times 
on the interfaces
in the physical and fiducial sectors,
and
$T_{(g,f)}(\tau_{(g,f)})<0$
and 
$R_{(g,f)}(\tau_{(g,f)}>0$
are smooth functions of 
$\tau_{(g,f)}$
representing the temporal and radial positions of the interfaces
in both the sectors, respectively.
The four velocities of the interfaces are, respectively, given by 
\be
\label{four_velo}
u_{(g)}^\mu=\left(\dot{T}_{(g)},\dot{R}_{(g)},0,0\right),\qquad
u_{(f)}^\mu=\left(\dot{T}_{(f)},\dot{R}_{(f)},0,0\right),
\ee
where the dot represents the derivative with respect to the proper time $\tau_{(g)}$ or $\tau_{(f)}$
in each sector, respectively.
Since matter is collapsing,
the interfaces are moving inward in the radial directions,
respectively,  
$\dot{R}_{(g)}<0$ and $\dot{R}_{(f)}<0$.
The trajectories of the interfaces are timelike 
$g_{\mu\nu}u_{(g)}^\mu u_{(g)}^\nu=f_{\mu\nu}u_{(f)}^\mu u_{(f)}^{\nu}=-1$,
which can be solved as 
\be
\dot{T}_{(g)}
&=&
\frac{1}{1-\left(N^r{}\right)^2}
\left[
 N^r \dot{R}_{(g)}
+
\sqrt{1-\left(N^r{}\right)^2+\dot{R}_{(g)}^2}
\right],
\label{dtaug}
\\
\dot{T}_{(f)}
&=&\frac{1}{C_0\left[b^2-(N_{(f)}^r{})^2\right]}
\left[
C_0 N_{(f)}^r \dot{R}_{(f)}
+
\sqrt{b^2-(N_{(f)}^r{})^2+C_0^2b^2\dot{R}_{(f)}^2}
\right].
\label{dtauf}
\ee
For the unique definition of the time derivatives \eqref{dtaug} and \eqref{dtauf}, 
$N^r$ and $N^r_{(f)}$ defined in the interior and exterior regions
have to coincide at the interfaces.

The unit normals to the interfaces satisfying
the normalization and orthogonality conditions
$g_{\mu\nu}n_{(g)}^\mu n_{(g)}^\nu=f_{\mu\nu}n^\mu_{(f)} n_{(f)}^{\nu}=1$ 
and 
$g_{\mu\nu}u_{(g)}^\mu n_{(g)}^\nu=f_{\mu\nu}u_{(f)}^\mu  n_{(f)}^\nu=0$
point outward in the radial direction.
The induced metrics on the interfaces are then given by
\be
&&
h_{ij}dy^i dy^j
=
g_{\mu\nu} e^{(g)\mu}{}_i e^{(g)\nu}{}_j 
dy^i dy^j
=
-d\tau^2
+R_{(g)}^2
\theta_{ab}
d\theta^ad\theta^b,
\label{hij}
\\
&&
k_{ij}dy^i dy^j
=
f_{\mu\nu} e^{(f)\mu}{}_i e^{(f)\nu}{}_j 
dy^i dy^j
=
-d\tau_{(f)}^2
+
C_0^2
R_{(f)}^2
\theta_{ab}
d\theta^a
d\theta^b,
\label{kij}
\ee
where the nonzero components of the projection tensors are given by
$e^{(g)t}   {}_\tau=\dot{T}_{(g)}$,
$e^{(g)r}   {}_\tau=\dot{R}_{(g)}$,
$e^{(g)a}   {}_b=\delta^a{}_b$,
$e^{(f)t}   {}_\tau=\dot{T}_{(f)}$,
$e^{(f)r}   {}_\tau=\dot{R}_{(f)}$,
and 
$e^{(f)a}  {}_b=\delta^a{}_b$ with the indices $a,b$ representing the angular directions.

The extrinsic curvature tensors on the interfaces in the physical and fiducial sectors
are given by 
\be
\label{ext_curv}
\tilde{K}^{(g)}_{ij}= e^{(g)\mu}{}_i e^{(g)\nu}{}_j \nabla^{(g)}_{\mu} n^{(g)}_{\nu},
\quad
\tilde{K}^{(f)}_{ij}
= e^{(f)\mu}  {}_i e^{(f)\nu}{}_j  \nabla^{(f)}_{\mu} n^{(f)}_{\nu},
\ee
whose nonzero components are given by
\be
\label{k_components}
\tilde{K}^{(g)}_{\tau\tau}
&=&
\frac{1}
       {\sqrt{1-\left(N^r{}\right)^2+\dot{R}_{(g)}^2}}
\left\{
-\ddot{R}_{(g)}
+
\frac{
\left(1-(N^r)^2\right)^2 N^r N^r_{,r}
-\left(N^r\dot{R}_{(g)}+\sqrt{1+\dot{R}_{(g)}^2-(N^r)^2}\right)^2N^r_{,t}}
{\left(1-(N^r)^2\right)^2}
\right\},
\nonumber\\
\tilde{K}^{(g)}_{ab}
&=&
R_{(g)}
\sqrt{1-\left(N^r{}\right)^2+\dot{R}_{(g)}^2}
\theta_{ab},
\nonumber\\
\tilde{K}^{(f)}_{\tau\tau}
&=&
\frac{1}
       {\sqrt{b^2-(N_{(f)}^r{})^2+b^2C_0^2\dot{R}_{(f)}^2}}
\nonumber\\
&\times&
\left\{
-bC_0 \ddot{R}_{(f)}
+
\frac{
\left(b^2-(N_{(f)}^r)^2\right)^2 N_{(f)}^r N^r_{(f),r}
-b^2\left(C_0 N_{(f)}^r\dot{R}_{(f)}+\sqrt{b^2+b^2C_0^2\dot{R}_{(f)}^2-(N_{(f)}^r)^2}\right)^2
N^r_{(f),t}}
{bC_0 \left(b^2-(N_{(f)}^r)^2\right)^2}
\right\},
\nonumber\\
\tilde{K}^{(f)}_{ab}
&=&
\frac{C_0 R_{(f)}}
       {b}
\sqrt{b^2-(N_{(f)}^r{})^2+b^2 C_0^2\dot{R}_{(f)}^2}
\theta_{ab}.
\ee
We consider the background solutions 
where Lagrange multipliers are trivial 
in both the interior and exterior regions 
and continuous across the interfaces
\be
\label{lambda_zero}
\lambda=\lambda^r=0,
\qquad
{\bar \lambda}={\bar\lambda}_0={\rm constant}.
\ee
Since the contributions of the constraints to the metric equations of motion
do not contain second-order derivatives of the metric functions,
in the case where the Lagrange multipliers are trivial and continuous across the interfaces,
the junction conditions in both the physical and fiducial sectors
are identical to those in the two copies of GR.

The first junction conditions correspond to the continuity of the induced metrics on the interfaces
 \be
\left[h_{ij}\right]
:= 
h^{(+)}_{ij}-h^{(-)}_{ij}=0,
\qquad 
\left[k_{ij}\right]
:= 
   k^{(+)}_{ij}
-k^{(-)}_{ij}=0,
\label{first}
\ee
where below $(+)$ and $(-)$ denote the regions outside and inside the interfaces,
which means the continuity of $N^r$ and $N^r_{(f)}$ across the interfaces
\be
N^{r(+)}\left(T_{(g)},R_{(g)}\right)=N^{r(-)}\left(T_{(g)},R_{(g)}\right),
\qquad 
N_{(f)}^{r(+)}\left(T_{(f)},R_{(f)}\right)=N_{(f)}^{r(-)}\left(T_{(f)},R_{(f)}\right).
\ee
The second junction conditions are the jump of the extrinsic curvature tensors
across the interfaces
\be
\label{junction}
&&
\left[{\tilde K}^{(g)}_{ij}\right]
:=
{\tilde K}^{(g)(+)}_{ij}
-
{\tilde K}^{(g)(-)}_{ij}
=
\kappa^2\left(S^{(g)}_{ij} - \frac{1}{2}S^{(g)} h_{ij} \right),
\nonumber\\
&&
\left[{\tilde K}^{(f)}_{ij}\right]
:=
{\tilde K}^{(f)(+)}_{ij}
-
{\tilde K}^{(f)(-)}_{ij}
=
\frac{\kappa^2}
        {{\tilde\alpha}^2}
 \left(S^{(f)}_{ij} -\frac{1}{2}S^{(f)}k_{ij} \right),
\ee
where $S^{(g)}_{ij}$ and $S^{(f)}_{ij}$ represent the surface energy-momentum tensors
of matter localized on the interfaces.

\subsubsection{The exterior and interior solutions}

For the spatially flat coordinates \eqref{generalg} and \eqref{generalf},
the solutions of the constraint equations~\eqref{cons2}-\eqref{const}
can be divided into the self-accelerating and normal branches \cite{Minamitsuji:2022vfv}.
We focus on the self-accelerating branch given by the condition
\be
\label{sa_cond}
C_0^2c_1+2 C_0c_2+c_3=0.
\ee
Note that the Schwarzschild solutions written in the Schwarzschild coordinates
cannot describe black hole solutions in MTBG,
as they are singular at the positions of the event horizon \cite{Minamitsuji:2022vfv}.

Outside the interfaces, $r>R_{(g)}$ and $r>R_{(f)}$, 
the metric solution is described by the Schwarzschild solutions \cite{Minamitsuji:2022vfv} 
written in the spatially flat coordinates \eqref{generalg} and \eqref{generalf},
\be
\label{outside}
&&
N^{r(+)}(r)=\sqrt{\frac{r_{(g)}}{r}},
\qquad 
N_{(f)}^{r(+)}(r)=\sqrt{\frac{r_{(f)}}{r}},
\ee
where $r_{(g)}$ and $r_{(f)}$ represent the Schwarzschild radii
in both the physical and fiducial sectors, respectively,
which exist under the conditions for asymptotically Minkowski spacetimes
\be
\label{sa_cond2}
&&
C_0^3 c_1+3C_0^2c_2+3C_0 c_3+c_4=0,
\\
&&
\label{sa_cond3}
C_0^3 c_0+3C_0^2c_1+3C_0 c_2+c_3=0,
\ee
together with Eq.\ \eqref{sa_cond}.
Note that the Schwarzschild solutions exist for an arbitrary $b>0$.

Inside the interfaces ($r<R_{(g)}$ and $r<R_{(f)}$), 
we assume that 
the metric solution is described by the spatially flat coordinates \eqref{generalg} and \eqref{generalf}
with 
\be
\label{inside}
&&
N^{r(-)}(t,r)=-r\frac{a_{,t}}{a(t)},
\qquad 
N_{(f)}^{r(-)}(t,r)=-r\frac{a_{,t}}{a(t)},
\ee
where $a_{,t}\equiv da/dt$.
Introducing the comoving radial coordinates
$t=t_c$, $r=r_c a(t)$,
the interior solutions can be expressed by the spatially flat FLRW metrics
\begin{eqnarray}
\label{generalfg2}
&&
g_{\mu\nu}
dx^\mu dx^\nu
=
-dt_c^2
+
a(t_c)^2
\left(
dr_c^2
+r_c^2 
\theta_{ab}
d\theta^a
d\theta^b
\right),
\\
&&
f_{\mu\nu}
dx^\mu dx^\nu
=
C_0^2
\left[
-b^2dt_c^2
+
a(t_c)^2
\left(dr_c^2
+r_c^2 
\theta_{ab}
d\theta^a
d\theta^b
\right)
\right].
\end{eqnarray}
The fact that for the two metrics the ratio of the two effective scale factors 
is a constant $C_0$, satisfying \eqref{sa_cond}, 
is a consequence of the constraints in MTBG for the self-accelerating branch. 
We assume that the spacetimes are filled with pressureless dust fluids whose energy-momentum tensors
are given by 
\be
\label{stress_energy_dust}
&&
T_{(g)}^{\mu\nu}=\rho_{(g)} v_{(g)}^\mu v_{(g)}^\nu,
\qquad
T_{(f)}^{\mu\nu}=\rho_{(f)} v_{(f)}^\mu v_{(f)}^\nu,
\ee 
where 
$\rho_{(g)}$ and $\rho_{(f)}$ represent the matter energy densities 
that behave as 
\be
\label{conservation}
\rho_{(g)}=\frac{\rho_{(g,0)}}{a(t)^3},
\qquad 
\rho_{(f)}=\frac{\rho_{(f,0)}}{a(t)^3},
\ee
with $\rho_{(g,0)}>0$ and $\rho_{(f,0)}>0$ being constants, 
and 
the four velocities of matter $v_{(g)}^\mu $ and $v_{(f)}^\mu$ 
are comoving with the Hubble flows of the spacetimes\footnote{In comoving coordinates, along the radial geodesics of matter, $r_c$ is constant. In this case, the normalized four-velocity is given by $v^\mu_{(g)}\partial_\mu=\partial_{t_c}$. For the change of variables between $(t_c,r_c)$ to $(t,r)$ one can show that $\partial_{t_c}=\partial_{t}+r a_{,t}/a\partial_r$, from which we obtain the first of Eq.\ (\ref{eq:v_u_g}). Along the same lines, one finds the second one of Eq.\ (\ref{eq:v_u_g}).}
\be
\label{matter_velocity}
v_{(g)}^\mu=\left(1,
r\frac{a_{,t}}{a},0,0\right),\label{eq:v_u_g}
\qquad
v_{(f)}^\mu=\frac{1}{b C_0}\left(1,
r\frac{a_{,t}} {a}
0,0\right).
\ee
The only nontrivial components of the gravitational equations in both the sectors of MTBG
set the following constraints:
\be
\frac{3a_{,t}^2}{a^2}=\kappa^2\frac{\rho_{(g,0)}}{a^3},
\qquad
b^2=\frac{\tilde\alpha^2}{C_0^2}\,\frac{\rho_{(g,0)}}{\rho_{(f,0)}}.
\label{eq:fixing_b}
\ee
This implies that the second Friedmann equation effectively sets the value of $b$ in terms of the matter content on both the sectors. 
Furthermore, this result states that the value of $b$ for the exterior metric, which was seen as a free parameter, is actually determined by the details of collapsing matter.
If matter is present only in one of the two sectors,
the value of the parameter $b$ is still fixed through the relation \eqref{eq:fixing_b},
although its value becomes either zero or divergent.
We note that in such a case
zero or divergent $b$ is obtained 
only if the effective cosmological constants 
$\Lambda_{(g)}$ and $\Lambda_{(f)}$ [see Eq.~\eqref{Lambdas}]
are tuned to zero. 
On the other hand, 
if $\Lambda_{(g)}$ and $\Lambda_{(f)}$ take positive values then 
the value of $b$ is fixed to be a nonzero finite value
even in the case where one or both of the matter energy densities vanishes \cite{Gumrukcuoglu:2015nua}.
On the other hand, if both $\Lambda_{(g)}$ and $\Lambda_{(f)}$ are exactly zero, then the value of $b$ in the limit of vanishing energy densities in both sectors depends on the way how the limit is taken.
The solution to the Friedmann equation is given by 
\be
\label{sol_flat_a}
a(t)=c_{(g)} (-t)^{\frac{2}{3}},
\ee
where we have supposed that $t<0$ (and $a_{,t}<0$) during collapse, 
which ends when $t=0$, 
and made the identification
\be
\rho_{(g,0)}=\frac{4c_{(g)}^3}{3\kappa^2}.
\ee
The solutions for the shift vectors \eqref{inside} are then given by 
\cite{Kanai:2010ae}
\be
\label{inside2}
&&
N^{r(-)}(t,r)
=
N_{(f)}^{r(-)}(t,r)
=\frac{2r}{3(-t)}.
\ee
The first junction condition \eqref{first} for \eqref{outside} and \eqref{inside}
determines the conditions on the trajectories of the interfaces
\be
\label{flat_matching}
r_{(g)}=\frac{4R_{(g)}^{3}}
                 {9(-T_{(g)})^2},
\qquad 
r_{(f)}=\frac{4R_{(f)}^{3}}
                 {9(-T_{(f)})^2},
\ee
and relate trajectories of the matter interfaces with the external Schwarzschild radii.
From Eq.\ \eqref{four_velo}, the four velocities of the matter interfaces
are given by 
\be
u_{(g)}^\mu=
\left(1,
-\sqrt{\frac{r_{(g)}}
                 {R_{(g)}}},
0,0\right),
\qquad 
u_{(f)}^\mu=\frac{1}{bC_0}\left(1,-\sqrt{\frac{r_{(f)}}{R_{(f)}}},0,0\right).\label{eq:uf}
\ee
From Eqs.\ \eqref{matter_velocity} \eqref{sol_flat_a}, and \eqref{flat_matching},
we find that 
$u_{(g)}^{\mu}=v_{(g)}^\mu$
and 
$u_{(f)}^{\mu}=v_{(f)}^\mu$
at the interfaces,
showing that 
the interfaces are comoving with geodesics of pressureless dust in both the sectors.

This result also shows that $\dot{R}_{(g)}=-\sqrt{r_{(g)}/R_{(g)}}$, 
which implies that collapse starts from infinity with zero velocity. 
This seems to imply that we are dealing with a star with an infinite initial radius.
However, we can interpret this result for a finite-radius star as follows: 
the model here describes that collapse has already started
before the time $t_i\leq t(<0)$, after which we consider the model
to be an approximate description of real stellar collapse.
In this case, our approximate model describes collapse in the time region
$t_i\leq t<0$.
We should notice that, at the beginning of collapse,
we have the following condition hold true: $\dot{R}_{(g)}(\tau_{(g)}=\tau_{(g)i})=-\sqrt{r_{(g)}/R_{(g)}(\tau_{(g)i})}$.
This corresponds to collapse which satisfies some particular initial conditions.
This will imply that the description in this section will qualitatively/approximately
describe collapse, which will have  $\dot{R}_{(g)}(\tau_{(g)}=\tau_{(g)i})\approx-\sqrt{r_{(g)}/R_{(g)}(\tau_{(g)i})}$,
at least as long as we are not too close to the final singularity.

On the other hand, when $\dot{R}_{(g)}(\tau_{(g)}=\tau_{(g)i})$ will not be well approximated by the previous condition,
then we should expect deviations from the description of collapse in this section.
It is not clear what we should expect in the latter case,
especially because the other known solutions for collapse,
as we will see later on, will also be (even more) fine-tuned.
We defer discussion on possible yet-unknown collapse solutions
to the discussion of the spatially closed case (Sec.\ \ref{sec3b}).
Furthermore, an analogous tuned initial condition also holds in
the fiducial sector
for $\dot{R}_{(f)}(\tau_{(f)}=\tau_{(f)i})$,
see Eq.\ (\ref{eq:uf}), since $C_0$ is defined by the parameters in the MTBG theory,
whereas $b$ is set by Eq.\ (\ref{eq:fixing_b}).
Nonetheless, fine-tuned initial conditions for collapse discussed 
here do not impose constraints on the matter content in them.
For instance, no fine-tuned condition has to be imposed on masses
of black holes in both the 
physical and fiducial sectors, namely, $r_{(g)}\neq r_{(f)}$ in general, as will happen,
in contrast, for the spatially closed case.
Because of this, we believe these spatially flat collapse solutions
to be less problematic than the spatially closed one, which will be discussed in Sec.\ \ref{sec3b}.

We can also show that the nontrivial components of the extrinsic
curvature tensors at the interfaces are given by
\be
\tilde{K}^{(g)(\pm)}_{\tau\tau}=0,\quad
\tilde{K}^{(g)(\pm)}_{ab}=R_{(g)}\theta_{ab},\quad
\tilde{K}^{(f)(\pm)}_{\tau\tau}=0,\quad \tilde{K}^{(f)
  (\pm)}_{ab}=C_0R_{(f)}\theta_{ab},
\ee respectively.  Thus, by imposing the continuity of the extrinsic curvature tensors
$[\tilde{K}^{(g)}_{\tau\tau}] = [\tilde{K}^{(g)}_{ab}] =
[\tilde{K}^{(f)}_{\tau\tau}] = [\tilde{K}^{(f)}_{ab}] = 0$, we can ensure that 
the surface energy-momentum tensors at the interfaces vanish,
 \be
S^{(g)}_{\tau\tau} = S^{(g)}_{ab} = S^{(f)}_{\tau\tau} = S^{(f)}_{ab}
=0, 
\ee
and the exterior Schwarzschild regions \eqref{outside} and
interior dust-dominated FLRW universes \eqref{inside} are smoothly
joined across the interfaces~\cite{Oppenheimer:1939ue,Blau}.
In the case where the interior regions are filled with matter fields with nonvanishing pressures, the propagation speed of sound waves is nonvanishing and thus nearby fluid elements do not evolve independently. In this case, in order to study the interior solution, one needs to solve a set of partial differential equations, instead of ordinary differential equations.

Before closing this subsection, 
we should mention that  it is straightforward to extend the collapsing solution discussed in this subsection
 (and the one in next subsection,
i.e., the Oppenheimer-Snyder model \cite{Oppenheimer:1939ue,Blau}) 
in the presence of the nonzero values of the effective cosmological constants [see Eq.~\eqref{Lambdas}]
as in the case of GR [see e.g., Refs. \cite{Markovic:1999di}],
where the exterior spacetimes are asymptotically de Sitter 
with the effective cosmological constants~\eqref{Lambdas}.
However, for gravitational collapse of astrophysical interest, 
the effect of the effective cosmological constants would be negligible. 
Thus, for our purpose,
it is sufficient to focus on the case of the asymptotically flat exterior solutions.

\subsection{Collapse in the spherically symmetric spacetimes in the spatially closed coordinates}
\label{sec3b}

We consider
the spherical collapse model of a pressureless dust where the spacetimes
inside the matter surfaces are described by the spatially closed FLRW
universes. In contrast to the spatially flat case, in the spatially closed
case gravitational collapse starts from a finite distance. We
note that the spatially closed case is an extension of the Oppenheimer-Snyder
model in GR \cite{Oppenheimer:1939ue} to the case of MTBG.

\subsubsection{The exterior solution}

In order to discuss gravitational collapse of the spatially closed matter surfaces
in both the physical and fiducial sectors,
we assume that the spacetime metrics in the exterior regions
are described by the Schwarzschild solutions written in the spatially closed coordinates
\be
&&
\label{closed_slicing_g}
g_{\mu\nu}^{(+)}dx^\mu dx^\nu
=
-dt^2
+ 
\frac{1}{1-q_{0,(g)}}
\left(
dr
+
\sqrt{
\frac{r_{(g)}}{r}
-q_{0,(g)}}
dt
\right)^2
+r^2 \theta_{ab}d\theta^a d\theta^b,
\\
&&
\label{closed_slicing_f}
f_{\mu\nu}^{(+)}dx^\mu dx^\nu
=
C_0^2
\left[
-b^2 dt^2
+ 
\frac{1}{1-q_{0,(f)}}
\left(
dr
+
\sqrt{
\frac{r_{(f)}}{r}
-
Q
q_{0,(f)}}
dt
\right)^2
+
r^2 
\theta_{ab}d\theta^a d\theta^b
\right].
\ee
where $q_{0,(g)}$ and $q_{0,(f)}$ are constants satisfying 
\be
&&
\label{rbound_g}
-1<-\frac{r_{(g)}}{r}<-q_{0,(g)}<0,
\\
&&
\label{rbound_f}
-1<-\frac{r_{(f)}}{Qr}<-q_{0,(f)}<0,
\ee
and $Q>0$ is also a constant.
We also assume that the Lagrange multipliers are given by Eq.\ \eqref{lambda_zero}.

In the self-accelerating branch,
the equations for $\lambda$ and ${\bar\lambda}$ are satisfied if
\be
q_{0,(f)}=q_{0,(g)},
\label{q_cond}
\ee
as well as the condition \eqref{sa_cond}.
The constraint equation for $\lambda^r$ is automatically satisfied.
The equations of motion for the metric variables finally reduce to 
Eqs.~\eqref{sa_cond2} and \eqref{sa_cond3}, and
\be
q_{0,(g)}{\tilde\alpha}^2 \left(Q-b^2\right)=0.
\ee
Thus, for the spatially closed coordinates $q_{0,(g)}\neq 0$,
we obtain
\be
\label{beq1}
Q=b^2.
\ee
Summarizing the above,
the exterior metrics~\eqref{closed_slicing_g}  and \eqref{closed_slicing_f} reduce to 
\be
&&
\label{closed_slicing_g2}
g_{\mu\nu}^{(+)}dx^\mu dx^\nu
=
-dt^2
+ 
\frac{1}{1-q_{0,(g)}}
\left(
dr
+
\sqrt{
\frac{r_{(g)}}{r}
-q_{0,(g)}}
dt
\right)^2
+r^2 \theta_{ab}d\theta^a d\theta^b,
\\
&&
\label{closed_slicing_f2}
f_{\mu\nu}^{(+)}dx^\mu dx^\nu
=
C_0^2
\left[
-
b^2
 dt^2
+ 
\frac{1}{1-q_{0,(g)}}
\left(
dr
+
\sqrt{
\frac{r_{(f)}}{r}
-b^2 q_{0,(g)}}
dt
\right)^2
+
r^2 
\theta_{ab}d\theta^a d\theta^b
\right].
\ee

\subsubsection{The interior solution}

We then consider
spatially closed FLRW 
metrics as the interior solution \cite{Blau,Kanai:2010ae},
written in comoving coordinates $(t_{c},r_{c})$,
which can be written as
\begin{align}
g^{(-)}_{\mu\nu} dx^\mu dx^\nu
& =-dt_{c}^{2}+a(t)^{2}\left[\frac{dr_{c}^{2}}{1-r_{c}^{2}}+r_{c}^{2}\theta_{ab}d\theta^{a}d\theta^{b}\right],\\
f^{(-)}_{\mu\nu} dx^\mu dx^\nu
& =-C_{0}^{2}
{\tilde b}(t)^{2}
dt_{c}^{2}+a_{f}(t)^{2}\left[\frac{dr_{c}^{2}}{1-r_{c}^{2}}+r_{c}^{2}\theta_{ab}d\theta^{a}d\theta^{b}\right],
\end{align}
and the time-reparametrization invariance allows us to change variables into
$t_{c}=t$ and $r_{c}=\frac{r}{a(t)}$,
as to bring the metric in the form similar to the one of GP coordinates as
\be
&&
g_{\mu\nu}^{(-)}dx^\mu dx^\nu
=
-dt^2
+ 
\frac{1}{1-\frac{r^2}{a(t)^2}}
\left(
dr
-r
\frac{\dot{a}(t)}
         {a(t)}
dt
\right)^2
+
r^2 
\theta_{ab}d\theta^a d\theta^b,
\\
&&
f_{\mu\nu}^{(-)}dx^\mu dx^\nu
=
C_0^2
\left[
-
\tilde{b}(t)^2
 dt^2
+ 
\frac{1}{1-\frac{r^2}{a(t)^2}}
\left(
dr
-
r
\frac{\dot{a}(t)}{a(t)}
dt
\right)^2
+r^2 
\theta_{ab}d\theta^a d\theta^b
\right],
\ee
where as for the spatially flat case, in MTBG, for the self-accelerating branch, the constraints impose that
\begin{equation}
a_{f}=C_{0}\,a.
\end{equation}
In the self-accelerating branch,
the constant $C_{0}$ is uniquely determined by the parameters of the Lagrangian,
as Eq \eqref{sa_cond}. 
In any case, either in comoving coordinates $(t_{c},r_{c})$
or in GP-like coordinates $(t,r)$, the Friedmann equations remain
the same. In particular, the (constant) proportionality between the
two scale factors leads to the following result 
\begin{equation}
H_{f}:=\frac{a_{f,t}}{C_{0}\tilde{b}(t)a_{f}}=\frac{a_{,t}}{C_{0}\tilde{b}(t)a}=\frac{H}{C_{0}\tilde{b}(t)}\,
\qquad{\rm or}\qquad \tilde{b}(t)=\frac{H}{C_{0}H_{f}}\,,
\end{equation}
where a $a_{,t}\equiv da/dt$, etc. If we use the Friedmann equations
of motion for both the metrics, we find
\be
&&
\label{friedmann_closed}
\frac{3(1+{a}_{,t}^2)}{a^2}
=
\kappa^2 
\frac{\rho_{(g,0)}}{a^3},
\qquad 
\frac{3(\tilde{b}^2+{a}_{,t}^2)}{\tilde{b}^2a^2}
=
\frac{C_0^2\kappa^2}{{\tilde\alpha}^2}
\frac{\rho_{(f,0)}}{a^3},
\nonumber
\\
&&
1+a_{,t}^2+2 a a_{,tt}=0,
\qquad 
\tilde{b}^3
 -2aa_{,t}{\tilde b}_{,t}
+\tilde{b} (a_{,t}^2+2aa_{,tt})=0,
\ee
so that $\tilde{b}(t)$ is, in general, a function of time, which can be written
as
\begin{equation}
\tilde{b}(t)
=\frac{1}{C_{0}}
\sqrt{\frac{\frac{\kappa^2\rho_{(g,0)}}{a^{3}}-\frac{3}{a^{2}}}
          {\frac{\kappa^2\rho_{(f,0)}}{\tilde{\alpha}^{2}a^{3}}-\frac{3}{C_{0}^{2}a^{2}}}}.\label{eq:M_cosmo-1}
\end{equation}
Since ${\tilde b}=H/(H_{f}C_{0})$, if we want to have a finite (and nonzero)
value of ${\tilde b}$ when $H$ (or $H_{f}$) vanishes, one needs to require
that the initial time of the collapse, usually set at $\dot{a}=0$,
is the same for both the metrics. This initial starting-from-rest collapse
time is then defined as $a_{,t}=0$ (or $a_{f,t}=0$) and $a=a_{{\rm max}}\neq0$.
In this case, we have
\begin{equation}
\kappa^2\rho_{(g,0)}-3a_{{\rm max}}=0=
\kappa^2\frac{C_{0}^{2}\rho_{(f,0)}}{\tilde{\alpha}^{2}}-3a_{{\rm max}}\,,
\end{equation}
indicating that
\be
\label{density_tuning}
\rho_{(f,0)}=\frac{\tilde{\alpha}^{2}}{C_{0}^{2}}\rho_{(g,0)}\,,
\ee
which, in general, corresponds to a fine-tuned value for the matter densities
in the fiducial sector with respect to the one in the physical sector, as $C_{0}$
and $\tilde{\alpha}$ are specified by the parameters of the theory,
as already stated above. 
In this case, we obtain
\begin{equation}
\tilde{b}(t)=\frac{1}{C_{0}}
\sqrt{\frac{\frac{\kappa^2\rho_{(g,0)}}{a^{3}}-\frac{3}{a^{2}}}{\frac{\kappa^2\rho_{(f,0)}}{\tilde{\alpha}^{2}a^{3}}-\frac{3}{C_{0}^{2}a^{2}}}}=1\,.\label{eq:M_cosmo_2-1}
\end{equation}
Thus, the value of ${\tilde b}$ is no longer a function of time but actually equal to unity.
On the other hand, we paid the price of a problematic fine-tuning
condition among matter contents in the different sectors, namely, Eq.~\eqref{density_tuning},
where, at least in the self-accelerating branch, the value $\tilde{\alpha}/C_{0}$
corresponds to a constant in 
the theory.
In particular, as was happening
in the spatially flat case, matter should be present in both the sectors.

The exact solution of Eq.~\eqref{friedmann_closed} is given by the parametrization
\be
t(\eta)
&=&
\frac{\kappa^2} {6}\rho_{0,(g)}
\left(
\eta+\sin\eta
\right),
\qquad 
a(\eta)
=
\frac{\kappa^2} {6}\rho_{0,(g)}
\left(1+\cos\eta\right),
\qquad 
{\tilde b}(\eta)
=
\frac{\tilde{\alpha}\sqrt{\rho_{0,(g)} (1-\cos\eta) } }
        {\sqrt{2C_0^2\rho_{0,(f)}  -\rho_{0,(g)}{\tilde \alpha}^2(1+\cos\eta)}},
\ee
where
$\eta$ corresponds to the conformal time $dt=a(\eta)d\eta$.
Since the exterior $b$ is constant, the interior $\tilde{b}$ has to be constant.
Imposing Eq.~\eqref{density_tuning},
we also obtain $\tilde{b}=1$.

This situation is much worse than that in the spatially flat case, in which
we only needed to require that $\rho_{(g,0)}>0$ and $\rho_{(f,0)}>0$.
This result of the spatially closed collapse might merely indicate
that a general configuration of matter in both the sectors will not follow
the kind of unique collapse described in this section. How to avoid
this situation? One possibility, still keeping the spatially closed
picture, would be that, similar to what we have encountered in the
spatially flat collapse, the spatially closed collapse does not start
from a configuration where $H$ and/or $H_{f}$ vanish, 
but collapse has already started some time before the particular time $t_{i}$
we start describing its dynamics by means of this model. 
Then, both $H$ and $H_{f}$ do not vanish and are both negative for $t_{i}\leq t$.
This possibility is viable, however, it leads, in general, to a time-dependence
of the function $\tilde{b}(t)$,
which cannot in general be matched with
a stationary configuration for the known exterior solutions of MTBG.
In fact, in MTBG, the Birkhoff theorem does not hold in general and,
at least in principle, we are bound to look for all the possible spherically
symmetric solutions allowed by the theory. Therefore, there could
be still unknown, in general time-dependent solutions that could
accommodate a matching with the $\tilde{b}(t)$ imposed by Eq.\ (\ref{eq:M_cosmo_2-1}).

In any case, in this section, we will only discuss $\tilde{b}(t)=1$ and describe
what this fine-tuned configuration leads to. Hence, on using the GP-like
coordinates described already for the spatially flat case, we can
consider the form of the two interior metrics given as follows:
\be
\label{metric_g_int}
&&
g_{\mu\nu}^{(-)}dx^\mu dx^\nu
=
-dt^2
+ 
\frac{1}{1-\frac{r^2}{a(t)^2}}
\left(
dr
-r
\frac{\dot{a}(t)}
         {a(t)}
dt
\right)^2
+
r^2 
\theta_{ab}d\theta^a d\theta^b,
\\
\label{metric_f_int}
&&
f_{\mu\nu}^{(-)}dx^\mu dx^\nu
=
C_0^2
\left[
-
 dt^2
+ 
\frac{1}{1-\frac{r^2}{a_f(t)^2}}
\left(
dr
-
r
\frac{\dot{a}_f(t)}{a_f(t)}
dt
\right)^2
+r^2 
\theta_{ab}d\theta^a d\theta^b
\right].
\ee
With the background equation \eqref{friedmann_closed},
the interior metrics \eqref{metric_g_int} and \eqref{metric_f_int} reduce to
\be
\label{metric_g_int2}
g_{\mu\nu}^{(-)}dx^\mu dx^\nu
&=&
-dt^2
+ 
\frac{1}{1-\frac{r^2}{a(t)^2}}
\left(
dr
+
\sqrt{
\frac{\kappa^2}{3}\frac{\rho_{0,(g)}}{a^3} r^2
-\frac{r^2}{a(t)^2}
}
dt
\right)^2
+r^2 
\theta_{ab}d\theta^a d\theta^b,
\\
\label{metric_f_int2}
f_{\mu\nu}^{(-)}dx^\mu dx^\nu
&=&
C_0^2
\left[
-
dt^2
+ 
\frac{1}{1-\frac{r^2}{a(t)^2}}
\left(
dr
+
\sqrt{
\frac{C_0^2\kappa^2}{3{\tilde \alpha}^2}
\frac{\rho_{0,(f)}}{a^3}
r^2-\frac{r^2}{a(t)^2}
}
dt
\right)^2
+r^2 
\theta_{ab}d\theta^a d\theta^b
\right].
\ee

\subsubsection{Matching conditions at the interfaces}

Matching of the temporal component of the exterior metric \eqref{closed_slicing_f2}
and interior metric \eqref{metric_f_int2} at the matter interfaces yields
\be
&&
b={\tilde b}=1.
\ee
We consider
the exterior metrics as in
\begin{align}
g^{(+)}_{\mu\nu}dx^{\mu}dx^{\nu} & =-dt^{2}+\frac{1}{1-q_{0,(g)}}\left(dr+\xi\,dt\right)^{2}+r^{2}\theta_{ab}d\theta^{a}d\theta^{b}\,,\\
f^{(+)}_{\mu\nu}dx^{\mu}dx^{\nu} & =C_{0}^{2}\left[-dt^{2}+\frac{1}{1-q_{0,(g)}}\left(dr+\xi_{2}\,dt\right)^{2}+r^{2}\theta_{ab}d\theta^{a}d\theta^{b}\right]\,,
\end{align}
where we have also defined
$N^r=
\xi :=\sqrt{\frac{r_{(g)}}{r}-q_{0,(g)}}$
and
$N^r_{(f)}=
\xi_{2}:=\sqrt{\frac{r_{(f)}}{r}-q_{0,(g)}}$.
This ansatz satisfies the MTBG equations of motion for the vanishing effective
cosmological constants in both the sectors that we have neglected in
these physical collapse scenarios. Let us now consider the four velocities
for both the sectors, at the interface, as given by
\be
u_{(g)}^{\mu}  =\left(\dot{T}_{(g)},\dot{R}_{(g)},0,0\right),
\qquad 
u_{(f)}^{\mu} =\left(\dot{T}_{(f)},\dot{R}_{(f)},0,0\right)\,,
\ee
where a dot, in this subsection, represents a derivative with respect
to the proper times $\tau_{(g)}$ (or $\tau_{(f)}$), e.g.\ $\dot{R}_{(g)}(\tau_{(g)})=R_{(g),\tau}=dR_{(g)}/d\tau_{(g)}$.
Their normalization conditions, $g_{\mu\nu}u_{(g)}^{\mu}u_{(g)}^{\nu}=-1$
and $f_{\mu\nu}u_{(f)}^{\mu}u_{(f)}^{\nu}=-1$, give
\begin{align}
\dot{T}_{(g)} & =\frac{\xi\dot{R}_{(g)}+\sqrt{(1-q_{0,(g)})(\dot{R}_{(g)}^{2}-\xi^{2}+1-q_{0,(g)})}}{1-\xi^{2}-q_{0,(g)}}>0\,,\label{eq:dotTg}\\
\dot{T}_{(f)} & =\frac{\xi_{2}\dot{R}_{(f)}C_{0}+\sqrt{(1-q_{0,(g)})(C_{0}^{2}\dot{R}_{(f)}^{2}-\xi_{2}^{2}+1-q_{0,(g)})}}{\left(1-\xi_{2}^{2}-q_{0,(g)}\right)C_{0}}>0\,,\label{eq:dotTf}
\end{align}
whereas we consider $\dot{R}_{(g)}<0$, $\dot{R}_{(f)}<0$. We will
make use of these equations later on again.

Next we define the normal to the interface as $n_{\mu}^{(g)}dx^{\mu}\propto(dr-R_{(g),t}\,dt)$,
which is normalized to $g_{\mu\nu}n_{(g)}^{\mu}n_{(g)}^{\nu}=1$,
and the interface equation is described by $r-R_{(g)}(t)=0$. Because
of $R_{(g),t}=dR_{(g)}/dt=\dot{R}_{(g)}/\dot{T}_{(g)}$, one can see
that $n_{\mu}^{(g)}u_{(g)}^{\mu}=0$ on the interface. Along the same
lines, we can introduce $n_{\mu}^{(f)}dx^{\mu}\propto(dr-R_{(f),t}\,dt)$,
which is now normalized to $f_{\mu\nu}n_{(f)}^{\mu}n_{(f)}^{\nu}=1$,
and the relative interface equation is $r-R_{(f)}(t)=0$. In this
case we have that $R_{(f),t}=dR_{(f)}/dt=\dot{R}_{(f)}/\dot{T}_{(f)}$. 

For the physical metric,
we define the projection tensors to the interface as $e_{(g)}^{\mu}{}_{\tau}\partial_{\mu}=\dot{T}_{(g)}\partial_{t}+\dot{R}_{(g)}\partial_{r}$
and 
$e_{(g)}^{\mu}{}_{a}\partial_{\mu}=\partial_{a}$,
satisfying $n_{\mu}^{(g)}e_{(g)}^{\mu}{}_{i}=0$, ($i=\tau,a$), and
calculate $h_{\mu\nu}^{(g)}=g_{\mu\nu}-n_{\mu}^{(g)}n_{\nu}^{(g)}$,
together with $K_{\mu\nu}^{(g)}=h_{\mu}^{(g)}{}^{\alpha}h_{\nu}^{(g)}{}^{\beta}n_{\alpha;\beta}^{(g)}$.
Finally, we can build $h_{ab}^{(g)}=\left.e_{(g)}^{\mu}{}_{a}e_{(g)}^{\nu}{}_{b}h_{\mu\nu}^{(g)}\right|_{t=T_{(g)},r=R_{(g)}}$
and $K_{ab}^{(g)}=\left.e_{(g)}^{\mu}{}_{a}e_{(g)}^{\nu}{}_{b}K_{\mu\nu}^{(g)}\right|_{t=T_{(g)},r=R_{(g)}}$.
At the interface, the nontrivial components to consider for the matching
conditions are the following ones:
\begin{align}
h_{\tau\tau}^{(g)} & =-1,
\qquad 
h_{ab}^{(g)} =R_{(g)}^{2}\theta_{ab}\,,\label{eq:h_g_out_zz}\\
K_{\tau\tau}^{(g)} &=\mathcal{K}(R_{(g)}\,,\dot{R}_{(g)},\ddot{R}_{(g)}),
\qquad 
K_{ab}^{(g)}  =R_{(g)}\theta_{ab}\sqrt{\dot{R}_{(g)}^{2}-\xi^{2}+1-q_{0,(g)}}\,,\label{eq:K_g_out_zz}
\end{align}
where in the expression of $\mathcal{K}$, the values of $\dot{T}_{(g)}$
and $\ddot{T}_{(g)}$, present because $\mathcal{K}$ depends also
on $R_{(g),tt}$ and $R_{(g),t}$,  have been replaced by using Eq.\ (\ref{eq:dotTg}).

For the 
fiducial metric, we can follow an analogous path, namely defining
the projectors $e_{(f)}^{\mu}{}_{\tau}\partial_{\mu}=\dot{T}_{(f)}\partial_{t}+\dot{R}_{(f)}\partial_{r}$
and $e_{(f)}^{\mu}{}_{a}\partial_{\mu}=\partial_{a}$,
satisfying $n_{\mu}^{(f)}e_{(f)}^{\mu}{}_{i}=0$, ($i=\tau,a$). After
calculating the expressions for $h_{\mu\nu}^{(f)}=f_{\mu\nu}-n_{\mu}^{(f)}n_{\nu}^{(f)}$,
and $K_{\mu\nu}^{(f)}=h_{\mu}^{(f)}{}^{\alpha}h_{\nu}^{(f)}{}^{\beta}n_{\alpha;\beta}^{(f)}$,
one can build $h_{ab}^{(f)}=\left.e_{(f)}^{\mu}{}_{a}e_{(f)}^{\nu}{}_{b}h_{\mu\nu}^{(f)}\right|_{t=T_{(f)},r=R_{(f)}}$
and $K_{ab}^{(f)}=\left.e_{(f)}^{\mu}{}_{a}e_{(f)}^{\nu}{}_{b}K_{\mu\nu}^{(f)}\right|_{t=T_{(f)},r=R_{(f)}}$
evaluating them on the interface. Then we find
\begin{align}
h_{\tau\tau}^{(f)} & =-1,\qquad 
h_{ab}^{(f)}  =C_{0}^{2}R_{(f)}^{2}\theta_{ab}\,,\\
K_{\tau\tau}^{(f)} & =\mathcal{K}_{f}(R_{(f)}\,,\dot{R}_{(f)},\ddot{R}_{(f)})\,,\label{eq:K_f_out_tt}
\qquad 
K_{ab}^{(f)}  =C_{0}R_{(f)}\theta_{ab}\sqrt{C_{0}^{2}\dot{R}_{(f)}^{2}-\xi_{2}^{2}+1-q_{0,(g)}}\,.
\end{align}
Notice that, at this point, we have still to set the junction conditions to match the interior/exterior solutions.

In fact, as for the internal metrics, we have
\begin{align}
g^{(-)}_{\mu\nu}dx^{\mu}dx^{\nu} & =-dt^{2}+\frac{1}{1-\frac{r^{2}}{a^{2}}}\left(dr-r\frac{a_{,t}}{a}\,dt\right)^{2}+r^2\theta_{ab}d\theta^{a}d\theta^{b}\,,\\
f^{(-)}_{\mu\nu}dx^{\mu}dx^{\nu} & =C_{0}^{2}\left[-dt^{2}
+\frac{1}{1-\frac{r^{2}}{a^{2}}}\left(dr-r\frac{a_{,t}}{a}\,dt\right)^{2}
+r^2\theta_{ab}d\theta^{a}d\theta^{b}\right],
\end{align}
where $a_{,t}=da/dt$. These expressions describe the case of a pair of spatially closed FLRW universes. 
For the case of $b=1$, 
as we have seen above,
$\rho_{(g,0)}$ and $\rho_{(f,0)}$
are proportional to each other,
so that the two Friedmann equations are equivalent to each other, giving effectively only one single constraint, which can be written as Eq.~\eqref{friedmann_closed}.
For these metrics, the four velocities of matter are given by\footnote{To see this, we can perform a coordinate change as $t_{c}=t$ and
$r_{c}=r/a$, that is, $dt_{c}=dt$ and $dr_{c}=dr/a-r\,(a_{,t}/a^{2})dt$.
Then in comoving coordinates $v_{(g)}^{\mu}\partial_{\mu}=\partial_{t_{c}}$.
However, we can see that $\partial_{t_{c}}=\partial_{t}+ra_{,t}/a\,\partial_{r}$
(and $\partial_{r_{c}}=a\partial_{r}$), giving the result in Eq.\ (\ref{eq:v_g_u}).}
\begin{align}
v_{(g)}^{\mu}\partial_{\mu} & =\partial_{t}+r\frac{a_{,t}}{a}\,\partial_{r}\,,\label{eq:v_g_u}\\
v_{(f)}^{\mu}\partial_{\mu} & =\frac{1}{C_{0}}\left(\partial_{t}+r\frac{a_{,t}}{a}\,\partial_{r}\right).
\end{align}
 We also define the normalized normal vector as $n_{\mu}^{(g)}\propto(dr-r_{c0}a_{,t}dt)$,
whereas $n_{\mu}^{(f)}\propto(dr-r_{c0f}a_{,t}dt)$. Then $v_{(g)}^{\mu}n_{\mu}^{(g)}=0$
on the interface defined by $r=r_{c0}a$,
where 
matter is moving on constant comoving coordinate $r_{c}$, 
and similarly by $r=r_{c0f}a$ in the fiducial sector.
Choosing $e_{(g)}^{\mu}{}_{\tau}\partial_{\mu}=\partial_{t}+r\frac{a_{,t}}{a}\,\partial_{r}$
and $e_{(g)}^{\mu}{}_{a}\partial_{\mu}=\partial_{a}$,
we can define $h_{ab}^{(g)}$ and $K_{ab}^{(g)}$ as was done for
the exterior metrics. Then we find
\begin{align}
h_{\tau\tau}^{(g)} & =-1,
\qquad 
h_{ab}^{(g)}  =r_{c0}^{2}a^{2}\theta_{ab}\,,\label{eq:h_in_g_zz}\\
K_{\tau\tau}^{(g)} & =0,
\qquad 
K_{ab}^{(g)}  =r_{c0}a\theta_{ab}\sqrt{1-r_{c0}^{2}}\,.\label{eq:K_in_g_zz}
\end{align}

We then discuss the junction conditions for the physical metric first.
In this case, we first find, from the $h_{ab}^{(g)}$ expressions
in Eqs.\ (\ref{eq:h_g_out_zz}) and (\ref{eq:h_in_g_zz}), that
$R_{(g)}=r_{c0}a\,$.
Then the  components of $K_{ab}^{(g)}$ given in Eqs.\ (\ref{eq:K_g_out_zz})
and (\ref{eq:K_in_g_zz}) lead to
$\dot{R}_{(g)}^{2}-\xi(R_{(g)})^{2}-q_{0,(g)}+r_{c0}^{2}=0$
or
\begin{equation}
\dot{R}_{(g)}=-\sqrt{\xi(R_{(g)})^{2}+q_{0,(g)}-r_{c0}^{2}}.\label{eq:R_dot_g}
\end{equation}
Using this relation, 
we find that $K_{\tau\tau}^{(g)}=\mathcal{K}=0$
also for the exterior solution.

Since we want $u_{(g)}^{\mu}$ to coincide at the interface
with $v_{(g)}^{\mu}$, then, from their zeroth component, we need to
set $\dot{T}_{(g)}=1$. 
Using this last condition, together with
the expression for $\dot{R}_{(g)}$ given in Eq.\ (\ref{eq:R_dot_g})
into Eq.\ (\ref{eq:dotTg}), we find that the condition $q_{0,(g)}=r_{c0}^{2}$
needs to be imposed. Furthermore, we have
\begin{equation}
\frac{da}{dt}=a_{,t}=\frac{1}{r_{c0}}\,R_{(g),t}=\frac{1}{r_{c0}}\,\frac{\dot{R}_{(g)}}{\dot{T}_{(g)}}=\frac{1}{r_{c0}}\,\dot{R}_{(g)}\,,
\end{equation}
which, once inserted into the Friedmann equation, together with the
expression for $\dot{R}_{(g)}$ given in Eq.\ (\ref{eq:R_dot_g}),
makes the Friedmann equation impose that $(\rho_{(g,0)}r_{c0}^{3}-3M_{P}^{2}r_{(g)})/R_{(g)}^{3}=0$
and sets the value of $r_{c0}$ in terms of the other parameters.
Furthermore, we can also see that, on the interface, $v_{(g)}^{r}=R_{(g)}\frac{a_{,t}}{a}=R_{(g),t}=\dot{R}_{(g)}=u_{(g)}^{r}$,
which holds true since $\dot{T}_{(g)}=1$.

Let us now focus on the fiducial sector.
For the fiducial metric, we choose
$e_{(f)}^{\mu}{}_{\tau}\partial_{\mu}=C_{0}^{-1}[\partial_{t}+r\frac{a_{,t}}{a}\,\partial_{r}]$
and
$e_{(f)}^{\mu}{}_{a}\partial_{\mu}=\partial_{a}$
as the projection tensors. On using the expression for $n_{(f)}^{\mu}$, we can
define $h_{ab}^{(f)}$ and $K_{ab}^{(f)}$ as done for the exterior
metrics. The results of these calculations can be written as
\begin{align}
h_{\tau\tau}^{(f)} & =-1,\qquad 
h_{ab}^{(f)}  =C_{0}^{2}r_{c0f}^{2}a^{2}\theta_{ab}\,,\\
K_{\tau\tau}^{(f)} & =0\,
\qquad
K_{ab}^{(f)}  =C_{0}r_{c0f}a\theta_{ab}\sqrt{1-r_{c0f}^{2}}\,,
\end{align}
which, on using the junction conditions for $h_{ab}^{(f)}$ and $K_{ab}^{(f)}$,
lead to
\begin{align}
R_{(f)} & =r_{c0f}a\,,\\
\dot{R}_{(f)} & =-C_{0}^{-1}\sqrt{\xi_{2}(R_{(f)})^{2}+q_{0,(g)}-r_{c0f}^{2}}\,.\label{eq:dot_R_f}
\end{align}
This relation, once inserted into Eq.\ (\ref{eq:K_f_out_tt}), gives
$K_{\tau\tau}^{(f)}=0$ for the exterior metric.

Following exactly the same procedure taken with the physical metric,
namely by setting $u_{(f)}^{\mu}$ to coincide with $v_{(f)}^{\mu}$
at the interface,~\footnote{As for the $r$-component, we have $u_{(f)}^{r}=\dot{R}_{(f)}$, whereas
$v_{(f)}^{r}=\left.\frac{r}{C_{0}}\frac{a_{,t}}{a}\right|_{t=T_{(f)},r=R_{(f)}}=\frac{R_{(f),t}}{C_{0}}=\frac{\dot{R}_{(f)}}{C_{0}\dot{T}_{f}}=\dot{R}_{(f)}$,
so that all the components of $u_{(f)}^{\mu}=v_{(f)}^{\mu}$ coincide
on the interface.} we need to set $\dot{T}_{(f)}=C_{0}^{-1}$, which, in turn, leads
to $q_{0,(g)}=r_{c0f}^{2}$ from Eqs.\ (\ref{eq:dotTf}) and (\ref{eq:dot_R_f}).
Then on using the result that $r_{c0}^{2}=q_{0,(g)}=r_{c0f}^{2}$,
we can see that $3r_{(g)}=\kappa^2\rho_{(g,0)}r_{c0}^{3}=\kappa^2\rho_{(g,0)}r_{c0f}^{3}=3r_{(f)}$
from the Friedmann equation given in Eq.\ (\ref{friedmann_closed}). Then
we are bound to conclude that $r_{(f)}=r_{(g)}$. Therefore, the fine-tuned
interior matter content leads to fine-tuned values for the Schwarzschild
radii for the exterior metrics.

\subsection{Absence of cusps of the constant-time hypersurfaces}

Before closing this section, 
we check the continuity of the normal vectors to the $t={\rm constant}$ hypersurfaces
across the matter interfaces
in both the spatially flat and closed cases and
in both the physical and fiducial sectors,
 i.e., the continuity of the derivatives of the St\"uckelberg fields along the normals
across the matter interfaces.
Because of the absence of the four-dimensional diffeomorphism invariance,
in general,
different time slicings would provide different solutions,
and hence this condition is not obvious and has to be checked for each solution.
If this condition fails to be satisfied, 
cusps would be formed on the interfaces, which would give rise to
extra boundary contributions to the evolution of the metric and matter fields.
In our cases, the nontrivial components of the normals to a $t={\rm constant}$ hypersurface 
\be
m_{\mu}^{(g)}dx^\mu=-dt,
\qquad 
m_{\mu}^{(f)}dx^\mu=-C_0 b dt,
\ee
where $b=1$ for the spatially closed case.
We then find that at the matter interfaces
\be
&&
n_{\alpha}^{(g)}m^\alpha_{(g)}
\Big|_{(T_{(g)},R_{(g)})}=0, 
\qquad 
n_{\alpha}^{(f)}m^\alpha_{(f)}
\Big|_{(T_{(f)},R_{(f)})}=0, 
\nonumber\\
&&
e^{\alpha}_{(g)\tau}m_\alpha^{(g)}
\Big|_{(T_{(g)},R_{(g)})}=-1, 
\qquad 
e^{\alpha}_{(f)\tau} m_\alpha^{(f)}
\Big|_{(T_{(f)},R_{(f)})}=-1, 
\nonumber\\
&&
e^{\alpha}_{(g)a}m_\alpha^{(g)}
\Big|_{(T_{(g)},R_{(g)})}=0, 
\qquad 
e^{\alpha}_{(f)a} m_\alpha^{(f)}
\Big|_{(T_{(f)},R_{(f)})}=0.
\ee
Hence, the normals to the  $t={\rm constant}$ hypersurfaces
are continuous across the matter interfaces in both the sectors.
This implies 
that there is no cusp formation
in the $t={\rm constant}$ hypersurfaces at the matter interfaces.

In summary,
the results in this section suggest that, in the self-accelerating
branch of MTBG with matter composed of two independent species \eqref{matter_action},
gravitational collapse proceeds as that in the two copies of GR.
Thus, in order to observe the deviation from the case of two copies
of GR, we should investigate the linear perturbations about the spherically
symmetric solutions. In the next section, we will study the odd-parity
perturbations of the Schwarzschild$-$de Sitter solutions written in the spatially flat
coordinates in the self-accelerating branch of MTBG.

\section{Odd-parity perturbations of Schwarzschild$-$de Sitter solutions in MTBG}
\label{sec4}

In this section, we investigate the linear perturbations of the
Schwarzschild$-$de Sitter solutions written in the spatially flat
coordinates in the self-accelerating branch of MTBG.  The linear
perturbations in the static and spherically symmetric black holes are
decomposed into the odd- and even-parity sectors.  In this
paper, we focus on the odd-parity sector of the perturbations.

\subsection{Odd-parity black hole perturbations in the spatially flat coordinates}
\label{sec5a}

The perturbed physical and fiducial metrics in the odd-parity sectors
about the Schwarzschild$-$de Sitter solutions
written in the spatially flat coordinates
are, respectively, given by 
\be
\label{metric_g}
g_{\mu\nu}dx^\mu dx^\nu
&=&
-
dt^2
+
\left(dr+N^r(r) dt\right)^2
+r^2\theta_{ab}d\theta^a d\theta^b
\nonumber\\
&+&
2
\sum_{\ell,m}
\left(
r^2
H_{t} (t,r)dt
+
r
H_{r}(t,r)
(dr+N^r(r) dt)
\right)
 E_a{}^b \partial_b Y_{\ell m}(\theta^c)  d\theta^a
\nonumber\\
&+&
r^2
\sum_{\ell,m}
H_{3}(t,r)
E_{(a}{}^c {\tilde\nabla}_{b)}{\tilde \nabla}_c Y_{\ell m}(\theta^c)
d\theta^a d\theta^b,
\\
\label{metric_f}
f_{\mu\nu}dx^\mu dx^\nu
&=&
C_0^2
\Big[
-
b^2
dt^2
+
\left(dr+N_{(f)}^r(r) dt\right)^2
+
r^2\theta_{ab}d\theta^a d\theta^b
\nonumber\\
&+&
2
\sum_{\ell,m}
\left(
r^2
K_{t}(t,r)
dt
+
r
K_{r}(t,r)
(dr+N_{(f)}^r(r) dt)
\right)
 E_a{}^b \partial_b Y_{\ell m}(\theta^c) d\theta^a
\nonumber\\
&+&
r^2
\sum_{\ell,m}
K_{3}(t,r)
E_{(a}{}^c {\tilde \nabla}_{b)} {\tilde\nabla}_c Y_{\ell m}(\theta^c)
d\theta^a d\theta^b
\Big],
\ee
where 
$N^r(r)$ and $N^r_{(f)}(r)$ are given by 
\be
\label{outside2}
&&
N^{r}(r)=\sqrt{ \frac{\Lambda_{(g)} r^{2}}{3}+\frac{r_{(g)}}{r}},
\qquad 
N_{(f)}^r(r)=\sqrt{\frac{r^{2} C_{0}^{2} b^{2} \Lambda_{(f)}}{3}+\frac{r_{(f)}}{r}},
\ee
$b>0$ is a constant,
$Y_{\ell m}(\theta^c)$ are spherical harmonics 
with the angular multipole and magnetic moments with $-\ell\leq m\leq \ell$,
respectively, 
${\tilde \nabla}_a$ represents  the covariant derivative with respect to 
the unit two-sphere with the metric $\theta_{ab}$,
and 
$E_{ab}$ represents the totally antisymmetric tensor on the two-sphere
satisfying ${\tilde\nabla}^a E_{ab}=0$.
$\Lambda_{(g)}$ and $\Lambda_{(f)}$
are the effective cosmological constants defined as $\Lambda_{(g)}=3H_{(g)}^2$ and $\Lambda_{(f)}=3H_{(f)}^2$ 
on a cosmological de Sitter expansion with Hubble parameters $H_{(g)}$
for the physical metric and $H_{(f)}$ for the fiducial one,
whose values can be rewritten as
\be
\label{Lambdas}
&&
\Lambda_{(g)}=\frac{m^{2} \left(c_4-2 C_{0}^{3} c_{1}-3 C_{0}^{2} c_{2}\right)}{2},
\qquad
\Lambda_{(f)}=\frac{\left(C_{0}^{2} c_{0}+2 C_{0} c_{1}+c_{2}\right) m^{2}}{2 C_{0}^{2} {\tilde\alpha}^{2}}.
\ee
For the existence of the background Schwarzschild$-$de Sitter solutions
written in the spatially flat coordinates in the self-accelerating branch,
the constants $c_0$, $c_1$, $c_2$, $c_3$, and $c_4$
satisfy Eq.\ \eqref{sa_cond},
 provided that we add to the right-hand side of Eq.~\eqref{sa_cond2}
the quantity $2\Lambda_{(g)}/m^2$ and the quantity $\frac{2 C_{0}^{3} \Lambda_{(f)} {\tilde\alpha}^{2}}{m^{2}}$ 
to the right-hand side of Eq.\ \eqref{sa_cond3}. This procedure gives the values written in Eq.\ \eqref{Lambdas}.
In the rest, 
with use of Eqs.\ \eqref{sa_cond} and \eqref{Lambdas},
the constants $c_0$, $c_3$, and $c_4$ are related to 
the other coupling constants $(c_1,c_2)$ and the constant $C_0$.
We also assume that $C_0c_1+c_2\neq 0$ and  $\beta C_0\neq 1$.
As noted in Sec.\ \ref{sec3a},
the constant $C_0b$ measures the ratio of the lapse functions between the fiducial and physical sectors.
As we will see, 
in the case of $b=1$
the odd-parity perturbations in the two sectors effectively become massless
and decoupled.

The perturbed three-dimensional metrics
in the odd-parity sector are, respectively, given by
\be
\gamma_{ij}dy^i dy^j
&=&
dr^2
+r^2\theta_{ab}d\theta^a d\theta^b
\nonumber\\
&+&
2
r
\sum_{\ell,m}
H_{r} (t,r)
 E_a{}^b \partial_b Y_{\ell m}(\theta^c)
dr
d\theta^a
+
r^2
\sum_{\ell,m}
H_{3}(t,r)
E_{(a}{}^c {\tilde\nabla}_{b)}{\tilde\nabla}_c Y_{\ell m}(\theta^c)
d\theta^a d\theta^b,
\\
\phi_{ij}dy^i dy^j
&=&
C_0^2
\Big[
dr^2
+
r^2\theta_{ab}d\theta^a d\theta^b
\nonumber\\
&+&
2r
\sum_{\ell,m}
K_{r}(t,r)
 E_a{}^b \partial_b Y_{\ell m}(\theta^c)
dr
d\theta^a
+
r^2
\sum_{\ell,m}
K_{3}(t,r)
E_{(a}{}^c {\tilde\nabla}_{b)}{\tilde \nabla}_c Y_{\ell m}(\theta^c)
d\theta^a d\theta^b
\Big].
\ee
The odd-parity perturbation of the Lagrange multipliers is given by 
\be
\lambda=0,
\quad
{\bar\lambda}=0,
\quad 
\lambda^r=0,
\quad
\lambda^a=\sum_{\ell,m}\Lambda(t,r) E^{ab}\partial_b Y_{\ell m}(\theta^c).
\ee
Note that in the odd-parity sector the $\ell=0$ mode is absent.
We first focus on the $\ell\geq 2$ modes, and then the dipolar modes $\ell =1$ separately.
For the dipolar mode $\ell=1$,
the $H_3$ and $K_3$ perturbations
are not present automatically.


\subsection{The higher-multipole modes $\ell \geq 2$}
\label{sec5b}

First, we focus on the modes $\ell \geq 2$.
Under the joint foliation-preserving spatial gauge transformation 
\be
t\to t,
\qquad
x^i\to x^i+\xi^i(t,x^i),
\ee
which for the odd-parity perturbations are explicitly given by 
\be
&&
\xi^r=0,
\qquad
\xi^a=
\sum_{\ell,m}
\Xi (t,r)
 E^{ab}  \partial_b Y_{\ell m}(\theta^c),
\ee
the metric perturbations are transformed as
\be
\label{gauge}
&&
{\bar \delta} H_{t}=
-\dot{\Xi}
+
N^r 
\Xi',
\qquad 
{\bar \delta} H_{r}=
-r\Xi',
\qquad 
{\bar \delta} H_{3}=-2\Xi,
\nonumber\\
&&
{\bar \delta} K_{t}=
-\dot{\Xi}
+
N_{(f)}^r 
 \Xi',
\qquad 
{\bar \delta} K_{r}=
-r\Xi',
\qquad
{\bar \delta} K_{3}=-2\Xi,
\ee
where in this section dots and primes  represent the derivatives with respect to $t$ and $r$, respectively.
Thus, the combinations of 
\be
H_{r}
-
K_r,
\qquad
 H_{3}
- K_3,
\ee
are gauge invariant.
For the convenience, 
we introduce the gauge-invariant variables
${\bar H}_t$, ${\bar H}_r$ ${\bar K}_t$, ${\bar K}_r$, and ${\bar K}_3$ by
\be
\label{gauge_invariants}
H_t
&=&
{\bar H}_t
+
\frac{1}{2}
\left(
\dot{H}_3
-N^r
H_3'
\right),
\nonumber
\\
H_r
&=&
{\bar H}_r
+
\frac{1}{2}
rH_3',
\nonumber\\
K_t
&=&
{\bar K}_t
+
\frac{1}{2}
\left(
\dot{K}_3
-N_{(f)}^r
K_3'
\right),
\nonumber
\\
K_r
&=&
{\bar K}_r
+
\frac{1}{2}
r K_3',
\nonumber
\\
K_3
&=&
{\bar K}_3+H_3.
\ee
The perturbation of the Lagrange multiplier is gauge invariant by itself.

Because of the degeneracy among the different $m$ modes
in the spherically symmetric backgrounds,
we may use the Legendre polynomials $P_\ell (\cos\theta)$ instead of  the spherical harmonics $Y_{\ell m}(\theta^c)$.
Expanding the MTBG action \eqref{action} on the Schwarzschild$-$de Sitter backgrounds
written in the spatially flat coordinates
with the algebraic conditions
\eqref{sa_cond} and \eqref{Lambdas}
up to the second order of the perturbations and integrating over the angular directions $\theta^a$,
we obtain the second-order action of the odd-parity perturbations
besides the $\ell=1$ mode,
\be
\label{Soriginal0}
\delta_{(2)} S
&=&
\sum_{\ell\neq 1} 
\frac{\ell (\ell+1)}
        {2\ell+1}
\int dt dr
L_\ell 
\left[
H_t, 
H_r,
H_3,
K_t,
K_r,
K_3,
\Lambda
\right].
\ee
Note that
at the level of the linearized perturbations
there is no mixing between different $\ell$ modes,
and so we minimize the second-order action for each $\ell$ mode.
With use of Eq.~\eqref{gauge_invariants},
the second-order action \eqref{Soriginal0}
can be rewritten 
in terms of the gauge-invariant variables as 
\be
\label{Soriginal}
\delta_{(2)} S
&=&
\sum_{\ell\neq 1} 
\frac{\ell (\ell+1)}
        {2\ell+1}
\int dt dr
L_\ell 
\left[
\bH_t, 
\bH_r,
\bK_t,
\bK_r,
\bK_3,
H_3,
\Lambda
\right].
\ee
Varying Eq.\ \eqref{Soriginal} 
under the algebraic conditions \eqref{sa_cond} and \eqref{Lambdas},
we obtain the equations of motion for 
the gauge-invariant quantities $\bH_t$, $\bH_r$, $\bK_t$, $\bK_r$, $\bK_3$, and $\Lambda$,
as Eqs.~\eqref{eom_ht}, \eqref{eom_hr}, \eqref{eom_kt}, \eqref{eom_kr}, \eqref{eom_k3}, and~\eqref{eom_lambda},
respectively.
We note that Eqs. \eqref{eom_ht}-\eqref{eom_lambda}
are not singular at the event horizons $r=r_{(g)}$ and $r=r_{(f)}$,
and the limit to the de Sitter metrics $r_{(g)}\to 0$ and $r_{(f)}\to 0$ can be taken smoothly.
The remaining mode $H_3$ corresponds to the gauge mode
and hence the equation of motion for $H_3$ becomes trivial.
Thus, without any loss of generality, we may set $H_3=0$,
and in Eq.~\eqref{Soriginal} 
$L_\ell 
\left[
\bH_t, 
\bH_r,
\bK_t,
\bK_r,
\bK_3,
H_3,
\Lambda
\right]
\to 
L_\ell 
\left[
\bH_t, 
\bH_r,
\bK_t,
\bK_r,
\bK_3,
\Lambda
\right]$.

In order to derive the master equations for the odd-parity perturbations,
we introduce the new variables $\chi_h$ and $\chi_k$ 
and define the new second-order action \cite{DeFelice:2011ka}\footnote{We follow here the same method used in the context of the $f(R,{\cal G})$ theories (${\cal G}$ being the Gauss-Bonnet scalar). Even though the method in this paper is equivalent to the one in \cite{DeFelice:2011ka}, the results will differ. In particular, in the $f(R,{\cal G})$ theories, the odd-mode master variable was sufficient in order to entirely describe the dynamics of all the odd-mode fields. In MTBG, we will see that this is not the case. In fact, we will see that the shadowy modes cannot be, in general, integrated out in terms of the introduced master variables and that the same shadowy modes will act as sources for the dynamics of the same master variables. This feature might be shared also by other theories that introduce shadowy modes, a phenomenon that could be worthy of further investigation.} by
\be
\label{Sprime}
\delta_{(2)}
S'
&=&
\delta_{(2)}
S
-
\sum_{\ell\neq 1} 
\frac{\ell (\ell+1)}
        {2\ell+1}
\int dt dr
\Big\{
-A_{1}(r)
\left(
\dot{\bH}_r
+
A_{2}(r)
\bH_r'
+
A_{3}(r)
\bH_t'
+
A_{4} (r)
\bH_r
-\frac{\chi_h}{r}
\right)^2
\nonumber
\\
&&
+
B_{1}(r)
\left(
\dot{\bK}_r
+
B_{2}(r)
\bK_r'
+
B_{3}(r)
\bK_t'
+
B_{4} (r)
\bK_r
-\frac{\chi_k}{r}
\right)^2
\Big\}
\nonumber\\
&=&
\sum_{\ell\neq 1} 
\frac{\ell (\ell+1)}
        {2\ell+1}
\int dt dr
\Big\{
L_\ell 
\left[
\bH_t, 
\bH_r,
\bK_t,
\bK_r,
\bK_3,
\Lambda
\right]
-A_{1}(r)
\left(
\dot{\bH}_r
+
A_{2}(r)
\bH_r'
+
A_{3}(r)
\bH_t'
+
A_{4} (r)
\bH_r
-\frac{\chi_h}{r}
\right)^2
\nonumber
\\
&&
-B_{1}(r)
\left(
\dot{\bK}_r
+
B_{2}(r)
\bK_r'
+
B_{3}(r)
\bK_t'
+
B_{4} (r)
\bK_r
-\frac{\chi_k}{r}
\right)^2
\Big\}.
\ee
Varying the new second-order action $\delta_{(2)}S'$,
Eq.~\eqref{Sprime},
with respect to $\chi_h$ and $\chi_k$,
we obtain
\be
\label{onshell_eq_h}
\dot{\bH}_r
+
A_{2}(r)
\bH_r'
+
A_{3}(r)
\bH_t'
+
A_{4} (r)
\bH_r
-\frac{\chi_h}{r}=0,
\\
\label{onshell_eq_k}
\dot{\bK}_r
+
B_{2}(r)
\bK_r'
+
B_{3}(r)
\bK_t'
+
B_{4} (r)
\bK_r
-\frac{\chi_k}{r}
=0.
\ee
Solving Eqs.\ \eqref{onshell_eq_h} and \eqref{onshell_eq_k}
in terms of $\chi_h$ and $\chi_k$
and then
substituting them back into Eq.\ \eqref{Sprime},
the new second-order action \eqref{Sprime}
reduces to the original one \eqref{Soriginal}.

Varying the new second-order action $\delta_{(2)}S'$,
Eq.\ \eqref{Sprime},
with respect to $\bH_t$, $\bH_r$, $\bK_t$, $\bK_r$,
we obtain the equations of motion for them.
Requiring that the derivatives of  $\bH_t$, $\bH_r$, $\bK_t$, $\bK_r$ vanish in their equations,
so that $\bH_t$, $\bH_r$, $\bK_t$, $\bK_r$ become auxiliary variables
of the second-order action \eqref{Sprime},
we determine the coefficients in Eq.\ \eqref{Sprime},
\be
\label{ABs}
&&
A_{1}(r)
=r^2,
\quad 
A_{2}(r)
=
-N^r,
\quad
A_{3}(r)
=-r,
\quad
A_{4}(r)
=
\frac{3}{2r}N^r
-\frac{\Lambda_{(g)}r}{2N^r},
\nonumber\\
&&
B_{1}(r)
=
\frac{{\tilde \alpha}^2C_0^2}
        {b}r^2,
\quad 
B_{2}(r)
=
-N_{(f)}^r,
\quad
B_{3}(r)
=-r,
\quad
B_{4}(r)
=
\frac{3}{2r}N_{(f)}^r
-\frac{\Lambda_{(f)}C_0^2b^2r}{2N_{(f)}^r}.
\ee
The equations of motion 
for $\bH_t$, $\bH_r$, $\bK_t$, and $\bK_r$ 
then relate them 
to the variables $\chi_h$ and $\chi_k$ as
\begin{align}
\label{HrHt_Sch}
\bH_t
&=
-\frac{r\chi_h'+2\chi_h}{(\ell+2)(\ell-1)r},
\nonumber
\\
\bH_r
&=
\frac{1}
       {2\left[(1-b)C_0^2 (C_0c_1+c_2)m^2r^2-2(\ell+2)(\ell-1)\right]}
\nonumber\\
&\times
\left[
C_0
\left(C_0c_1+c_2\right)m^2r^2
\left\{
(1-b)C_0\left(r\bK_3'+2\bK_r\right)
+2r (1-C_0\beta)\Lambda'
\right\}
-
4
\left(
N^r\left(
2\chi_h + r\chi_h'
\right)
-
r
\dot{\chi}_h
\right)
\right],
\nonumber
\\
\bK_t
&=
-\frac{r\chi_k'+2\chi_k}{(\ell+2)(\ell-1)r},
\nonumber
\\
\bK_r
&=
\frac{1}
       {2b C_0 \left[(1-b)(C_0c_1+c_2)m^2 r^2-2 (\ell+2)(\ell-1)b{\tilde\alpha}^2\right]}
\nonumber\\
&\times
\left[
b \left(C_0c_1+c_2\right)m^2r^2
\left\{
\left(
   (1-b)C_0 \left(-r \bK_3'+2\bH_r\right)
-2r(1-C_0\beta)\Lambda'
\right)
\right\}
-
4C_0{\tilde\alpha}^2
\left(
N_{(f)}^r\left(
2\chi_k
+r\chi_k'
\right)
-r\dot{\chi}_k
\right)
\right].
\nonumber\\
\end{align}
Since in the above expression \eqref{HrHt_Sch} $\bH_r$ and $\bK_r$ are coupled,
we still need to solve for each of them,
although we omit to show them explicitly.
$\chi_h$ and $\chi_k$
can be interpreted as the master variables
for the odd-parity perturbations for $\ell \geq 2$.
Substituting Eq.\ \eqref{ABs} into Eqs.\ \eqref{onshell_eq_h} and \eqref{onshell_eq_k},
we obtain
\be
\label{master_eq_h}
&&
\chi_h
-\left(\frac{3}{2}N^r-\frac{r^2\Lambda_{(g)}}{2N^r}\right)
\bH_r
+r
\left(
N^r\bH_r'
+r \bH_t'
-
\dot{\bH}_{r}
\right)=0,
\\
\label{master_eq_k}
&&
\chi_k
-\left(\frac{3}{2}N_{(f)}^r-\frac{C_0^2b^2r^2\Lambda_{(f)}}{2N_{(f)}^r}\right)\bK_r
+r
\left(
N_{(f)}^r\bK_r'
+r \bK_t'
-\dot{\bK}_{r}
\right)=0,
\ee
and moreover substituting Eq.\ \eqref{HrHt_Sch} into Eqs.~\eqref{master_eq_h} and \eqref{master_eq_k},
we obtain the master equations for $\chi_h$ and $\chi_k$.

Varying the action \eqref{Sprime} with Eq.\ \eqref{ABs}
with respect to $\Lambda$ and $K_3$
leads to the equations of motion for $\Lambda$ and  $\bK_3$, which 
are given by 
\be
\label{eq_sol_Lambda}
&&
r^2\Lambda''
+
4r \Lambda'
-
\left(\ell+2\right)\left(\ell-1\right)
\Lambda
=0,
\\
\label{eq_sol_K3}
&&
r^2\bK_3''
+
4r \bK_3'
-
\left(\ell+2\right)\left(\ell-1\right)
\bK_3
=
-2r
\left[
\left(
\bK_r'-\bH_r'
\right)
+
\frac{3}{r}
\left(
\bK_r-\bH_r
\right)
\right],
\ee
which are also obtained by combining Eqs.~\eqref{eom_k3} and \eqref{eom_lambda}.
Eliminating $\bH_r$ and $\bK_r$ in Eq.\ \eqref{eq_sol_K3} with Eq.\ \eqref{HrHt_Sch},
the equation \eqref{eq_sol_K3}
reduces to the form
\begin{align}
\label{eq_sol_K3_2}
r^2\bK_3''
&+\left\{4r-\frac{2 r^{3} \left({\tilde\alpha}^{2} C_{0}^{2} b +1\right) m^{2} \left(b -1\right) \left(C_{0} c_{1}+c_{2}\right)}{( C_{0}^{2} {\tilde\alpha}^{2} b +1 )\,r^2(C_{0} c_{1}+c_{2}) (b -1) m^{2}+2b (\ell^{2}+\ell -2) {\tilde\alpha}^{2}}
\right\}\bK_3'
\nonumber\\
&-\left\{\left(\ell +2\right) \left(\ell -1\right)+
\frac{m^{2} r^{2} \left(b -1\right) \left({\tilde\alpha}^{2} b C_{0}^{2}+1\right) \left(C_{0} c_{1}+c_{2}\right)}{2 {\tilde\alpha}^{2} b}\right\}
\bK_3\nonumber\\
&=
{\cal F}
\left[
\dot{\chi}_h',
\chi_h'',
\dot{\chi}_h,
\chi_h',
\chi_h;
\dot{\chi}_k',
\chi_k'',
\dot{\chi}_k,
\chi_k',
\chi_k;
\Lambda'', 
\Lambda', 
\Lambda
\right],
\end{align}
where ${\cal F}$ represents a linear combination of the variables shown
in the arguments, which is not explicitly written.  Since
Eqs.~\eqref{eq_sol_Lambda} and \eqref{eq_sol_K3_2} are elliptic
equations, $\Lambda$ and $\bK_3$ are fixed under the given spatial
boundary conditions.  Thus, $\Lambda$ and $\bK_3$ correspond to the
shadowy modes \cite{DeFelice:2018ewo, DeFelice:2021hps}.
In Eq.~\eqref{eq_sol_Lambda}, $\Lambda$ is not coupled
to the other variables and can be solved as \be
\label{sol_Lambda}
\Lambda
=
g_1(t) r^{-2-\ell}
+
g_2(t) r^{-1+\ell},
\ee
where $g_1(t)$ and $g_2(t)$ are integration constants.
For instance, requiring the regularity at the spatial infinity $r\to \infty$
for the $\ell\geq 2$ modes, we may impose $g_2(t)=0$.
In the rest, however, 
we leave $g_1(t)$ and $g_2(t)$ as the general functions of the time $t$.

\subsubsection{The case of $b=1$}

In the case of $b=1$,  
the substitution of Eq.~\eqref{HrHt_Sch} into Eqs.\ \eqref{master_eq_h} and \eqref{master_eq_k}
yields the master equations
\be
\label{epsilonh}
&&
\ddot{\chi}_{h} - 2 N^{r} \dot{\chi}_{h}'+[\left(N^{r}\right)^{2}-1] \chi_{h }''
-\frac{\Lambda_{(g)} r^{2}+3 \left(N^{r}\right)^{2} }{2 r \,N^{r}}\dot{\chi}_{h}
\nonumber\\
&&
+\frac{\Lambda_{(g)} r^{2}+\left(N^{r}\right)^{2}-2 }{r}\chi_{h}'
+\frac{2 \Lambda_{(g)} r^{2}-4 \left(N^{r}\right)^{2}+\ell^{2}+\ell  }{r^{2}}\chi_{h}\nonumber\\
&&=\frac{\left(C_{0} c_{1}+c_{2}\right) m^{2} r \left(1-\beta  C_{0}\right) C_{0}
  \{[\Lambda_{(g)} r^{2}+3 \left(N^{r}\right)^{2}] \Lambda'
  +2 r \,N^{r} (\Lambda'' N^{r}-\dot{\Lambda}')\} }{4 N^{r}},
\\
\label{epsilonk}
&&
\ddot{\chi}_{k}-2 N_{(f)}^{r} \dot{\chi}_{k}'
+[(N_{(f)}^{r})^{2}-1]\chi_{k}''
-\frac{C_{0}^{2} \Lambda_{(f)} r^{2}+3 (N_{(f)}^{r})^{2}}{2 r N_{(f)}^{r}}\dot{\chi}_{k}
\nonumber\\
&&
+\frac{C_{0}^{2} \Lambda_{(f)} r^{2}+(N_{(f)}^{r})^{2}-2 }{r}\chi_{k}'
+\frac{2 C_{0}^{2} \Lambda_{(f)} r^{2}-4 (N_{(f)}^{r})^{2}+\ell^2+\ell  }{r^{2}}\chi_{k}\nonumber\\
&&
=-\frac{m^{2} \left(C_{0} c_{1}+c_{2}\right) (1-\beta  C_{0}) r \left(\Lambda' C_{0}^{2} \Lambda_{(f)} r^{2}+2 (N_{(f)}^{r})^{2} r \Lambda''
  +3 \Lambda' (N_{(f)}^{r})^{2}-2 N_{(f)}^{r} r \dot{\Lambda}'\right)}{4 N_{(f)}^{r} C_{0} {\tilde\alpha}^{2}}.
\ee
The left-hand sides of Eqs.\ \eqref{epsilonh} and \eqref{epsilonk}
correspond to the operators for the master equations in 
the two copies of GR.
In our case, 
Eqs.\ \eqref{epsilonh} and \eqref{epsilonk},
together with Eqs.~\eqref{eq_sol_Lambda} and \eqref{eq_sol_K3_2},
form a closed system.
In Eqs.\ \eqref{epsilonh} and \eqref{epsilonk},
the two dynamical modes $\chi_h$ and $\chi_k$ are decoupled
and sourced by the shadowy mode $\Lambda$.
From Eqs.\ \eqref{epsilonh} and \eqref{epsilonk},
we find that
in both the sectors the odd-parity perturbations propagate with the speed of light.

With the explicit form of the solution for $\Lambda$,
Eq.\ \eqref{sol_Lambda}, 
the equations for the odd-parity perturbations reduce to
\be
\label{epsilonh_2}
&&
\ddot{\chi}_{h} - 2 N^{r} \dot{\chi}_{h}'+[\left(N^{r}\right)^{2}-1] \chi_{h }''
-\frac{\Lambda_{(g)} r^{2}+3 \left(N^{r}\right)^{2} }{2 r \,N^{r}}\dot{\chi}_{h}
+\frac{\Lambda_{(g)} r^{2}+\left(N^{r}\right)^{2}-2 }{r}\chi_{h}'
+\frac{2 \Lambda_{(g)} r^{2}-4 \left(N^{r}\right)^{2}+\ell^{2}+\ell  }{r^{2}}\chi_{h}\nonumber\\
&&=\frac{m^{2} \left(C_{0} c_{1}+c_{2}\right) (1-\beta  C_{0}) C_{0} r}{2 N^{r}}
\left\{g_{1} \left(\ell +\frac{3}{2}\right) \left(N^{r}\right)^{2} \left(\ell +2\right) r^{-\ell -3}
+\dot{g}_{1} N^{r} \left(\ell +2\right) r^{-\ell -2}-\frac{\Lambda_{(g)} g_{1} \left(\ell +2\right) r^{-\ell -1}}{2}\right.\nonumber\\
&&+\left.\left(\ell -1\right) \left(g_{2} \left(N^{r}\right)^{2} \left(\ell -\frac{1}{2}\right) r^{\ell -2}
+\frac{\Lambda_{(g)} g_{2} r^{\ell}}{2}-r^{\ell -1} \dot{g}_{2} N^{r}\right)\right\},
\\
\label{epsilonk_2}
&&
\ddot{\chi}_{k}-2 N_{(f)}^{r} \dot{\chi}_{k}'
+[(N_{(f)}^{r})^{2}-1]\chi_{k}''
-\frac{C_{0}^{2} \Lambda_{(f)} r^{2}+3 (N_{(f)}^{r})^{2}}{2 r N_{(f)}^{r}}\dot{\chi}_{k}
\nonumber\\
&&
+\frac{C_{0}^{2} \Lambda_{(f)} r^{2}+(N_{(f)}^{r})^{2}-2 }{r}\chi_{k}'
+\frac{2 C_{0}^{2} \Lambda_{(f)} r^{2}-4 (N_{(f)}^{r})^{2}+\ell^2+\ell  }{r^{2}}\chi_{k}\nonumber\\
&&
=-\frac{m^{2}\left(C_{0} c_{1}+c_{2}\right) \left(1-\beta  C_{0}\right) r}{2 N_{(f)}^{r} C_{0} {\tilde\alpha}^{2}}
\left\{(N_{(f)}^{r})^{2} \left(\ell +\frac{3}{2}\right) \left(\ell +2\right) g_{1} r^{-\ell -3}+\dot{g}_{1} N_{(f)}^{r} \left(\ell +2\right) r^{-\ell -2}-\frac{C_{0}^{2} \Lambda_{(f)} g_{1} \left(\ell +2\right) r^{-\ell -1}}{2}\right.\nonumber\\
  &&
  +\left.\left(\ell -1\right) \left(g_{2} (N_{(f)}^{r})^{2} \left(\ell -\frac{1}{2}\right) r^{\ell -2}+\frac{C_{0}^{2} \Lambda_{(f)} g_{2} r^{\ell}}{2}-r^{\ell -1} \dot{g}_{2} N_{(f)}^{r}\right)\right\}.
\ee

\subsubsection{The case of $b\neq 1$}

In the case of $b\neq 1$,
after eliminating $\bH_t$, $\bH_r$, $\bK_t$, and $\bK_r$
in Eqs.~\eqref{master_eq_h} and \eqref{master_eq_k} with Eq.\ \eqref{HrHt_Sch}
and combining them appropriately,
we obtain the master equations
where the kinetic terms are diagonalized.
Although the master equations are too long to be shown explicitly,
their general structure is given by
\be
\label{eqchih}
&&
 \ddot{\chi}_h
+
M_1(r)
\chi_h
+
M_2(r)
\chi_k
+
L_{1}(r)
\dot{\chi}_h'
+
L_2(r)
\chi_h''
+
L_3(r)
\dot{\chi}_h
+
L_4(r)
\chi_h'
+
P_{1}(r)
\dot{\chi}_k'
+
P_{2}(r)
\chi_k''
+
P_3(r)
\dot{\chi}_k
+
P_4(r)
\chi_k'
\nonumber\\
&&
=
Q_1(r)
\dot{\bK}_3'
+
Q_2(r)
\bK_3''
+
Q_3(r)
\bK_3'
+
S_1(r)
\dot{\Lambda}'
+
S_2(r)
\Lambda''
+
S_3(r)
\Lambda',
\\
\label{eqchik}
&&
\ddot{\chi}_k
+
{\tilde M}_1(r)
\chi_k
+
{\tilde M}_2(r)
\chi_h
+
\tilde{L}_{1}(r)
\dot{\chi}_k'
+
\tilde{L}_{2}(r)
\chi_k''
+
\tilde{L}_{3}(r)
\dot{\chi}_k
+
\tilde{L}_{4}(r)
\chi_k'
+
{\tilde P}_1(r)
\dot{\chi}_h'
+
{\tilde P}_2(r)
\chi_h''
+
{\tilde P}_3(r)
\dot{\chi}_h
+
{\tilde P}_4(r)
\chi_h'
\nonumber\\
&&
=
{\tilde Q}_1(r)
\dot{\bK}_3'
+
{\tilde Q}_2(r)
\bK_3''
+
{\tilde Q}_3(r)
\bK_3'
+
{\tilde S}_1(r)
\dot{\Lambda}'
+
{\tilde S}_2(r)
\Lambda''
+
{\tilde S}_3(r)
\Lambda',
\ee
where the background-dependent coefficients
$M_{1,2}(r)$, $L_{1,2,3,4}(r)$, $P_{1,2,3,4}(r)$, $Q_{1,2,3}(r)$, $S_{1,2,3}(r)$,
${\tilde M}_{1,2}(r)$, ${\tilde L}_{1,2,3,4}(r)$, ${\tilde P}_{1,2,3,4}(r)$, ${\tilde Q}_{1,2,3}(r)$, $\tilde{S}_{1,2,3}(r)$
are the pure functions of $r$.
In the limit of $b\to 1$,
$M_2(r)$,
$P_{1,2,3,4}(r)$,
$Q_{1,2,3}(r)$,
${\tilde M}_2(r)$, 
${\tilde P}_{1,2,3,4}(r)$,
and 
${\tilde Q}_{1,2,3}(r)$
vanish,
and Equations.\ \eqref{eqchih} and \eqref{eqchik}
reduce to Eqs.\ \eqref{epsilonh} and \eqref{epsilonk}, respectively.

On the other hand, 
in the general case of $b\neq 1$
the above coefficients do not vanish,
and
the two dynamical modes $\chi_h$ and $\chi_k$ are coupled to each other,
respectively.
Eqs.\ \eqref{eqchih} and \eqref{eqchik},
together with Eqs.~\eqref{eq_sol_Lambda} and \eqref{eq_sol_K3_2},
form the closed system.
While the two modes $\chi_h$ and $\chi_k$ are dynamical,
the remaining ones $\bK_3$ and $\Lambda$ are shadowy.
From Eqs.\eqref{epsilonh} and \eqref{epsilonk},
the dynamical $\chi_h$ and $\chi_k$ are coupled
and sourced by the shadowy modes $\Lambda$ and $\bK_3$.
From Eqs.\ \eqref{eqchih} and \eqref{eqchik},
we define the mass matrix by 
\be
{\cal M}=
\begin{pmatrix}
M_1(r) & M_2 (r)
\\
{\tilde M}_2 (r) & {\tilde M}_1(r)
\end{pmatrix}.
\ee In the large distance limit $r\gg {\rm max} (r_{(g)},r_{(f)})$ and
assuming astrophysical scales, i.e.,\ neglecting the cosmological
constants terms, we find \be M_1(r) &=& M_{1,0} +{\cal O} \left(
\frac{1}{r} \right), \qquad M_2(r)= - M_{1,0} +{\cal O} \left(
\frac{1}{r} \right), \nonumber \\ {\tilde M}_1(r) &=& {\tilde M}_{1,0}
+{\cal O} \left( \frac{1}{r} \right), \qquad {\tilde M}_2(r) = -
{\tilde M}_{1,0} +{\cal O} \left( \frac{1}{r} \right), \ee where \be
M_{1,0} := \frac{m^2C_0^2(C_0c_1+c_2)(b-1)\ell (\ell+1)}
{2(\ell+2)(\ell-1)}, \qquad {\tilde M}_{1,0} :=
\frac{m^2(C_0c_1+c_2)b(b-1)\ell (\ell+1)}
     {2(\ell+2)(\ell-1){\tilde\alpha}^2}.  \ee This squared mass is of
     order of $m^2\sim H_0^2$ for cosmological implementation of this
     MTBG. Only for astrophysical purposes we
     can neglect its contribution to the dynamics of the fields.

So far, we have considered the parameter $b$ as a free parameter, 
as it is not fixed by any background equation of motion
for the spatially flat coordinates. 
However, as we have seen in Sec. \ref{sec3}, 
the collapse fixes the value of $b$ at least for the cases we have found a solution. 
Indeed,  in the spatially closed case, $b$ was set to unity,
whereas for the spatially flat case, $b$ was determined by the amount of matter 
energy densities
in the physical and fiducial sectors. 
However, we do not know all the collapse solutions,
and therefore there could be other possibilities for both the interior and exterior metrics.
In particular, as we have already discussed in the spatially closed collapsing case,
a time dependence on $b$ might appear that could determine a deviation from a stationary configuration for the exterior metric too.

Furthermore, we might wonder what boundary conditions we should impose
on the shadowy modes.  We believe that these topics have a nontrivial
answer, mostly because it may depend on the system we consider.  In
fact, we know that the theory possesses shadowy modes, and as such, we
should expect that the behavior of the solution to depend on the
boundary conditions which we fix on them.  These boundary conditions
might differ depending on the environment of the solution itself.  For
instance, if we were to consider a typical astrophysical system,
before reaching cosmological scales, we would encounter the
inhomogeneities induced by the matter distribution, such as the
baryonic matter contribution coming from the galaxy and then the dark
matter halo distribution. 
Therefore, in this case we would need to
match the astrophysical scale solution with an environment filled with
matter. On top of that, we would need to fix boundary conditions for
the shadowy modes at these same scales.

Instead if we were to describe an approximately spherically symmetric
large cluster of galaxies, then it would make sense to choose
cosmological boundary conditions to impose not only for the shadowy
modes but also for the value of $b$. Because of these possibilities, we
will leave the parameter $b$ as a free parameter.

\subsection{Conditions for the absence of ghost and gradient instabilities}

In this subsection, we are going to study the conditions 
for the absence of the ghost and gradient instabilities for the odd modes.
After eliminating $\bH_t$, $\bH_r$, $\bK_t$, and $\bK_r$ with Eq.\ \eqref{HrHt_Sch},
the second-order action \eqref{Sprime} can be rewritten as the functional of 
the two dynamical modes $\chi_h$, $\chi_k$ 
and two shadowy modes $\bK_3$, $\Lambda$.
The shadowy modes cannot be integrated out, in general, and they source the dynamical fields $\chi_{h}$
and $\chi_{k}$. 
So, in the following, we will also assume that the
shadowy modes $\Lambda$ and $\bK_{3}$ will not be given boundary conditions
as to drastically change the dynamics of the equations of motion for $\chi_{h}$ and $\chi_{k}$.
On neglecting a possible strong backreaction of the shadowy modes on $\chi_{h,k}$ 
assuming appropriate boundary conditions are imposed, 
we can find the kinetic matrix for $\chi_{h,k}$.

The no-ghost conditions can be obtained
by imposing the positivity of the eigenvalues 
of the kinetic matrix,~\footnote{To study the sign of the eigenvalues, it is sufficient to diagonalize the symmetric kinetic matrix via a simple congruence diagonalization method using a unitriangular matrix (see e.g.\ Ref.~\cite{lipschutz68}), which corresponds to make a field redefinition, with determinant equal to unity, for the perturbation variables. Then, the positivity conditions on the sign of eigenvalues are equivalent to imposing the diagonal elements to be positive.}
which is constituted by the coefficients of the $\dot{\chi}_h^2$, $\dot{\chi}_k^2$, and  $\dot{\chi}_h\dot{\chi}_k$ terms 
in the reduced second-order action.
In the following, 
we will also consider the case of the large $\omega$-$k_{r}$-$\ell$
case, so as to assume that time and $r$ derivatives give large contributions
as well as $\ell\gg1$, e.g.,\ $r|\chi'_{h}|\gg|\chi_{h}|$, etc.
When evaluated in the high-$\ell$ regime,
the off-diagonal components of the kinetic matrix are suppressed by $1/\ell^2$,
and the no-ghost conditions in this regime are given by 
\begin{equation}
\frac{C_{0}^{2}\tilde{\alpha}^{2}r^{2}}{4b^{3}}>0\,,\qquad\frac{r^{2}}{4}>0\,,
\end{equation}
which are both trivially satisfied.

As for the speed of propagation, we can consider the leading contribution
in the equations of motion for $\chi_{h}$ and $\chi_{k}$, once more,
on neglecting backreactions from the shadowy modes and assuming $m\simeq H_{0}$.
In this case the equations of motion simplify to
\begin{align}
\ddot{\chi}_{h} & -2N^{r}\dot{\chi}'_{h}+[(N^{r})^{2}-1]\,\chi''_{h}+\frac{\ell(\ell+1)}{r^{2}}\,\chi_{h}\simeq0\,,\\
\ddot{\chi}_{k} & -2N_{(f)}^{r}\dot{\chi}'_{k}+[(N_{(f)}^{r})^{2}-b^{2}]\,\chi''_{k}+\frac{b^{2}\ell(\ell+1)}{r^{2}}\,\chi_{k}\simeq0\,.\label{eq:chi_k_eq}
\end{align}
The first case is simple, as it reduces to the leading contributions
for a massless scalar propagating in the physical metric and the speed
of propagation is evidently $c_{s,g}^{2}=1$. As for the second equation,
in order to make the analysis simpler, one can perform a coordinate
transformation, namely,
\begin{align}
dt  =d\tau,
\qquad
dr  =N_{(f)}^{r}\,(d\rho-d\tau)\,,
\end{align}
for which $r=r[\rho-\tau]$. This leads to a metric in the so-called
Lema{\^i}tre coordinates as in
\begin{equation}
ds_{f}^{2}=C_{0}^{2}\,[-b^{2}d\tau^{2}+(N_{(f)}^{r})^{2}\,d\rho^{2}+r [\rho-\tau]^{2}\,\theta_{ab}d\theta^{a}d\theta^{b}]\,.
\end{equation}
After little algebra, one finds
\begin{align}
\frac{\partial}{\partial t}  =\frac{\partial}{\partial\tau}+\frac{\partial}{\partial\rho}\,,
\qquad
\frac{\partial}{\partial r}  =\frac{1}{N_{(f)}^{r}}\,\frac{\partial}{\partial\rho}\,,
\end{align}
where we have used the conditions holding for a coordinate basis,
$dt(\partial_{t})=1$, $dt(\partial_{r})=0$, etc. In this case Eq.\ (\ref{eq:chi_k_eq})
reduces to
\begin{equation}
\frac{\partial^{2}\chi_{k}}{\partial\tau^{2}}-\frac{b^{2}}{(N_{(f)}^{r})^{2}}\,\frac{\partial^{2}\chi_{k}}{\partial\rho^{2}}+\frac{b^{2}\ell(\ell+1)}{r^{2}}\,\chi_{k}\simeq0\,.
\end{equation}
If we were assuming the mode traveling on the spacetime of the metric
$ds_{f}^{2}=-d\tau^{2}+(N_{(f)}^{r})^{2}\,d\rho^{2}+r[\rho-\tau]^{2}\,\theta_{ab}d\theta^{a}d\theta^{b}$,
then the speed of both the radial and angular propagations is $c_{f}^{2}=b^{2}$,  the proper radial
distance being $(N_{(f)}^{r})\,d\rho$ \cite{Takahashi:2021bml},
which does not lead to any new condition to impose.

\subsection{The dipolar modes}
\label{sec5d}

\subsubsection{The general solution}

For the dipolar mode $\ell=1$, $K_{3}$ and $H_{3}$ automatically vanish.
Then, the second-order action is a functional of $H_{t},H_{r},K_{t}$ and $K_{r}$ together
with $\Lambda$.
However, the method employed in Sec.\ \ref{sec5b} can be implemented.

The second-order action for the $\ell=1$ mode is given by
\be
\label{Soriginal_L1}
\delta_{(2)} S_{\ell=1}
&=&
\frac{2}
        {3}
\int dt dr
L_1
\left[
H_t, 
H_r,
K_t,
K_r,
\Lambda
\right].
\ee
As in Sec.\ \ref{sec5b}, we introduce the new variables $\chi_h$ and $\chi_k$,
\be
\label{Sprime_L1}
\delta_{(2)}
S'_{\ell=1}
&=&
\frac{2}{3}
\int dt dr
\Big\{
L_1
\left[
H_t, 
H_r,
K_t,
K_r,
\Lambda
\right]
-A_{1}(r)
\left(
\dot{H}_r
+
A_{2}(r)
H_r'
+
A_{3}(r)
H_t'
+
A_{4} (r)
H_r
-\frac{\chi_h}{r}
\right)^2
\nonumber
\\
&&
-B_{1}(r)
\left(
\dot{K}_r
+
B_{2}(r)
K_r'
+
B_{3}(r)
K_t'
+
B_{4} (r)
K_r
-\frac{\chi_k}{r}
\right)^2
\Big\}.
\ee
Varying the new second-order action $\delta_{(2)}S_{\ell=1}'$
with respect to $\chi_h$ and $\chi_k$,
we obtain
\be
\label{onshell_eq_h_L1}
\dot{H}_r
+
A_{2}(r)
H_r'
+
A_{3}(r)
H_t'
+
A_{4} (r)
H_r
-\frac{\chi_h}{r}=0,
\\
\label{onshell_eq_k_L1}
\dot{K}_r
+
B_{2}(r)
K_r'
+
B_{3}(r)
K_t'
+
B_{4} (r)
K_r
-\frac{\chi_k}{r}
=0.
\ee
Varying $\delta_{(2)}S_{\ell=1}'$
with respect to $H_t$, $H_r$, $K_t$, $K_r$,
we obtain the equations of motion for them.
Requiring that the derivatives of  $H_t$, $H_r$, $K_t$, $K_r$ vanish in their equations,
the coefficients in Eq.\ \eqref{Sprime_L1} are given by Eq.~\eqref{ABs}.

In the case of $b\neq1$ as well as $C_{0}c_{1}+c_{2}\neq0$,
we can integrate out $H_{r}$ using its own algebraic
equation of motion, as in
\begin{eqnarray}
\label{hrkr}
H_{r}=\frac{r\left(\beta C_{0}-1\right)\Lambda'}{C_{0}\left(b-1\right)}+K_{r}+\frac{\left(2N^{r}\chi'_{h}-2\dot{\chi}_{h}\right)r+4N^{r}\chi_{h}}{r^{2}C_{0}^{2}\left(b-1\right)\left(C_{0}c_{1}+c_{2}\right)m^{2}}\,.\label{eq:H_sz_l_eq_1}
\end{eqnarray}
After inserting this expression into Eq.~\eqref{Sprime_L1}, we have that
 $H_{t}$, $K_{t}$ and $K_{r}$ are all Lagrange multipliers
that enter only linearly as to set constraints on the other fields.
In any case, we can now find all the relevant equations of motion and
try to solve them. In particular, the constraint for $H_{t}$ reads
\begin{equation}
\label{fh}
r\chi_{h}'+2\chi_{h}=0\,,\qquad\chi_{h}=\frac{F_{h}(t)}{r^{2}}\,,
\end{equation}
whereas the constraint for $K_{t}$ can be written and solved as
\begin{equation}
\label{fk}
r\chi_{k}'+2\chi_{k}=0\,,\qquad\chi_{k}=\frac{F_{k}(t)}{r^{2}}\,.
\end{equation}
Using these solutions, the constraint for the field $K_{r}$ reads
\begin{equation}
C_{0}^{2}\tilde{\alpha}^{2}\dot{F}_{k}+b\dot{F}_{h}=0\,,\qquad{\rm where}\qquad F_{k}=F_{\mathit{k}0}-\frac{bF_{h}}{C_{0}^{2}\tilde{\alpha}^{2}}\,.
\end{equation}.
On using these solutions in the equation of motion for $\Lambda$,
we find
\begin{equation}
r^{2}\Lambda''+4r\Lambda'=0\,,\qquad\Lambda=g_{2}(t)+\frac{g_{1}(t)}{r^{3}}\,.
\end{equation}
We can evaluate the following gauge-invariant combinations
for the $\ell=1$ mode:
\begin{align}
\label{gr}
G_{r}&:= H_{r}-K_{r},
\\
\label{gt}
G_{t} & := H_{t}-K_{t}+\frac{N^{r}}{r}\,H_{r}-\frac{N_{(f)}^{r}}{r}\,K_{r}\,,
\\
\label{gh}
G_h& :=H'_{t}-\frac{\dot{H}_{r}}{r}+\left(\frac{N^{r}H_{r}}{r}\right)'\,,
\\
\label{gk}
G_{k} & := K'_{t}-\frac{\dot{K}_{r}}{r}+\left(\frac{N_{(f)}^{r}K_{r}}{r}\right)'\,.
\end{align}
From Eqs.~\eqref{hrkr}, \eqref{fh} and \eqref{fk}, we obtain
\be
\label{gr2}
G_r=\frac{3\left(1-\beta C_{0}\right)g_{1}}{C_{0}\left(b-1\right)r^{3}}-\frac{2\dot{F}_{h}}{r^{3}C_{0}^{2}\left(b-1\right)\left(C_{0}c_{1}+c_{2}\right)m^{2}}.
\ee
Considering a linear combination of the equations of motion for
$\chi_{h}$ and $\chi_{k}$, we find
\begin{equation}
G'_{t}=\frac{F_{\mathit{k0}}}{r^{4}}-\frac{\left(\tilde{\alpha}^{2}C_{0}^{2}+b\right)F_{h}}{\tilde{\alpha}^{2}r^{4}C_{0}^{2}}+\frac{3\left(1-\beta C_{0}\right)\dot{g}_{1}}{C_{0}r^{4}\left(b-1\right)}-\frac{2\ddot{F}_{h}}{C_{0}^{2}r^{4}m^{2}\left(C_{0}c_{1}+c_{2}\right)\left(b-1\right)}\,,
\end{equation}
which can be easily integrated. Finally, the equation of motion for
$\chi_{k}$ gives
\begin{equation}
G_{k}=\frac{bF_{h}}{r^{4}C_{0}^{2}\tilde{\alpha}^{2}}-\frac{F_{\mathit{k}0}}{r^{4}}\,.
\end{equation}
We can also find  
\begin{equation}
G_{h}=-\frac{F_{h}}{r^{4}}\,.
\end{equation}

\subsubsection{The slow-rotation limit of the Kerr$-$de Sitter solution}

By introducing a function of time $k_0(t)$ by
\be
\label{hrkr2}
G_r
=
-\frac{k_0(t)}{r^3}, 
\ee
with Eq.~\eqref{gr2}, 
$g_1(t)$ can be eliminated as
\be
g_1(t)=\frac{(b-1)C_0^2(C_0c_1+c_2)m^2 k_0(t)-2\dot{F}_h(t)}
                 {3C_0(C_0c_1+c_2)(C_0\beta-1)m^2}.
\ee
As the solution for the physical sector,
we consider the slow-rotation limit of the Kerr$-$de Sitter solution
\be
H_t
=
-
\frac{1}{1-(N^r)^2}
\left(
\omega_0+\frac{2J_{(g)0}}{r^3}
\right),
\qquad 
H_r
=
\frac{r N^r}{1-(N^r)^2}
\left(
\omega_0+\frac{2J_{(g)0}}{r^3}
\right),\label{eq:Kerr_fields}
\ee
where $\omega_0$ and $J_{(g)0}$ are constants,
which satisfies $H_r+rN^r H_t=0$.
This solution describing the slow-rotation limit of the Kerr$-$de Sitter solution
has been found as follows.
First, we consider the limit for small rotations in Boyer-Lindquist coordinates for the Kerr$-$de Sitter metric,
which reduces to the Schwarzschild$-$de Sitter contribution 
written in the Schwarzschild coordinates plus a term of the form $ds^2\ni-2a(N^r)^2(1-z^2)dt_sd\varphi_s$,
where $a$ is the spin angular momentum per unit mass, $z:=\cos\theta$, and $t_s$
is the time of the Schwarzschild coordinates. 
Since we want the metric in the GP coordinates,
we make the coordinate transformation $dt_s=dt+F'\,dr$ with $F'=-N^r/(1-(N^r)^2)$ together with a time-dependent shift in the angular variable, as in $\varphi_s=\varphi+\omega_0\,t$, describing a reference system rotating uniformly, not due to gravitational effects.
These transformations lead to a metric that can be expanded in terms of both $\omega_0$ and $a$. At lowest order, we find the Schwarzschild$-$de Sitter metric written in GP coordinates, as expected.
At linear level, we find two contributions that can be written as a contribution to the shift, $g_{t\varphi}=(\omega_0r^2-a\,(N^r)^2)(1-z^2)$, and an element $g_{r\varphi}=a\xi^3(1-z^2)/(1-(N^r)^2)$.
In turn, matching these elements to the perturbation variables for the dipolar modes in the physical sector
gives the modes $\tilde{H}_t=\frac{\left(-\omega_0  (N^{r})^{2}+\omega_0 \right) r^{3}+2 J_{(g)0}}{\left(({N^{r}})^{2}-1\right) r^{3}}$ and $\tilde{H}_r=-\frac{2 J_{(g)0} N^{r}}{r^{2} \left(({N^{r}})^{2}-1\right)}$,
where we have replaced the spin angular momentum per unit mass $a$ with $-2J_{(g)0}/r_{(g)}$.
On performing an $r$-dependent gauge transformation defined by $\Xi'(r)=\frac{N^{r} \omega_0}{({N^{r}})^{2}-1}$,
 we find the fields\footnote{It should be noted that the combination $H_t+N^r H_r/r$ for an $r$-dependent gauge transformation becomes a gauge-invariant quantity; as such, it is also equal to ${\tilde H}_t+N^r {\tilde H}_r/r=-\omega_{0}-\frac{2 J}{r^{3}}$.} in the form given in Eq.\ (\ref{eq:Kerr_fields}).

The constant $J_{(g)0}$ is then linked to the angular momentum 
in the slow-rotation limit of a Kerr$-$de Sitter black hole.
With Eq.\ \eqref{hrkr2}, the solution for $K_r$ is then
\be
K_r=
\left(
\omega_0+\frac{2J_{(g)0}}{r^3}
\right)
\frac{r N^r}{1-(N^r)^2}
+\frac{k_0(t)}{r^3}.
\ee
The equation of motion for $\chi_h$ provides
\be
F_h=-6J_{(g)0}.
\ee
The equation of motion for $\chi_k$ can then be solved as
\be
K_t
=
-
\frac{2J_{(f)0}}
        {r^3}
-
\frac{N^r N^r_{(f)}}{1-(N^r)^2}
\left(
\omega_0+\frac{2J_{(g)0}}{r^3}
\right)
-\frac{1}{r^3}
\left(
\frac{1}{r}
N_{(f)}^r
k_0(t)
+
\frac{\dot{k}_0(t)}{3}
\right)
-k_3(t),
\ee
where the constant $J_{(f)0}$ is introduced by fixing
\be
F_{k0}
=
-
\frac{6}{C_0^2{\tilde\alpha}^2}
\left(
b J_{(g)0}
+
C_0^2{\tilde\alpha}^2
J_{(f)0}
\right),
\ee
and $k_3(t)$ is an integration constant.
In the limit of $r\to \infty$, the perturbed metrics behave as 
\be
&&
g_{ta}
\to 
-\omega_0 r^2-\frac{2J_{(g)0}}{r},
\qquad 
g_{ra}
\to 
-\sqrt{\frac{3}{\Lambda_{(g)}}}
\omega_0 r,
\\
&&
f_{ta}
\to 
-k_3(t) r^2
-\frac{1}{r}
\left(
2J_{(f)0}
+
\frac{k_0'(t)}
        {3}
\right),
\qquad 
f_{ra}
\to 
-\sqrt{\frac{3}{\Lambda_{(g)}}}
\omega_0 r.
\ee
For the above solution, the gauge-invariant quantities \eqref{gr}-\eqref{gk}
reduce to 
\be
&&
G_r
=-\frac{k_0}{r^3},
\qquad
G_t
=
\frac{6(J_{(f)0}-J_{(g)0})+k_0'(t)}{3r^3}
-
\left(
\omega_0-k_3(t)
\right),
\qquad 
G_h=
\frac{6J_{(g)0}}{r^4},
\qquad 
G_k=
\frac{6J_{(f)0}}{r^4}.
\ee
The leading terms
$\omega_0$ and $k_3(t)$
are fixed by the boundary conditions at the spatial infinity.
Then, the constants $J_{(g)0}$ and $J_{(f)0}$ correspond to the angular momenta in the 
physical and fiducial sectors.
Thus, the general solution for the $\ell=1$ mode in the fiducial sector
deviates from the slow-rotation approximation of the Kerr$-$de Sitter metric.
Instead, 
in the case where the fiducial metric is given by the slow-rotation limit of 
the Kerr$-$de Sitter metric, 
the general solution in the physical sector
deviates from the slow-rotation limit of the Kerr$-$de Sitter metric.

By setting $k_0=0$, then $G_r=0$, we obtain
\be
&&
K_t=
H_t
+
\frac{2}{r^3}
\left(
\frac{1-N^r N^r_{(f)}}
        {1-(N^r)^2}
J_{(g)0}
-J_{(f)0}
\right)
+
\left(
\frac{1-N^r N^r_{(f)}}
        {1-(N^r)^2}
\omega_0
-k_3(t)
\right).
\ee
In the special case where 
$N^r_{(f)}=N^r$
and $J_{(f)0}=J_{(g)0}$,
\be
K_t=
H_t
+
\left(
\omega_0
-k_3(t)
\right).
\ee
which coincide up to the difference between $\omega_0$ and $k_3(t)$
fixed by the boundary conditions at the spatial infinity.
Namely, 
in this case,
we obtain the identical Kerr$-$de Sitter metrics in the slow-rotation limit in both the sectors.

\section{Conclusions}
\label{sec5}

As a continuation of our previous work \cite{Minamitsuji:2022vfv}, 
we have studied the dynamical processes of spherically symmetric systems in MTBG.
In particular, we have focused on gravitational collapse of pressureless dust 
and odd-parity perturbations of static and spherically symmetric 
Schwarzschild$-$de Sitter black holes.
Throughout the paper,
we have focused on the self-accelerating branch satisfying the condition \eqref{sa_cond}.

In Sec.~\ref{sec3}, we have found the exact solutions describing
gravitational collapse of pressureless dust, where the interior
spacetime geometries are written with the spatially flat and closed
FLRW universes, respectively, and the exterior vacuum solutions
are described by the Schwarzschild solutions with specific time slicings.
For simplicity, we foliated the interior geometry by homogeneous and isotropic spacetimes. 
For a spatially flat interior universe, we foliated the exterior geometry by a time-independent spatially flat space,
while for a spatially curved interior universe,
we foliated the exterior geometry by a time-independent spacetime with deficit solid angle. 
Despite the rather restrictive choice of foliations, 
we have successfully found interesting classes of exact solutions that represent gravitational collapse in MTBG.

While in the spatially flat case the
matter surfaces start to collapse from the spatial infinities with zero initial velocities
or from finite distances with nonzero initial velocities,
in the spatially closed case they started from finite
radii with vanishing initial velocities.  The spatially closed case
corresponds to an extension of the Oppenheimer-Snyder model to the
case of MTBG.  Since the Lagrange multipliers are trivial and
continuous across the interfaces of the collapsing matter in both the
physical and fiducial sectors, the junction conditions across them
remain the same as those in the two copies of GR.

In the spatially flat case, under the above-mentioned tuning 
of the initial conditions of the collapse, we have obtained 
solutions in which gravitational collapse in the physical and
fiducial sectors proceed independently and the gravitational radii in
the two sectors may be different. On the other hand, 
in the spatially closed case, under certain tuning of the matter energy densities and 
Schwarzschild radii between the two sectors, we have found exact solutions of gravitational collapse. Needless to say, these tunings reflect the fact that we restrict our considerations to  
a rather specific choice of time slicings. It is expected that relaxing the restriction on the choice of time slicings,
e.g.\ to those with time-dependent spatial metrics, 
should allow for more general solutions.
Furthermore, since in general the Birkhoff theorem does not hold in MTBG, 
yet-unknown non-Schwarzschild exterior solutions may be found and then even wider classes of global solutions representing gravitational collapse may be constructed. It is also worthwhile studying spherically-symmetric inhomogeneous dust collapse, i.e., analog of the Lema\^{i}tre-Tolman-Bondi model \cite{Lemaitre:1933gd,Tolman:1934za,Bondi:1947fta}, 
in the context of MTBG and investigating the homogeneous limit. 
Moreover, 
it would be also important to investigate deviations from the spherical GR metric solutions
in the weak-field limit
and relate them to the parametrized post-Newtonian (PPN) parameters (see e.g. Ref. \cite{Will:2014kxa}).
Since the absence of the Birkhoff theorem 
is due to the presence of the effective energy-momentum tensor 
arising from the variation of the interaction and constraint terms of the Lagrangian \eqref{lag} and \eqref{lag2},
in the weak-field limit the deviation from the GR solution 
would be proportional to $m^2$,
which may give rise to a new bound on the graviton mass parameter $m$ with the Solar System experiments.
We leave the explicit PPN formulation for future work.

In Sec.\ \ref{sec4}, we studied odd-parity perturbations of the Schwarzschild$-$de Sitter black holes in the self-accelerating branch of  MTBG. 
As the background solutions, 
we considered the Schwarzschild$-$de Sitter solutions written in the spatially flat coordinates.
In the equations of motion for the odd-parity perturbations, 
there were the contributions of the interaction terms and the Lagrange multipliers
in addition to those from the two copies of the Einstein-Hilbert terms.
We distinguished the dipolar mode with the angular multipole moment $\ell=1$
and the higher-multipolar modes $\ell\geq 2$. 
In order to discuss the odd-parity perturbations with $\ell\geq 2$,
we introduced the master variables.
For the modes $\ell\geq 2$, the system of the odd-parity perturbations
in the self-accelerating branch 
is composed of the two dynamical modes and the two shadowy modes.
The dynamical modes correspond to the two master variables
of the metric perturbations,
while the shadowy modes
correspond to the gauge-invariant perturbation of the Lagrange multiplier
and one of the two metric perturbations in the angular directions.
We note that the other odd-parity metric perturbation in the angular directions
is a gauge mode and can be set to be zero.
The equation of motion for one of the two shadowy modes
can be easily integrated as it is not coupled to the other modes,
and the other shadowy mode
is sourced by the two dynamical modes and the first shadowy mode.
The two dynamical modes are also coupled to each other
and sourced by the shadowy modes.

In the case where the ratio of the lapse functions between the 
physical and fiducial sectors is equal to $C_0$, which is a constant determined by the
parameters of the theory, the two dynamical modes are decoupled.
Since they are still sourced by one of the shadowy modes, their
behavior is different from that in the two copies of GR and depends on
the boundary conditions of the shadowy mode.
In the case where the ratio of the lapse functions between the
physical and fiducial sectors differs from $C_0$, the two dynamical modes are coupled to
each other and sourced by both the shadowy modes.  In the high
frequency and short wavelength limits, we showed that the
perturbations do not suffer from ghost or gradient instabilities.
On the other hand,
in the dipolar sector of $\ell=1$, 
we found that the two copies of the slow-rotation limit of the Kerr$-$de Sitter metrics, in general, are not a solution in the self-accelerating branch of MTBG,
unless the masses and angular momenta of black holes and
the effective cosmological constants are tuned to be the same
in both the sectors.
Therefore, deviation from GR is expected for rotating black holes in the self-accelerating branch of MTBG.
Readers should also refer to the last paragraph in Sec.\ \ref{sec4} of \cite{DeFelice:2019jxs} for a similar statement in the context of MTMG.

The analysis of the even-parity black hole perturbations in the self-accelerating branch
will be left for future work. 
The analysis of the same issues in the normal branch, 
i.e., gravitational collapse and black hole perturbations,
will also be left for future work.
We expect that the spacetime dynamics in the normal branch 
would differ from that in the case of the two copies of GR
at both the background and perturbation levels.
We hope to come back to these issues in a future publication.

\section*{ACKNOWLEDGMENTS}
The work of A.D.F.\ was supported by Japan Society for the Promotion of Science Grants-in-Aid for Scientific Research No.~20K03969.
M.M.~was supported by the Portuguese national fund through the Funda\c{c}\~{a}o para a Ci\^encia e a Tecnologia in the scope of the framework of the Decree-Law 57/2016 of August 29, changed by Law 57/2017 of July 19, and the Centro de Astrof\'{\i}sica e Gravita\c c\~ao through the Project~No.~UIDB/00099/2020.
M.M.~also would like to thank Yukawa Institute for Theoretical Physics for their hospitality under the Visitors Program of FY2022.
The work of S.M.\ was supported in part by Japan Society for the Promotion of Science Grants-in-Aid for Scientific Research No.\ 17H02890, No.\ 17H06359, and by World Premier International Research Center Initiative, MEXT, Japan. M.O.~would like to thank Yukawa Institute for Theoretical Physics for its institutional support.

\appendix

\section{Equations of motion for the odd-parity perturbations in the spatially flat coordinates in MTBG}
\label{app_c}

We show the equations of motion for the odd-parity perturbations
about the Schwarzschild$-$de Sitter solutions
written in the spatially flat coordinates
in the self-accelerating branch of MTBG.

\subsection{The $\ell\geq 2$ modes}

The equations of motion for the gauge invariants $\bH_t$, $\bH_r$, $\bK_t$, $\bK_r$, and $\bK_3$
in the odd-parity perturbations
for the $\ell\geq 2$ modes
are given by
\be
\label{eom_ht}
0
&=&
\dot{\bH}_r'
-
N^r(r)
  \bH_{r}''
-r \bH_{t}''
+\frac{3}{r}\dot{\bH}_r
\nonumber\\
&&
-
\frac{2}{r} \left(N^r(r)+r N^r{}'(r)\right)
  \bH_r'
-4 \bH_t'
+\frac{1}{r}\left(\ell-1\right)\left(\ell+2\right)
  \bH_t
+
\left(
\frac{2N^r(r)}{r^2}
-
\frac{2N^r{}'(r)}{r}
-N^r{}''(r)
\right)
  \bH_{r},
\\
\label{eom_hr}
0&=&
\ddot{\bH}_r
-
2N^r(r)
\dot{\bH}_r'
+
\left(N^r(r)\right)^2
\bH_r''
-r \dot{\bH}_t'
+
r N^r(r)
\bH_t''
-
\left(
N^r{}'(r)
+\frac{2}{r}
N^r (r)
\right)
\dot{\bH}_r
\nonumber\\
&&
+
\frac{2N^r(r)}{r}
\left(
N^r(r)+rN^r{}'(r)
\right)
\bH_r'
+
4
N^r(r)
\bH_t'
\nonumber\\
&&
+
\left[
\frac{(\ell +2)(\ell-1)}{r^2}
+(2C_0^3c_1+3C_0^2c_2-c_4)m^2
+
N^r(r)
\left(
N^r{}''(r)
+
\frac{6}{r}
N^r{}'(r)
\right)
\right]
\bH_r
\nonumber\\
&&
+\frac{m^2  C_0^2 \left(b-1\right)\left(C_0c_1+c_2\right)}{2}
\left(
\bH_r-\bK_r-\frac{r}{2}\bK_3'
\right)
+\frac{m^2C_0\left(C_0 c_1 +c_2\right)\left(1-\beta C_0 \right)}{2}
r\Lambda',
\\
\label{eom_kt}
0
&=&
\dot{\bK}_r'
-
 N^r_{(f)}(r)
  \bK_{r}''
-r \bK_{t}''
+\frac{3}{r}\dot{\bK}_r
\nonumber\\
&&
-
\frac{2}{r} \left(N_{(f)}^r(r)+r N_{(f)}^r{}'(r)\right)
  \bK_r'
-4 \bK_t'
+\frac{1}{r}\left(\ell-1\right)\left(\ell+2\right)
  \bK_t
+
\left(
\frac{2N_{(f)}^r(r)}{r^2}
-
\frac{2N_{(f)}^r{}'(r)}{r}
-N_{(f)}^r{}''(r)
\right)
  \bK_{r},
\\
\label{eom_kr}
0
&=&
\ddot{\bK}_r
-
2N^r_{(f)}(r)
\dot{\bK}_r'
+
\left(N^r_{(f)}(r)\right)^2
\bK_r''
-r \dot{\bK}_t'
+
rN^r_{(f)}(r)
\bK_t''
-
\left(
\frac{2N^r_{(f)}(r)}
        {r}
+N^r_{(f)}{}'(r)
\right)
\dot{\bK}_r
\nonumber\\
&&
+
\frac{2N^r_{(f)}}{r}
\left(
N^r_{(f)}(r)
+
r
N^r_{(f)} {}'(r)
\right)
\bK_r'
+4
N^r_{(f)}(r)
\bK_t'
\nonumber\\
&&
+
\left[
\frac{b^2(\ell+2) (\ell-1)}{r^2}
-b^2
\frac{(c_0 C_0^2+2C_0c_1+c_2)m^2}
        {{\tilde\alpha}^2}
+
N^r_{(f)}(r)
\left(
N^r_{(f)}{}''(r)
+
\frac{6}{r}
N^r_{(f)}{}'(r)
\right)
\right]
\bK_r
\nonumber\\
&&
-
\frac{m^2 b\left(b-1\right) (C_0c_1+c_2)}{2{\tilde \alpha}^2} 
\left(
\bH_r
-\bK_r
-\frac{1}{2}r \bK_3'
\right)
-
\frac{m^2b \left(c_1 C_0 +c_2\right)\left(1-\beta C_0 \right)}{2C_0{\tilde \alpha}^2} 
r\Lambda',
\\
\label{eom_k3}
0&=&
m^2
\left[
(1-b)C_0
\left(
r^2 {\bK_3}''
+4  r  {\bK_3}'
-(\ell+2)(\ell-1)\bK_3
+2 r \left(\bK_r'-\bH_r'\right)
+6\left(\bK_r-\bH_r\right)
\right)
\right.
\nonumber\\
&&
\left.
+
2(1-C_0\beta)
\left(
r^2\Lambda''
+
4r \Lambda'
-
\left(\ell+2\right)\left(\ell-1\right)
\Lambda
\right)
\right],
\\
\label{eom_lambda}
0&=&
m^2
\left[
r^2 {\bK_3}''
+4 r  {\bK_3}'
-(\ell+2)(\ell-1)\bK_3
+2r \left(\bK_r'-\bH_r'\right)
+6 \left(\bK_r-\bH_r\right)
\right].
\ee

\subsection{The dipolar mode}

The equations of motion for the odd-parity perturbations 
for the dipolar $\ell =1$ mode are given by 
\be
\label{eom_ht_ell1}
0
&=&
\dot{H}_r'
-
N^r(r)
  H_{r}''
-r H_{t}''
+\frac{3}{r}\dot{H}_r
\nonumber\\
&&
-\frac{2}{r} \left(N^r(r)+r N^r{}'(r)\right)
  H_r'
-4 H_t'
+
\left(
\frac{2N^r(r)}{r^2}
-
\frac{2N^r{}'(r)}{r}
-N^r{}''(r)
\right)
  H_{r},
\\
\label{eom_hr_ell1}
0&=&
\ddot{H}_r
-
2N^r(r)
\dot{H}_r'
+
\left(N^r(r)\right)^2
H_r''
-r \dot{H}_t'
+%
r N^r(r)
H_t''
-
\left(
N^r{}'(r)
+\frac{2}{r}
N^r (r)
\right)
\dot{H}_r
\nonumber\\
&&
+
\frac{2N^r(r)}{r}
\left(
N^r(r)+rN^r{}'(r)
\right)
H_r'
+
4
N^r(r)
H_t'
\nonumber\\
&&
+
\left[
(2C_0^3c_1+3C_0^2c_2-c_4)m^2
+
N^r(r)
\left(
N^r{}''(r)
+
\frac{6}{r}
N^r{}'(r)
\right)
\right]
H_r
\nonumber\\
&&
+\frac{m^2  C_0^2 \left(b-1\right)\left(C_0c_1+c_2\right)}{2}
\left(
H_r-K_r
\right)
+\frac{m^2C_0\left(C_0 c_1 +c_2\right)\left(1-\beta C_0 \right)}{2}
r\Lambda',
\\
\label{eom_kt_ell1}
0&=&
\dot{\bK}_r'
-
 N^r_{(f)}(r)
  K_{r}''
-r K_{t}''
+\frac{3}{r}\dot{K}_r
\nonumber\\
&&
-
\frac{2}{r} \left(N_{(f)}^r(r)+r N_{(f)}^r{}'(r)\right)
  K_r'
-4 K_t'
+
\left(
\frac{2N_{(f)}^r(r)}{r^2}
-
\frac{2N_{(f)}^r{}'(r)}{r}
-N_{(f)}^r{}''(r)
\right)
  K_{r},
\\
\label{eom_kr_ell1}
0
&=&
\ddot{K}_r
-
2N^r_{(f)}(r)
\dot{K}_r'
+
\left(N^r_{(f)}(r)\right)^2
K_r''
-r \dot{K}_t'
+
rN^r_{(f)}(r)
K_t''
-
\left(
\frac{2N^r_{(f)}(r)}
        {r}
+N^r_{(f)}{}'(r)
\right)
\dot{K}_r
\nonumber\\
&&
+
\frac{2N^r_{(f)}}{r}
\left(
N^r_{(f)}(r)
+
r
N^r_{(f)} {}'(r)
\right)
K_r'
+4
N^r_{(f)}(r)
K_t'
\nonumber\\
&&
+
\left[
-b^2
\frac{(c_0 C_0^2+2C_0c_1+c_2)m^2}
        {{\tilde\alpha}^2}
+
N^r_{(f)}(r)
\left(
N^r_{(f)}{}''(r)
+
\frac{6}{r}
N^r_{(f)}{}'(r)
\right)
\right]
K_r
\nonumber\\
&&
-
\frac{m^2 b\left(b-1\right) (C_0c_1+c_2)}{2{\tilde \alpha}^2} 
\left(
H_r
-K_r
\right)
-
\frac{m^2b \left(c_1 C_0 +c_2\right)\left(1-\beta C_0 \right)}{2C_0{\tilde \alpha}^2} 
r\Lambda',
\\
\label{eom_lambda_ell1}
0&=&
m^2
\left(C_0c_1 +c_2\right)
\left(1-\beta C_0 \right)
\left[
r\left(H_r'-K_r'\right)
+3\left(H_r-K_r\right)
\right].
\ee

\bibliography{refs}

\begin{thebibliography}{44}%
\makeatletter
\providecommand \@ifxundefined [1]{%
 \@ifx{#1\undefined}
}%
\providecommand \@ifnum [1]{%
 \ifnum #1\expandafter \@firstoftwo
 \else \expandafter \@secondoftwo
 \fi
}%
\providecommand \@ifx [1]{%
 \ifx #1\expandafter \@firstoftwo
 \else \expandafter \@secondoftwo
 \fi
}%
\providecommand \natexlab [1]{#1}%
\providecommand \enquote  [1]{``#1''}%
\providecommand \bibnamefont  [1]{#1}%
\providecommand \bibfnamefont [1]{#1}%
\providecommand \citenamefont [1]{#1}%
\providecommand \href@noop [0]{\@secondoftwo}%
\providecommand \href [0]{\begingroup \@sanitize@url \@href}%
\providecommand \@href[1]{\@@startlink{#1}\@@href}%
\providecommand \@@href[1]{\endgroup#1\@@endlink}%
\providecommand \@sanitize@url [0]{\catcode `\\12\catcode `\$12\catcode
  `\&12\catcode `\#12\catcode `\^12\catcode `\_12\catcode `\%12\relax}%
\providecommand \@@startlink[1]{}%
\providecommand \@@endlink[0]{}%
\providecommand \url  [0]{\begingroup\@sanitize@url \@url }%
\providecommand \@url [1]{\endgroup\@href {#1}{\urlprefix }}%
\providecommand \urlprefix  [0]{URL }%
\providecommand \Eprint [0]{\href }%
\providecommand \doibase [0]{https://doi.org/}%
\providecommand \selectlanguage [0]{\@gobble}%
\providecommand \bibinfo  [0]{\@secondoftwo}%
\providecommand \bibfield  [0]{\@secondoftwo}%
\providecommand \translation [1]{[#1]}%
\providecommand \BibitemOpen [0]{}%
\providecommand \bibitemStop [0]{}%
\providecommand \bibitemNoStop [0]{.\EOS\space}%
\providecommand \EOS [0]{\spacefactor3000\relax}%
\providecommand \BibitemShut  [1]{\csname bibitem#1\endcsname}%
\let\auto@bib@innerbib\@empty
\bibitem [{\citenamefont {Clifton}\ \emph {et~al.}(2012)\citenamefont
  {Clifton}, \citenamefont {Ferreira}, \citenamefont {Padilla},\ and\
  \citenamefont {Skordis}}]{Clifton:2011jh}%
  \BibitemOpen
  \bibfield  {author} {\bibinfo {author} {\bibfnamefont {T.}~\bibnamefont
  {Clifton}}, \bibinfo {author} {\bibfnamefont {P.~G.}\ \bibnamefont
  {Ferreira}}, \bibinfo {author} {\bibfnamefont {A.}~\bibnamefont {Padilla}},\
  and\ \bibinfo {author} {\bibfnamefont {C.}~\bibnamefont {Skordis}},\
  }\bibfield  {title} {\bibinfo {title} {{Modified Gravity and Cosmology}},\
  }\href {https://doi.org/10.1016/j.physrep.2012.01.001} {\bibfield  {journal}
  {\bibinfo  {journal} {Phys. Rept.}\ }\textbf {\bibinfo {volume} {513}},\
  \bibinfo {pages} {1} (\bibinfo {year} {2012})},\ \Eprint
  {https://arxiv.org/abs/1106.2476} {arXiv:1106.2476 [astro-ph.CO]}
  \BibitemShut {NoStop}%
\bibitem [{\citenamefont {Berti}\ \emph {et~al.}(2015)\citenamefont {Berti}
  \emph {et~al.}}]{Berti:2015itd}%
  \BibitemOpen
  \bibfield  {author} {\bibinfo {author} {\bibfnamefont {E.}~\bibnamefont
  {Berti}} \emph {et~al.},\ }\bibfield  {title} {\bibinfo {title} {{Testing
  General Relativity with Present and Future Astrophysical Observations}},\
  }\href {https://doi.org/10.1088/0264-9381/32/24/243001} {\bibfield  {journal}
  {\bibinfo  {journal} {Class. Quant. Grav.}\ }\textbf {\bibinfo {volume}
  {32}},\ \bibinfo {pages} {243001} (\bibinfo {year} {2015})},\ \Eprint
  {https://arxiv.org/abs/1501.07274} {arXiv:1501.07274 [gr-qc]} \BibitemShut
  {NoStop}%
\bibitem [{\citenamefont {Will}(2014)}]{Will:2014kxa}%
  \BibitemOpen
  \bibfield  {author} {\bibinfo {author} {\bibfnamefont {C.~M.}\ \bibnamefont
  {Will}},\ }\bibfield  {title} {\bibinfo {title} {{The Confrontation between
  General Relativity and Experiment}},\ }\href
  {https://doi.org/10.12942/lrr-2014-4} {\bibfield  {journal} {\bibinfo
  {journal} {Living Rev. Rel.}\ }\textbf {\bibinfo {volume} {17}},\ \bibinfo
  {pages} {4} (\bibinfo {year} {2014})},\ \Eprint
  {https://arxiv.org/abs/1403.7377} {arXiv:1403.7377 [gr-qc]} \BibitemShut
  {NoStop}%
\bibitem [{\citenamefont {Berti}\ \emph
  {et~al.}(2018{\natexlab{a}})\citenamefont {Berti}, \citenamefont {Yagi},
  \citenamefont {Yang},\ and\ \citenamefont {Yunes}}]{Berti:2018vdi}%
  \BibitemOpen
  \bibfield  {author} {\bibinfo {author} {\bibfnamefont {E.}~\bibnamefont
  {Berti}}, \bibinfo {author} {\bibfnamefont {K.}~\bibnamefont {Yagi}},
  \bibinfo {author} {\bibfnamefont {H.}~\bibnamefont {Yang}},\ and\ \bibinfo
  {author} {\bibfnamefont {N.}~\bibnamefont {Yunes}},\ }\bibfield  {title}
  {\bibinfo {title} {{Extreme Gravity Tests with Gravitational Waves from
  Compact Binary Coalescences: (II) Ringdown}},\ }\href
  {https://doi.org/10.1007/s10714-018-2372-6} {\bibfield  {journal} {\bibinfo
  {journal} {Gen. Rel. Grav.}\ }\textbf {\bibinfo {volume} {50}},\ \bibinfo
  {pages} {49} (\bibinfo {year} {2018}{\natexlab{a}})},\ \Eprint
  {https://arxiv.org/abs/1801.03587} {arXiv:1801.03587 [gr-qc]} \BibitemShut
  {NoStop}%
\bibitem [{\citenamefont {Berti}\ \emph
  {et~al.}(2018{\natexlab{b}})\citenamefont {Berti}, \citenamefont {Yagi},\
  and\ \citenamefont {Yunes}}]{Berti:2018cxi}%
  \BibitemOpen
  \bibfield  {author} {\bibinfo {author} {\bibfnamefont {E.}~\bibnamefont
  {Berti}}, \bibinfo {author} {\bibfnamefont {K.}~\bibnamefont {Yagi}},\ and\
  \bibinfo {author} {\bibfnamefont {N.}~\bibnamefont {Yunes}},\ }\bibfield
  {title} {\bibinfo {title} {{Extreme Gravity Tests with Gravitational Waves
  from Compact Binary Coalescences: (I) Inspiral-Merger}},\ }\href
  {https://doi.org/10.1007/s10714-018-2362-8} {\bibfield  {journal} {\bibinfo
  {journal} {Gen. Rel. Grav.}\ }\textbf {\bibinfo {volume} {50}},\ \bibinfo
  {pages} {46} (\bibinfo {year} {2018}{\natexlab{b}})},\ \Eprint
  {https://arxiv.org/abs/1801.03208} {arXiv:1801.03208 [gr-qc]} \BibitemShut
  {NoStop}%
\bibitem [{\citenamefont {Fierz}\ and\ \citenamefont
  {Pauli}(1939)}]{Fierz:1939ix}%
  \BibitemOpen
  \bibfield  {author} {\bibinfo {author} {\bibfnamefont {M.}~\bibnamefont
  {Fierz}}\ and\ \bibinfo {author} {\bibfnamefont {W.}~\bibnamefont {Pauli}},\
  }\bibfield  {title} {\bibinfo {title} {{On relativistic wave equations for
  particles of arbitrary spin in an electromagnetic field}},\ }\href
  {https://doi.org/10.1098/rspa.1939.0140} {\bibfield  {journal} {\bibinfo
  {journal} {Proc. Roy. Soc. Lond. A}\ }\textbf {\bibinfo {volume} {173}},\
  \bibinfo {pages} {211} (\bibinfo {year} {1939})}\BibitemShut {NoStop}%
\bibitem [{\citenamefont {Boulware}\ and\ \citenamefont
  {Deser}(1972)}]{Boulware:1972yco}%
  \BibitemOpen
  \bibfield  {author} {\bibinfo {author} {\bibfnamefont {D.~G.}\ \bibnamefont
  {Boulware}}\ and\ \bibinfo {author} {\bibfnamefont {S.}~\bibnamefont
  {Deser}},\ }\bibfield  {title} {\bibinfo {title} {{Can gravitation have a
  finite range?}},\ }\href {https://doi.org/10.1103/PhysRevD.6.3368} {\bibfield
   {journal} {\bibinfo  {journal} {Phys. Rev. D}\ }\textbf {\bibinfo {volume}
  {6}},\ \bibinfo {pages} {3368} (\bibinfo {year} {1972})}\BibitemShut
  {NoStop}%
\bibitem [{\citenamefont {de~Rham}\ \emph {et~al.}(2011)\citenamefont
  {de~Rham}, \citenamefont {Gabadadze},\ and\ \citenamefont
  {Tolley}}]{deRham:2010kj}%
  \BibitemOpen
  \bibfield  {author} {\bibinfo {author} {\bibfnamefont {C.}~\bibnamefont
  {de~Rham}}, \bibinfo {author} {\bibfnamefont {G.}~\bibnamefont {Gabadadze}},\
  and\ \bibinfo {author} {\bibfnamefont {A.~J.}\ \bibnamefont {Tolley}},\
  }\bibfield  {title} {\bibinfo {title} {{Resummation of Massive Gravity}},\
  }\href {https://doi.org/10.1103/PhysRevLett.106.231101} {\bibfield  {journal}
  {\bibinfo  {journal} {Phys.Rev.Lett.}\ }\textbf {\bibinfo {volume} {106}},\
  \bibinfo {pages} {231101} (\bibinfo {year} {2011})},\ \Eprint
  {https://arxiv.org/abs/1011.1232} {arXiv:1011.1232 [hep-th]} \BibitemShut
  {NoStop}%
\bibitem [{\citenamefont {Hassan}\ and\ \citenamefont
  {Rosen}(2012)}]{Hassan:2011zd}%
  \BibitemOpen
  \bibfield  {author} {\bibinfo {author} {\bibfnamefont {S.}~\bibnamefont
  {Hassan}}\ and\ \bibinfo {author} {\bibfnamefont {R.}~\bibnamefont {Rosen}},\
  }\bibfield  {title} {\bibinfo {title} {Bimetric gravity from ghost-free
  massive gravity},\ }\href {https://doi.org/10.1007/JHEP02(2012)126}
  {\bibfield  {journal} {\bibinfo  {journal} {J. High Energy Phys.}\ }\textbf
  {\bibinfo {volume} {2012}}\bibfield  {number} {\bibinfo  {number} { (02)},\
  \bibinfo {eid} {126}},\ }\Eprint {https://arxiv.org/abs/1109.3515}
  {arXiv:1109.3515 [hep-th]} \BibitemShut {NoStop}%
\bibitem [{\citenamefont {Yamashita}\ \emph {et~al.}(2014)\citenamefont
  {Yamashita}, \citenamefont {De~Felice},\ and\ \citenamefont
  {Tanaka}}]{Yamashita:2014fga}%
  \BibitemOpen
  \bibfield  {author} {\bibinfo {author} {\bibfnamefont {Y.}~\bibnamefont
  {Yamashita}}, \bibinfo {author} {\bibfnamefont {A.}~\bibnamefont
  {De~Felice}},\ and\ \bibinfo {author} {\bibfnamefont {T.}~\bibnamefont
  {Tanaka}},\ }\bibfield  {title} {\bibinfo {title} {{Appearance of
  Boulware\textendash{}Deser ghost in bigravity with doubly coupled matter}},\
  }\href {https://doi.org/10.1142/S0218271814430032} {\bibfield  {journal}
  {\bibinfo  {journal} {Int. J. Mod. Phys. D}\ }\textbf {\bibinfo {volume}
  {23}},\ \bibinfo {pages} {1443003} (\bibinfo {year} {2014})},\ \Eprint
  {https://arxiv.org/abs/1408.0487} {arXiv:1408.0487 [hep-th]} \BibitemShut
  {NoStop}%
\bibitem [{\citenamefont {de~Rham}\ \emph {et~al.}(2014)\citenamefont
  {de~Rham}, \citenamefont {Heisenberg},\ and\ \citenamefont
  {Ribeiro}}]{deRham:2014fha}%
  \BibitemOpen
  \bibfield  {author} {\bibinfo {author} {\bibfnamefont {C.}~\bibnamefont
  {de~Rham}}, \bibinfo {author} {\bibfnamefont {L.}~\bibnamefont
  {Heisenberg}},\ and\ \bibinfo {author} {\bibfnamefont {R.~H.}\ \bibnamefont
  {Ribeiro}},\ }\bibfield  {title} {\bibinfo {title} {{Ghosts and matter
  couplings in massive gravity, bigravity and multigravity}},\ }\href
  {https://doi.org/10.1103/PhysRevD.90.124042} {\bibfield  {journal} {\bibinfo
  {journal} {Phys. Rev. D}\ }\textbf {\bibinfo {volume} {90}},\ \bibinfo
  {pages} {124042} (\bibinfo {year} {2014})},\ \Eprint
  {https://arxiv.org/abs/1409.3834} {arXiv:1409.3834 [hep-th]} \BibitemShut
  {NoStop}%
\bibitem [{\citenamefont {Gumrukcuoglu}\ \emph {et~al.}(2015)\citenamefont
  {Gumrukcuoglu}, \citenamefont {Heisenberg}, \citenamefont {Mukohyama},\ and\
  \citenamefont {Tanahashi}}]{Gumrukcuoglu:2015nua}%
  \BibitemOpen
  \bibfield  {author} {\bibinfo {author} {\bibfnamefont {A.~E.}\ \bibnamefont
  {Gumrukcuoglu}}, \bibinfo {author} {\bibfnamefont {L.}~\bibnamefont
  {Heisenberg}}, \bibinfo {author} {\bibfnamefont {S.}~\bibnamefont
  {Mukohyama}},\ and\ \bibinfo {author} {\bibfnamefont {N.}~\bibnamefont
  {Tanahashi}},\ }\bibfield  {title} {\bibinfo {title} {{Cosmology in bimetric
  theory with an effective composite coupling to matter}},\ }\href
  {https://doi.org/10.1088/1475-7516/2015/04/008} {\bibfield  {journal}
  {\bibinfo  {journal} {JCAP}\ }\textbf {\bibinfo {volume} {04}},\ \bibinfo
  {pages} {008}},\ \Eprint {https://arxiv.org/abs/1501.02790} {arXiv:1501.02790
  [hep-th]} \BibitemShut {NoStop}%
\bibitem [{\citenamefont {De~Felice}\ and\ \citenamefont
  {Mukohyama}(2016{\natexlab{a}})}]{DeFelice:2015hla}%
  \BibitemOpen
  \bibfield  {author} {\bibinfo {author} {\bibfnamefont {A.}~\bibnamefont
  {De~Felice}}\ and\ \bibinfo {author} {\bibfnamefont {S.}~\bibnamefont
  {Mukohyama}},\ }\bibfield  {title} {\bibinfo {title} {{Minimal theory of
  massive gravity}},\ }\href {https://doi.org/10.1016/j.physletb.2015.11.050}
  {\bibfield  {journal} {\bibinfo  {journal} {Phys. Lett. B}\ }\textbf
  {\bibinfo {volume} {752}},\ \bibinfo {pages} {302} (\bibinfo {year}
  {2016}{\natexlab{a}})},\ \Eprint {https://arxiv.org/abs/1506.01594}
  {arXiv:1506.01594 [hep-th]} \BibitemShut {NoStop}%
\bibitem [{\citenamefont {De~Felice}\ and\ \citenamefont
  {Mukohyama}(2016{\natexlab{b}})}]{DeFelice:2015moy}%
  \BibitemOpen
  \bibfield  {author} {\bibinfo {author} {\bibfnamefont {A.}~\bibnamefont
  {De~Felice}}\ and\ \bibinfo {author} {\bibfnamefont {S.}~\bibnamefont
  {Mukohyama}},\ }\bibfield  {title} {\bibinfo {title} {{Phenomenology in
  minimal theory of massive gravity}},\ }\href
  {https://doi.org/10.1088/1475-7516/2016/04/028} {\bibfield  {journal}
  {\bibinfo  {journal} {JCAP}\ }\textbf {\bibinfo {volume} {04}},\ \bibinfo
  {pages} {028}},\ \Eprint {https://arxiv.org/abs/1512.04008} {arXiv:1512.04008
  [hep-th]} \BibitemShut {NoStop}%
\bibitem [{\citenamefont {Gumrukcuoglu}\ \emph {et~al.}(2012)\citenamefont
  {Gumrukcuoglu}, \citenamefont {Lin},\ and\ \citenamefont
  {Mukohyama}}]{Gumrukcuoglu:2011zh}%
  \BibitemOpen
  \bibfield  {author} {\bibinfo {author} {\bibfnamefont {A.~E.}\ \bibnamefont
  {Gumrukcuoglu}}, \bibinfo {author} {\bibfnamefont {C.}~\bibnamefont {Lin}},\
  and\ \bibinfo {author} {\bibfnamefont {S.}~\bibnamefont {Mukohyama}},\
  }\bibfield  {title} {\bibinfo {title} {{Cosmological perturbations of
  self-accelerating universe in nonlinear massive gravity}},\ }\href
  {https://doi.org/10.1088/1475-7516/2012/03/006} {\bibfield  {journal}
  {\bibinfo  {journal} {JCAP}\ }\textbf {\bibinfo {volume} {03}},\ \bibinfo
  {pages} {006}},\ \Eprint {https://arxiv.org/abs/1111.4107} {arXiv:1111.4107
  [hep-th]} \BibitemShut {NoStop}%
\bibitem [{\citenamefont {De~Felice}\ \emph {et~al.}(2012)\citenamefont
  {De~Felice}, \citenamefont {G{\"u}mr{\"u}k{\c c}{\"u}o{\u g}lu},\ and\
  \citenamefont {Mukohyama}}]{DeFelice:2012mx}%
  \BibitemOpen
  \bibfield  {author} {\bibinfo {author} {\bibfnamefont {A.}~\bibnamefont
  {De~Felice}}, \bibinfo {author} {\bibfnamefont {A.~E.}\ \bibnamefont
  {G{\"u}mr{\"u}k{\c c}{\"u}o{\u g}lu}},\ and\ \bibinfo {author} {\bibfnamefont
  {S.}~\bibnamefont {Mukohyama}},\ }\bibfield  {title} {\bibinfo {title}
  {{Massive gravity: nonlinear instability of the homogeneous and isotropic
  universe}},\ }\href {https://doi.org/10.1103/PhysRevLett.109.171101}
  {\bibfield  {journal} {\bibinfo  {journal} {Phys.Rev.Lett.}\ }\textbf
  {\bibinfo {volume} {109}},\ \bibinfo {pages} {171101} (\bibinfo {year}
  {2012})},\ \Eprint {https://arxiv.org/abs/1206.2080} {arXiv:1206.2080
  [hep-th]} \BibitemShut {NoStop}%
\bibitem [{\citenamefont {Fasiello}\ and\ \citenamefont
  {Tolley}(2012)}]{Fasiello:2012rw}%
  \BibitemOpen
  \bibfield  {author} {\bibinfo {author} {\bibfnamefont {M.}~\bibnamefont
  {Fasiello}}\ and\ \bibinfo {author} {\bibfnamefont {A.~J.}\ \bibnamefont
  {Tolley}},\ }\bibfield  {title} {\bibinfo {title} {{Cosmological
  perturbations in Massive Gravity and the Higuchi bound}},\ }\href
  {https://doi.org/10.1088/1475-7516/2012/11/035} {\bibfield  {journal}
  {\bibinfo  {journal} {JCAP}\ }\textbf {\bibinfo {volume} {11}},\ \bibinfo
  {pages} {035}},\ \Eprint {https://arxiv.org/abs/1206.3852} {arXiv:1206.3852
  [hep-th]} \BibitemShut {NoStop}%
\bibitem [{\citenamefont {De~Felice}\ and\ \citenamefont
  {Mukohyama}(2017)}]{DeFelice:2016ufg}%
  \BibitemOpen
  \bibfield  {author} {\bibinfo {author} {\bibfnamefont {A.}~\bibnamefont
  {De~Felice}}\ and\ \bibinfo {author} {\bibfnamefont {S.}~\bibnamefont
  {Mukohyama}},\ }\bibfield  {title} {\bibinfo {title} {{Graviton mass might
  reduce tension between early and late time cosmological data}},\ }\href
  {https://doi.org/10.1103/PhysRevLett.118.091104} {\bibfield  {journal}
  {\bibinfo  {journal} {Phys. Rev. Lett.}\ }\textbf {\bibinfo {volume} {118}},\
  \bibinfo {pages} {091104} (\bibinfo {year} {2017})},\ \Eprint
  {https://arxiv.org/abs/1607.03368} {arXiv:1607.03368 [astro-ph.CO]}
  \BibitemShut {NoStop}%
\bibitem [{\citenamefont {Hagala}\ \emph {et~al.}(2021)\citenamefont {Hagala},
  \citenamefont {Felice}, \citenamefont {Mota},\ and\ \citenamefont
  {Mukohyama}}]{Hagala:2020eax}%
  \BibitemOpen
  \bibfield  {author} {\bibinfo {author} {\bibfnamefont {R.}~\bibnamefont
  {Hagala}}, \bibinfo {author} {\bibfnamefont {A.~D.}\ \bibnamefont {Felice}},
  \bibinfo {author} {\bibfnamefont {D.~F.}\ \bibnamefont {Mota}},\ and\
  \bibinfo {author} {\bibfnamefont {S.}~\bibnamefont {Mukohyama}},\ }\bibfield
  {title} {\bibinfo {title} {{Non-linear dynamics of the minimal theory of
  massive gravity}},\ }\href {https://doi.org/10.1051/0004-6361/202040018}
  {\bibfield  {journal} {\bibinfo  {journal} {Astron. Astrophys.}\ }\textbf
  {\bibinfo {volume} {653}},\ \bibinfo {pages} {A148} (\bibinfo {year}
  {2021})},\ \Eprint {https://arxiv.org/abs/2011.14697} {arXiv:2011.14697
  [astro-ph.CO]} \BibitemShut {NoStop}%
\bibitem [{\citenamefont {De~Felice}\ \emph
  {et~al.}(2021{\natexlab{a}})\citenamefont {De~Felice}, \citenamefont
  {Mukohyama},\ and\ \citenamefont {Pookkillath}}]{DeFelice:2021trp}%
  \BibitemOpen
  \bibfield  {author} {\bibinfo {author} {\bibfnamefont {A.}~\bibnamefont
  {De~Felice}}, \bibinfo {author} {\bibfnamefont {S.}~\bibnamefont
  {Mukohyama}},\ and\ \bibinfo {author} {\bibfnamefont {M.~C.}\ \bibnamefont
  {Pookkillath}},\ }\bibfield  {title} {\bibinfo {title} {{Minimal theory of
  massive gravity and constraints on the graviton mass}},\ }\href
  {https://doi.org/10.1088/1475-7516/2021/12/011} {\bibfield  {journal}
  {\bibinfo  {journal} {JCAP}\ }\textbf {\bibinfo {volume} {12}}\bibfield
  {number} {\bibinfo  {number} { (12)},\ \bibinfo {pages} {011}},\ }\Eprint
  {https://arxiv.org/abs/2110.01237} {arXiv:2110.01237 [astro-ph.CO]}
  \BibitemShut {NoStop}%
\bibitem [{\citenamefont {De~Felice}\ \emph
  {et~al.}(2018{\natexlab{a}})\citenamefont {De~Felice}, \citenamefont
  {Larrouturou}, \citenamefont {Mukohyama},\ and\ \citenamefont
  {Oliosi}}]{DeFelice:2018vza}%
  \BibitemOpen
  \bibfield  {author} {\bibinfo {author} {\bibfnamefont {A.}~\bibnamefont
  {De~Felice}}, \bibinfo {author} {\bibfnamefont {F.}~\bibnamefont
  {Larrouturou}}, \bibinfo {author} {\bibfnamefont {S.}~\bibnamefont
  {Mukohyama}},\ and\ \bibinfo {author} {\bibfnamefont {M.}~\bibnamefont
  {Oliosi}},\ }\bibfield  {title} {\bibinfo {title} {{Black holes and stars in
  the minimal theory of massive gravity}},\ }\href
  {https://doi.org/10.1103/PhysRevD.98.104031} {\bibfield  {journal} {\bibinfo
  {journal} {Phys. Rev. D}\ }\textbf {\bibinfo {volume} {98}},\ \bibinfo
  {pages} {104031} (\bibinfo {year} {2018}{\natexlab{a}})},\ \Eprint
  {https://arxiv.org/abs/1808.01403} {arXiv:1808.01403 [gr-qc]} \BibitemShut
  {NoStop}%
\bibitem [{\citenamefont {De~Felice}\ \emph
  {et~al.}(2021{\natexlab{b}})\citenamefont {De~Felice}, \citenamefont
  {Larrouturou}, \citenamefont {Mukohyama},\ and\ \citenamefont
  {Oliosi}}]{DeFelice:2020ecp}%
  \BibitemOpen
  \bibfield  {author} {\bibinfo {author} {\bibfnamefont {A.}~\bibnamefont
  {De~Felice}}, \bibinfo {author} {\bibfnamefont {F.}~\bibnamefont
  {Larrouturou}}, \bibinfo {author} {\bibfnamefont {S.}~\bibnamefont
  {Mukohyama}},\ and\ \bibinfo {author} {\bibfnamefont {M.}~\bibnamefont
  {Oliosi}},\ }\bibfield  {title} {\bibinfo {title} {{Minimal Theory of
  Bigravity: construction and cosmology}},\ }\href
  {https://doi.org/10.1088/1475-7516/2021/04/015} {\bibfield  {journal}
  {\bibinfo  {journal} {JCAP}\ }\textbf {\bibinfo {volume} {04}},\ \bibinfo
  {pages} {015}},\ \Eprint {https://arxiv.org/abs/2012.01073} {arXiv:2012.01073
  [gr-qc]} \BibitemShut {NoStop}%
\bibitem [{\citenamefont {Garcia-Saenz}\ \emph {et~al.}(2021)\citenamefont
  {Garcia-Saenz}, \citenamefont {Held},\ and\ \citenamefont
  {Zhang}}]{Garcia-Saenz:2021uyv}%
  \BibitemOpen
  \bibfield  {author} {\bibinfo {author} {\bibfnamefont {S.}~\bibnamefont
  {Garcia-Saenz}}, \bibinfo {author} {\bibfnamefont {A.}~\bibnamefont {Held}},\
  and\ \bibinfo {author} {\bibfnamefont {J.}~\bibnamefont {Zhang}},\ }\bibfield
   {title} {\bibinfo {title} {{Destabilization of Black Holes and Stars by
  Generalized Proca Fields}},\ }\href
  {https://doi.org/10.1103/PhysRevLett.127.131104} {\bibfield  {journal}
  {\bibinfo  {journal} {Phys. Rev. Lett.}\ }\textbf {\bibinfo {volume} {127}},\
  \bibinfo {pages} {131104} (\bibinfo {year} {2021})},\ \Eprint
  {https://arxiv.org/abs/2104.08049} {arXiv:2104.08049 [gr-qc]} \BibitemShut
  {NoStop}%
\bibitem [{\citenamefont {Silva}\ \emph {et~al.}(2022)\citenamefont {Silva},
  \citenamefont {Coates}, \citenamefont {Ramazano\u{g}lu},\ and\ \citenamefont
  {Sotiriou}}]{Silva:2021jya}%
  \BibitemOpen
  \bibfield  {author} {\bibinfo {author} {\bibfnamefont {H.~O.}\ \bibnamefont
  {Silva}}, \bibinfo {author} {\bibfnamefont {A.}~\bibnamefont {Coates}},
  \bibinfo {author} {\bibfnamefont {F.~M.}\ \bibnamefont {Ramazano\u{g}lu}},\
  and\ \bibinfo {author} {\bibfnamefont {T.~P.}\ \bibnamefont {Sotiriou}},\
  }\bibfield  {title} {\bibinfo {title} {{Ghost of vector fields in compact
  stars}},\ }\href {https://doi.org/10.1103/PhysRevD.105.024046} {\bibfield
  {journal} {\bibinfo  {journal} {Phys. Rev. D}\ }\textbf {\bibinfo {volume}
  {105}},\ \bibinfo {pages} {024046} (\bibinfo {year} {2022})},\ \Eprint
  {https://arxiv.org/abs/2110.04594} {arXiv:2110.04594 [gr-qc]} \BibitemShut
  {NoStop}%
\bibitem [{\citenamefont {Demirbo\u{g}a}\ \emph {et~al.}(2022)\citenamefont
  {Demirbo\u{g}a}, \citenamefont {Coates},\ and\ \citenamefont
  {Ramazano\u{g}lu}}]{Demirboga:2021nrc}%
  \BibitemOpen
  \bibfield  {author} {\bibinfo {author} {\bibfnamefont {E.~S.}\ \bibnamefont
  {Demirbo\u{g}a}}, \bibinfo {author} {\bibfnamefont {A.}~\bibnamefont
  {Coates}},\ and\ \bibinfo {author} {\bibfnamefont {F.~M.}\ \bibnamefont
  {Ramazano\u{g}lu}},\ }\bibfield  {title} {\bibinfo {title} {{Instability of
  vectorized stars}},\ }\href {https://doi.org/10.1103/PhysRevD.105.024057}
  {\bibfield  {journal} {\bibinfo  {journal} {Phys. Rev. D}\ }\textbf {\bibinfo
  {volume} {105}},\ \bibinfo {pages} {024057} (\bibinfo {year} {2022})},\
  \Eprint {https://arxiv.org/abs/2112.04269} {arXiv:2112.04269 [gr-qc]}
  \BibitemShut {NoStop}%
\bibitem [{\citenamefont {Manita}\ \emph {et~al.}(2022)\citenamefont {Manita},
  \citenamefont {Aoki}, \citenamefont {Fujita},\ and\ \citenamefont
  {Mukohyama}}]{Manita:2022tkl}%
  \BibitemOpen
  \bibfield  {author} {\bibinfo {author} {\bibfnamefont {Y.}~\bibnamefont
  {Manita}}, \bibinfo {author} {\bibfnamefont {K.}~\bibnamefont {Aoki}},
  \bibinfo {author} {\bibfnamefont {T.}~\bibnamefont {Fujita}},\ and\ \bibinfo
  {author} {\bibfnamefont {S.}~\bibnamefont {Mukohyama}},\ }\bibfield  {title}
  {\bibinfo {title} {{Spin-2 dark matter from anisotropic Universe in
  bigravity}},\ }\href@noop {} {\  (\bibinfo {year} {2022})},\ \Eprint
  {https://arxiv.org/abs/2211.15873} {arXiv:2211.15873 [gr-qc]} \BibitemShut
  {NoStop}%
\bibitem [{\citenamefont {De~Felice}\ \emph
  {et~al.}(2018{\natexlab{b}})\citenamefont {De~Felice}, \citenamefont
  {Langlois}, \citenamefont {Mukohyama}, \citenamefont {Noui},\ and\
  \citenamefont {Wang}}]{DeFelice:2018ewo}%
  \BibitemOpen
  \bibfield  {author} {\bibinfo {author} {\bibfnamefont {A.}~\bibnamefont
  {De~Felice}}, \bibinfo {author} {\bibfnamefont {D.}~\bibnamefont {Langlois}},
  \bibinfo {author} {\bibfnamefont {S.}~\bibnamefont {Mukohyama}}, \bibinfo
  {author} {\bibfnamefont {K.}~\bibnamefont {Noui}},\ and\ \bibinfo {author}
  {\bibfnamefont {A.}~\bibnamefont {Wang}},\ }\bibfield  {title} {\bibinfo
  {title} {{Generalized instantaneous modes in higher-order scalar-tensor
  theories}},\ }\href {https://doi.org/10.1103/PhysRevD.98.084024} {\bibfield
  {journal} {\bibinfo  {journal} {Phys. Rev. D}\ }\textbf {\bibinfo {volume}
  {98}},\ \bibinfo {pages} {084024} (\bibinfo {year} {2018}{\natexlab{b}})},\
  \Eprint {https://arxiv.org/abs/1803.06241} {arXiv:1803.06241 [hep-th]}
  \BibitemShut {NoStop}%
\bibitem [{\citenamefont {De~Felice}\ \emph {et~al.}(2020)\citenamefont
  {De~Felice}, \citenamefont {Doll},\ and\ \citenamefont
  {Mukohyama}}]{DeFelice:2020eju}%
  \BibitemOpen
  \bibfield  {author} {\bibinfo {author} {\bibfnamefont {A.}~\bibnamefont
  {De~Felice}}, \bibinfo {author} {\bibfnamefont {A.}~\bibnamefont {Doll}},\
  and\ \bibinfo {author} {\bibfnamefont {S.}~\bibnamefont {Mukohyama}},\
  }\bibfield  {title} {\bibinfo {title} {{A theory of type-II minimally
  modified gravity}},\ }\href {https://doi.org/10.1088/1475-7516/2020/09/034}
  {\bibfield  {journal} {\bibinfo  {journal} {JCAP}\ }\textbf {\bibinfo
  {volume} {09}},\ \bibinfo {pages} {034}},\ \Eprint
  {https://arxiv.org/abs/2004.12549} {arXiv:2004.12549 [gr-qc]} \BibitemShut
  {NoStop}%
\bibitem [{\citenamefont {De~Felice}\ \emph
  {et~al.}(2021{\natexlab{c}})\citenamefont {De~Felice}, \citenamefont
  {Mukohyama},\ and\ \citenamefont {Takahashi}}]{DeFelice:2021hps}%
  \BibitemOpen
  \bibfield  {author} {\bibinfo {author} {\bibfnamefont {A.}~\bibnamefont
  {De~Felice}}, \bibinfo {author} {\bibfnamefont {S.}~\bibnamefont
  {Mukohyama}},\ and\ \bibinfo {author} {\bibfnamefont {K.}~\bibnamefont
  {Takahashi}},\ }\bibfield  {title} {\bibinfo {title} {{Nonlinear definition
  of the shadowy mode in higher-order scalar-tensor theories}},\ }\href
  {https://doi.org/10.1088/1475-7516/2021/12/020} {\bibfield  {journal}
  {\bibinfo  {journal} {JCAP}\ }\textbf {\bibinfo {volume} {12}}\bibfield
  {number} {\bibinfo  {number} { (12)},\ \bibinfo {pages} {020}},\ }\Eprint
  {https://arxiv.org/abs/2110.03194} {arXiv:2110.03194 [gr-qc]} \BibitemShut
  {NoStop}%
\bibitem [{\citenamefont {Minamitsuji}\ \emph {et~al.}(2022)\citenamefont
  {Minamitsuji}, \citenamefont {De~Felice}, \citenamefont {Mukohyama},\ and\
  \citenamefont {Oliosi}}]{Minamitsuji:2022vfv}%
  \BibitemOpen
  \bibfield  {author} {\bibinfo {author} {\bibfnamefont {M.}~\bibnamefont
  {Minamitsuji}}, \bibinfo {author} {\bibfnamefont {A.}~\bibnamefont
  {De~Felice}}, \bibinfo {author} {\bibfnamefont {S.}~\bibnamefont
  {Mukohyama}},\ and\ \bibinfo {author} {\bibfnamefont {M.}~\bibnamefont
  {Oliosi}},\ }\bibfield  {title} {\bibinfo {title} {{Static and spherically
  symmetric general relativity solutions in minimal theory of bigravity}},\
  }\href {https://doi.org/10.1103/PhysRevD.105.123026} {\bibfield  {journal}
  {\bibinfo  {journal} {Phys. Rev. D}\ }\textbf {\bibinfo {volume} {105}},\
  \bibinfo {pages} {123026} (\bibinfo {year} {2022})},\ \Eprint
  {https://arxiv.org/abs/2204.08217} {arXiv:2204.08217 [gr-qc]} \BibitemShut
  {NoStop}%
\bibitem [{\citenamefont {De~Felice}\ \emph
  {et~al.}(2021{\natexlab{d}})\citenamefont {De~Felice}, \citenamefont
  {Mukohyama},\ and\ \citenamefont {Pookkillath}}]{DeFelice:2020cpt}%
  \BibitemOpen
  \bibfield  {author} {\bibinfo {author} {\bibfnamefont {A.}~\bibnamefont
  {De~Felice}}, \bibinfo {author} {\bibfnamefont {S.}~\bibnamefont
  {Mukohyama}},\ and\ \bibinfo {author} {\bibfnamefont {M.~C.}\ \bibnamefont
  {Pookkillath}},\ }\bibfield  {title} {\bibinfo {title} {{Addressing $H_0$
  tension by means of VCDM}},\ }\href
  {https://doi.org/10.1016/j.physletb.2021.136201} {\bibfield  {journal}
  {\bibinfo  {journal} {Phys. Lett. B}\ }\textbf {\bibinfo {volume} {816}},\
  \bibinfo {pages} {136201} (\bibinfo {year} {2021}{\natexlab{d}})},\ \Eprint
  {https://arxiv.org/abs/2009.08718} {arXiv:2009.08718 [astro-ph.CO]}
  \BibitemShut {NoStop}%
\bibitem [{\citenamefont {De~Felice}\ \emph
  {et~al.}(2021{\natexlab{e}})\citenamefont {De~Felice}, \citenamefont {Doll},
  \citenamefont {Larrouturou},\ and\ \citenamefont
  {Mukohyama}}]{DeFelice:2020onz}%
  \BibitemOpen
  \bibfield  {author} {\bibinfo {author} {\bibfnamefont {A.}~\bibnamefont
  {De~Felice}}, \bibinfo {author} {\bibfnamefont {A.}~\bibnamefont {Doll}},
  \bibinfo {author} {\bibfnamefont {F.}~\bibnamefont {Larrouturou}},\ and\
  \bibinfo {author} {\bibfnamefont {S.}~\bibnamefont {Mukohyama}},\ }\bibfield
  {title} {\bibinfo {title} {{Black holes in a type-II minimally modified
  gravity}},\ }\href {https://doi.org/10.1088/1475-7516/2021/03/004} {\bibfield
   {journal} {\bibinfo  {journal} {JCAP}\ }\textbf {\bibinfo {volume} {03}},\
  \bibinfo {pages} {004}},\ \Eprint {https://arxiv.org/abs/2010.13067}
  {arXiv:2010.13067 [gr-qc]} \BibitemShut {NoStop}%
\bibitem [{\citenamefont {De~Felice}\ \emph {et~al.}(2022)\citenamefont
  {De~Felice}, \citenamefont {Maeda}, \citenamefont {Mukohyama},\ and\
  \citenamefont {Pookkillath}}]{DeFelice:2022uxv}%
  \BibitemOpen
  \bibfield  {author} {\bibinfo {author} {\bibfnamefont {A.}~\bibnamefont
  {De~Felice}}, \bibinfo {author} {\bibfnamefont {K.-i.}\ \bibnamefont
  {Maeda}}, \bibinfo {author} {\bibfnamefont {S.}~\bibnamefont {Mukohyama}},\
  and\ \bibinfo {author} {\bibfnamefont {M.~C.}\ \bibnamefont {Pookkillath}},\
  }\bibfield  {title} {\bibinfo {title} {{Comparison of two theories of
  Type-IIa minimally modified gravity}},\ }\href
  {https://doi.org/10.1103/PhysRevD.106.024028} {\bibfield  {journal} {\bibinfo
   {journal} {Phys. Rev. D}\ }\textbf {\bibinfo {volume} {106}},\ \bibinfo
  {pages} {024028} (\bibinfo {year} {2022})},\ \Eprint
  {https://arxiv.org/abs/2204.08294} {arXiv:2204.08294 [gr-qc]} \BibitemShut
  {NoStop}%
\bibitem [{\citenamefont {Oppenheimer}\ and\ \citenamefont
  {Snyder}(1939)}]{Oppenheimer:1939ue}%
  \BibitemOpen
  \bibfield  {author} {\bibinfo {author} {\bibfnamefont {J.~R.}\ \bibnamefont
  {Oppenheimer}}\ and\ \bibinfo {author} {\bibfnamefont {H.}~\bibnamefont
  {Snyder}},\ }\bibfield  {title} {\bibinfo {title} {{On Continued
  gravitational contraction}},\ }\href {https://doi.org/10.1103/PhysRev.56.455}
  {\bibfield  {journal} {\bibinfo  {journal} {Phys. Rev.}\ }\textbf {\bibinfo
  {volume} {56}},\ \bibinfo {pages} {455} (\bibinfo {year} {1939})}\BibitemShut
  {NoStop}%
\bibitem [{\citenamefont {Kanai}\ \emph {et~al.}(2011)\citenamefont {Kanai},
  \citenamefont {Siino},\ and\ \citenamefont {Hosoya}}]{Kanai:2010ae}%
  \BibitemOpen
  \bibfield  {author} {\bibinfo {author} {\bibfnamefont {Y.}~\bibnamefont
  {Kanai}}, \bibinfo {author} {\bibfnamefont {M.}~\bibnamefont {Siino}},\ and\
  \bibinfo {author} {\bibfnamefont {A.}~\bibnamefont {Hosoya}},\ }\bibfield
  {title} {\bibinfo {title} {{Gravitational collapse in Painleve-Gullstrand
  coordinates}},\ }\href {https://doi.org/10.1143/PTP.125.1053} {\bibfield
  {journal} {\bibinfo  {journal} {Prog. Theor. Phys.}\ }\textbf {\bibinfo
  {volume} {125}},\ \bibinfo {pages} {1053} (\bibinfo {year} {2011})},\ \Eprint
  {https://arxiv.org/abs/1008.0470} {arXiv:1008.0470 [gr-qc]} \BibitemShut
  {NoStop}%
\bibitem [{\citenamefont {Blau}(2017)}]{Blau}%
  \BibitemOpen
  \bibfield  {author} {\bibinfo {author} {\bibfnamefont {M.}~\bibnamefont
  {Blau}},\ }\href@noop {} {\bibinfo {title} {Lecture notes on general
  relativity}} (\bibinfo {year} {2017})\BibitemShut {NoStop}%
\bibitem [{\citenamefont {De~Felice}\ \emph {et~al.}(2019)\citenamefont
  {De~Felice}, \citenamefont {Larrouturou}, \citenamefont {Mukohyama},\ and\
  \citenamefont {Oliosi}}]{DeFelice:2019jxs}%
  \BibitemOpen
  \bibfield  {author} {\bibinfo {author} {\bibfnamefont {A.}~\bibnamefont
  {De~Felice}}, \bibinfo {author} {\bibfnamefont {F.}~\bibnamefont
  {Larrouturou}}, \bibinfo {author} {\bibfnamefont {S.}~\bibnamefont
  {Mukohyama}},\ and\ \bibinfo {author} {\bibfnamefont {M.}~\bibnamefont
  {Oliosi}},\ }\bibfield  {title} {\bibinfo {title} {{On the absence of
  conformally flat slicings of the Kerr spacetime}},\ }\href
  {https://doi.org/10.1103/PhysRevD.100.124044} {\bibfield  {journal} {\bibinfo
   {journal} {Phys. Rev. D}\ }\textbf {\bibinfo {volume} {100}},\ \bibinfo
  {pages} {124044} (\bibinfo {year} {2019})},\ \Eprint
  {https://arxiv.org/abs/1908.03456} {arXiv:1908.03456 [gr-qc]} \BibitemShut
  {NoStop}%
\bibitem [{\citenamefont {Markovic}\ and\ \citenamefont
  {Shapiro}(2000)}]{Markovic:1999di}%
  \BibitemOpen
  \bibfield  {author} {\bibinfo {author} {\bibfnamefont {D.}~\bibnamefont
  {Markovic}}\ and\ \bibinfo {author} {\bibfnamefont {S.}~\bibnamefont
  {Shapiro}},\ }\bibfield  {title} {\bibinfo {title} {{Gravitational collapse
  with a cosmological constant}},\ }\href
  {https://doi.org/10.1103/PhysRevD.61.084029} {\bibfield  {journal} {\bibinfo
  {journal} {Phys. Rev. D}\ }\textbf {\bibinfo {volume} {61}},\ \bibinfo
  {pages} {084029} (\bibinfo {year} {2000})},\ \Eprint
  {https://arxiv.org/abs/gr-qc/9912066} {arXiv:gr-qc/9912066} \BibitemShut
  {NoStop}%
\bibitem [{\citenamefont {De~Felice}\ \emph {et~al.}(2011)\citenamefont
  {De~Felice}, \citenamefont {Suyama},\ and\ \citenamefont
  {Tanaka}}]{DeFelice:2011ka}%
  \BibitemOpen
  \bibfield  {author} {\bibinfo {author} {\bibfnamefont {A.}~\bibnamefont
  {De~Felice}}, \bibinfo {author} {\bibfnamefont {T.}~\bibnamefont {Suyama}},\
  and\ \bibinfo {author} {\bibfnamefont {T.}~\bibnamefont {Tanaka}},\
  }\bibfield  {title} {\bibinfo {title} {{Stability of Schwarzschild-like
  solutions in f(R,G) gravity models}},\ }\href
  {https://doi.org/10.1103/PhysRevD.83.104035} {\bibfield  {journal} {\bibinfo
  {journal} {Phys. Rev. D}\ }\textbf {\bibinfo {volume} {83}},\ \bibinfo
  {pages} {104035} (\bibinfo {year} {2011})},\ \Eprint
  {https://arxiv.org/abs/1102.1521} {arXiv:1102.1521 [gr-qc]} \BibitemShut
  {NoStop}%
\bibitem [{\citenamefont {Lipschutz}(2013)}]{lipschutz68}%
  \BibitemOpen
  \bibfield  {author} {\bibinfo {author} {\bibfnamefont {S.}~\bibnamefont
  {Lipschutz}},\ }\href@noop {} {\emph {\bibinfo {title} {Schaum's Outline of
  Theory and Problems of Linear Algebra}}},\ \bibinfo {edition} {5th}\ ed.\
  (\bibinfo  {publisher} {McGraw-Hill},\ \bibinfo {address} {New York},\
  \bibinfo {year} {2013})\BibitemShut {NoStop}%
\bibitem [{\citenamefont {Takahashi}\ and\ \citenamefont
  {Motohashi}(2021)}]{Takahashi:2021bml}%
  \BibitemOpen
  \bibfield  {author} {\bibinfo {author} {\bibfnamefont {K.}~\bibnamefont
  {Takahashi}}\ and\ \bibinfo {author} {\bibfnamefont {H.}~\bibnamefont
  {Motohashi}},\ }\bibfield  {title} {\bibinfo {title} {{Black hole
  perturbations in DHOST theories: master variables, gradient instability, and
  strong coupling}},\ }\href {https://doi.org/10.1088/1475-7516/2021/08/013}
  {\bibfield  {journal} {\bibinfo  {journal} {JCAP}\ }\textbf {\bibinfo
  {volume} {08}},\ \bibinfo {pages} {013}},\ \Eprint
  {https://arxiv.org/abs/2106.07128} {arXiv:2106.07128 [gr-qc]} \BibitemShut
  {NoStop}%
\bibitem [{\citenamefont {Lemaitre}(1933)}]{Lemaitre:1933gd}%
  \BibitemOpen
  \bibfield  {author} {\bibinfo {author} {\bibfnamefont {G.}~\bibnamefont
  {Lemaitre}},\ }\bibfield  {title} {\bibinfo {title} {{The expanding
  universe}},\ }\href {https://doi.org/10.1023/A:1018855621348} {\bibfield
  {journal} {\bibinfo  {journal} {Annales Soc. Sci. Bruxelles A}\ }\textbf
  {\bibinfo {volume} {53}},\ \bibinfo {pages} {51} (\bibinfo {year}
  {1933})}\BibitemShut {NoStop}%
\bibitem [{\citenamefont {Tolman}(1934)}]{Tolman:1934za}%
  \BibitemOpen
  \bibfield  {author} {\bibinfo {author} {\bibfnamefont {R.~C.}\ \bibnamefont
  {Tolman}},\ }\bibfield  {title} {\bibinfo {title} {{Effect of imhomogeneity
  on cosmological models}},\ }\href {https://doi.org/10.1073/pnas.20.3.169}
  {\bibfield  {journal} {\bibinfo  {journal} {Proc. Nat. Acad. Sci.}\ }\textbf
  {\bibinfo {volume} {20}},\ \bibinfo {pages} {169} (\bibinfo {year}
  {1934})}\BibitemShut {NoStop}%
\bibitem [{\citenamefont {Bondi}(1947)}]{Bondi:1947fta}%
  \BibitemOpen
  \bibfield  {author} {\bibinfo {author} {\bibfnamefont {H.}~\bibnamefont
  {Bondi}},\ }\bibfield  {title} {\bibinfo {title} {{Spherically symmetrical
  models in general relativity}},\ }\href
  {https://doi.org/10.1093/mnras/107.5-6.410} {\bibfield  {journal} {\bibinfo
  {journal} {Mon. Not. Roy. Astron. Soc.}\ }\textbf {\bibinfo {volume} {107}},\
  \bibinfo {pages} {410} (\bibinfo {year} {1947})}\BibitemShut {NoStop}%
\end{thebibliography}%
\end{document}